\documentclass[aps,prd,eqsecnum,nofootinbib,twocolumn,showpacs,preprintnumbers,superscriptaddress,floatfix]{revtex4-1}

\usepackage{amsmath,amssymb}
\usepackage{bm} 
\usepackage{calculator}
\usepackage{color}
\usepackage{dsfont} 
\usepackage{CJK}
\usepackage{textcomp,gensymb} 
\usepackage{graphicx}
\usepackage{natbib}
\usepackage{slashed}
\usepackage[caption=false]{subfig}
\usepackage{pbox}
\usepackage[pdftex,colorlinks=true,
    linkcolor=blue,
    filecolor=blue,
    urlcolor=blue,
    citecolor=blue,
    plainpages=false,
    bookmarksopen=true]{hyperref} 

\newcommand{\order}{\ensuremath{\mathrm{O}}}

\newcommand{\onehalf}{\ensuremath{\frac{1}{2}}}
\newcommand{\threehalves}{\ensuremath{\frac{3}{2}}}

\newcommand{\Dfour}{\ensuremath{[(\text{SU}(2)\times\mathds{Z}_4)/\mathds{Z}_2]_{\text{D}_4}}}

\newcommand{\pndme}{Bhattacharya:2013ehc,*Bhattacharya:2015esa,*Bhattacharya:2015wna,*Bhattacharya:2016zcn,*Gupta:2018lvp,%
*Gupta:2018qil,Lin:2018obj,Rajan:2017lxk,*Jang:2019jkn}
\newcommand{\callat}{Bouchard:2016heu,*Berkowitz:2017gql,*Chang:2018uxx,Berkowitz:2017opd,Nicholson:2018mwc}
\newcommand{\allmixed}{Bhattacharya:2013ehc,*Bhattacharya:2015esa,*Bhattacharya:2015wna,*Bhattacharya:2016zcn,*Gupta:2018lvp,%
*Gupta:2018qil,Lin:2018obj,Rajan:2017lxk,*Jang:2019jkn,%
Bouchard:2016heu,*Berkowitz:2017gql,*Chang:2018uxx,Berkowitz:2017opd,Nicholson:2018mwc}

\allowdisplaybreaks

\newcommand{\fermilab}{Fermi National Accelerator Laboratory, Batavia, IL 60510, USA}

\newcommand{\chicago}{University of Chicago, Department of Physics, Chicago, IL 60637, USA}
\newcommand{\bnl}{Brookhaven National Laboratory, Upton, NY 11973, USA}

\usepackage{xcolor}
\definecolor{red}{HTML}{A00000}
\definecolor{pink}{HTML}{FF88FF}
\definecolor{green}{HTML}{307722}
\definecolor{mocha}{HTML}{885500}
\definecolor{cayenne}{HTML}{941100}

\begin{document}

\begin{CJK*}{UTF8}{bsmi} 
	
\title{Nucleon Mass with Highly Improved Staggered Quarks}

\author{{Yin~\surname{Lin}} (林胤)}
\email{yin01@uchicago.edu}
\affiliation{\chicago}\affiliation{\fermilab}

\author{Aaron~S.~\surname{Meyer}}
\email{asmeyer.physics@gmail.com}
\affiliation{\chicago}
\affiliation{\fermilab}
\affiliation{\bnl}

\author{Ciaran \surname{Hughes}}
\email{chughes@fnal.gov}
\affiliation{\fermilab}

\author{{Andreas~S.~\surname{Kronfeld}}}
\email{ask@fnal.gov}
\affiliation{\fermilab}

\author{James~N.~\surname{Simone}}
\email{simone@fnal.gov}
\affiliation{\fermilab}

\author{Alexei~\surname{Strelchenko}}
\email{astrel@fnal.gov}
\affiliation{\fermilab}

\collaboration{Fermilab Lattice Collaboration}
\noaffiliation

\date{\today}

\preprint{FERMILAB-PUB-19-422-T}

\begin{abstract}

We present the first computation in a program of lattice-QCD baryon physics using staggered fermions for sea \emph{and} valence
quarks.
For this initial study, we present a calculation of the nucleon mass, obtaining $964\pm16$~MeV with all sources of statistical and
systematic errors controlled and accounted~for.
This result is the most precise determination to date of the nucleon mass from first~principles.
We use the highly-improved staggered quark action, which is computationally efficient.
Three gluon ensembles are employed, which have approximate lattice spacings $a\approx0.09$~fm, $0.12$~fm, and $0.15$~fm, each with
equal-mass $u$/$d$, $s$, and $c$ quarks in the sea.
Further, all ensembles have the light valence and sea $u$/$d$ quarks tuned to reproduce the physical pion mass, avoiding
complications from chiral extrapolations.
Our work opens a new avenue for precise calculations of baryon properties, which are both feasible and relevant to experiments in
particle and nuclear physics.

\end{abstract}

\maketitle
\end{CJK*}

\newcommand{\cube}{
\draw[densely dotted,lightgray] (0,0,-0) -- (0,0,-1);
\draw[densely dotted,lightgray] (1,0,-1) -- (0,0,-1);
\draw[densely dotted,lightgray] (0,1,-1) -- (0,0,-1);
\draw (1,1,-1) -- (1,1,-0) -- (0,1,-0) -- (0,1,-1);
\draw (1,1,-0) -- (1,0,-0) -- (0,0,-0) -- (0,1,-0);
\draw (1,0,-0) -- (1,0,-1);
\draw (1,0,-1) -- (1,1,-1) -- (0,1,-1);
}

\def\asz{0.1}
\newcommand{\triplet}[3]{
\INTEGERDIVISION{#1}{2}{\tmpa}{\xa}
\INTEGERDIVISION{\tmpa}{2}{\tmpA}{\yA}
\INTEGERDIVISION{\tmpA}{2}{\tmpa}{\za}
\INTEGERDIVISION{#2}{2}{\tmpb}{\xb}
\INTEGERDIVISION{\tmpb}{2}{\tmpB}{\yB}
\INTEGERDIVISION{\tmpB}{2}{\tmpb}{\zb}
\INTEGERDIVISION{#3}{2}{\tmpc}{\xc}
\INTEGERDIVISION{\tmpc}{2}{\tmpC}{\yC}
\INTEGERDIVISION{\tmpC}{2}{\tmpc}{\zc}
\MULTIPLY{\yA}{-1}{\ya}
\MULTIPLY{\yB}{-1}{\yb}
\MULTIPLY{\yC}{-1}{\yc}
\draw (\xa,\za,\ya) circle (\asz);
\draw (\xb-\asz,\zb,\yb) -- (\xb,\zb-\asz,\yb)
   -- (\xb+\asz,\zb,\yb) -- (\xb,\zb+\asz,\yb) -- cycle;
\draw (\xc-\asz,\zc-\asz,\yc) -- (\xc-\asz,\zc+\asz,\yc)
   -- (\xc+\asz,\zc+\asz,\yc) -- (\xc+\asz,\zc-\asz,\yc) -- cycle;
}

\makeatletter
\newcommand\makebig[2]{%
  \@xp\newcommand\@xp*\csname#1\endcsname{\bBigg@{#2}}%
  \@xp\newcommand\@xp*\csname#1l\endcsname{\@xp\mathopen\csname#1\endcsname}%
  \@xp\newcommand\@xp*\csname#1r\endcsname{\@xp\mathclose\csname#1\endcsname}%
}
\makeatother

\def\rs{1.65em}
\makebig{biggg} {3.0}

\newcommand{\GBeigtSon}{
{\cal O}^{S,8,1}_{ \vec{D},\vec{A}\vec{B}\vec{C}}&=
 \left(
  \delta_{\vec{A},\vec{D}}               \delta_{\vec{B},\vec{D}}               \delta_{\vec{C},\vec{D}}               
\right)
\label{eq:GBeigtSon}
}

\newcommand{\GBeigtStw}{
{\cal O}^{S,8,2}_{ \vec{D},\vec{A}\vec{B}\vec{C}}&=
 \frac{1}{\sqrt{3}}\left(
  \delta_{\vec{A},\vec{D}}               \delta_{\vec{B},\vec{D}+\widehat{x}}   \delta_{\vec{C},\vec{D}+\widehat{x}}   
+ \delta_{\vec{A},\vec{D}}               \delta_{\vec{B},\vec{D}+\widehat{y}}   \delta_{\vec{C},\vec{D}+\widehat{y}}   
+ \delta_{\vec{A},\vec{D}}               \delta_{\vec{B},\vec{D}+\widehat{z}}   \delta_{\vec{C},\vec{D}+\widehat{z}}   
\right)
\label{eq:GBeigtStw}
}

\newcommand{\GBeigtSth}{
{\cal O}^{S,8,3}_{ \vec{D},\vec{A}\vec{B}\vec{C}}&=
 \frac{1}{\sqrt{3}}\left(
  \delta_{\vec{A},\vec{D}}               \delta_{\vec{B},\vec{D}+\widehat{yz}}  \delta_{\vec{C},\vec{D}+\widehat{yz}}  
+ \delta_{\vec{A},\vec{D}}               \delta_{\vec{B},\vec{D}+\widehat{xz}}  \delta_{\vec{C},\vec{D}+\widehat{xz}}  
+ \delta_{\vec{A},\vec{D}}               \delta_{\vec{B},\vec{D}+\widehat{xy}}  \delta_{\vec{C},\vec{D}+\widehat{xy}}  
\right)
\label{eq:GBeigtSth}
}

\newcommand{\GBeigtSfv}{
{\cal O}^{S,8,5}_{ \vec{D},\vec{A}\vec{B}\vec{C}}&=
 \left(
  \delta_{\vec{A},\vec{D}+\widehat{xyz}} \delta_{\vec{B},\vec{D}+\widehat{xyz}} \delta_{\vec{C},\vec{D}}               
\right)
\label{eq:GBeigtSfv}
}

\newcommand{\GBeigtSsx}{
{\cal O}^{S,8,6}_{ \vec{D},\vec{A}\vec{B}\vec{C}}&=
 \frac{1}{\sqrt{3}}\left(
  \delta_{\vec{A},\vec{D}+\widehat{xyz}} \delta_{\vec{B},\vec{D}+\widehat{yz}}  \delta_{\vec{C},\vec{D}+\widehat{x}}   
- \delta_{\vec{A},\vec{D}+\widehat{xyz}} \delta_{\vec{B},\vec{D}+\widehat{xz}}  \delta_{\vec{C},\vec{D}+\widehat{y}}   
+ \delta_{\vec{A},\vec{D}+\widehat{xyz}} \delta_{\vec{B},\vec{D}+\widehat{xy}}  \delta_{\vec{C},\vec{D}+\widehat{z}}   
\right)
\label{eq:GBeigtSsx}
}

\newcommand{\GBeprmSfr}{
{\cal O}^{S,8',4}_{ \vec{D},\vec{A}\vec{B}\vec{C}}&=
 \frac{1}{\sqrt{3}}\left(
  \delta_{\vec{A},\vec{D}+\widehat{yz}}  \delta_{\vec{B},\vec{D}+\widehat{z}}   \delta_{\vec{C},\vec{D}+\widehat{y}}   
- \delta_{\vec{A},\vec{D}+\widehat{xz}}  \delta_{\vec{B},\vec{D}+\widehat{z}}   \delta_{\vec{C},\vec{D}+\widehat{x}}   
+ \delta_{\vec{A},\vec{D}+\widehat{xy}}  \delta_{\vec{B},\vec{D}+\widehat{y}}   \delta_{\vec{C},\vec{D}+\widehat{x}}   
\right)
\label{eq:GBeprmSfr}
}

\newcommand{\GBeprmSsv}{
{\cal O}^{S,8',7}_{ \vec{D},\vec{A}\vec{B}\vec{C}}&=
 \left(
  \delta_{\vec{A},\vec{D}+\widehat{yz}}  \delta_{\vec{B},\vec{D}+\widehat{xz}}  \delta_{\vec{C},\vec{D}+\widehat{xy}}  
\right)
\label{eq:GBeprmSsv}
}

\newcommand{\GBsxtpStw}{
{\cal O}^{S,16,2}_{+\vec{D},\vec{A}\vec{B}\vec{C}}&=
 \frac{1}{\sqrt{6}}\left(
  \delta_{\vec{A},\vec{D}}               \delta_{\vec{B},\vec{D}+\widehat{x}}   \delta_{\vec{C},\vec{D}+\widehat{x}}   
+ \delta_{\vec{A},\vec{D}}               \delta_{\vec{B},\vec{D}+\widehat{y}}   \delta_{\vec{C},\vec{D}+\widehat{y}}   
-2\delta_{\vec{A},\vec{D}}               \delta_{\vec{B},\vec{D}+\widehat{z}}   \delta_{\vec{C},\vec{D}+\widehat{z}}   
\right)
\label{eq:GBsxtpStw}
}

\newcommand{\GBsxtdStw}{
{\cal O}^{S,16,2}_{-\vec{D},\vec{A}\vec{B}\vec{C}}&=
 \frac{1}{\sqrt{2}}\left(
  \delta_{\vec{A},\vec{D}}               \delta_{\vec{B},\vec{D}+\widehat{x}}   \delta_{\vec{C},\vec{D}+\widehat{x}}   
- \delta_{\vec{A},\vec{D}}               \delta_{\vec{B},\vec{D}+\widehat{y}}   \delta_{\vec{C},\vec{D}+\widehat{y}}   
\right)
\label{eq:GBsxtdStw}
}

\newcommand{\GBsxtpSth}{
{\cal O}^{S,16,3}_{+\vec{D},\vec{A}\vec{B}\vec{C}}&=
 \frac{1}{\sqrt{6}}\left(
  \delta_{\vec{A},\vec{D}}               \delta_{\vec{B},\vec{D}+\widehat{yz}}  \delta_{\vec{C},\vec{D}+\widehat{yz}}  
+ \delta_{\vec{A},\vec{D}}               \delta_{\vec{B},\vec{D}+\widehat{xz}}  \delta_{\vec{C},\vec{D}+\widehat{xz}}  
-2\delta_{\vec{A},\vec{D}}               \delta_{\vec{B},\vec{D}+\widehat{xy}}  \delta_{\vec{C},\vec{D}+\widehat{xy}}  
\right)
\label{eq:GBsxtpSth}
}

\newcommand{\GBsxtdSth}{
{\cal O}^{S,16,3}_{-\vec{D},\vec{A}\vec{B}\vec{C}}&=
 \frac{1}{\sqrt{2}}\left(
  \delta_{\vec{A},\vec{D}}               \delta_{\vec{B},\vec{D}+\widehat{yz}}  \delta_{\vec{C},\vec{D}+\widehat{yz}}  
- \delta_{\vec{A},\vec{D}}               \delta_{\vec{B},\vec{D}+\widehat{xz}}  \delta_{\vec{C},\vec{D}+\widehat{xz}}  
\right)
\label{eq:GBsxtdSth}
}

\newcommand{\GBsxtpSfr}{
{\cal O}^{S,16,4}_{+\vec{D},\vec{A}\vec{B}\vec{C}}&=
 \frac{1}{\sqrt{2}}\left(
 -\delta_{\vec{A},\vec{D}+\widehat{yz}}  \delta_{\vec{B},\vec{D}+\widehat{z}}   \delta_{\vec{C},\vec{D}+\widehat{y}}   
- \delta_{\vec{A},\vec{D}+\widehat{xz}}  \delta_{\vec{B},\vec{D}+\widehat{z}}   \delta_{\vec{C},\vec{D}+\widehat{x}}   
\right)
\label{eq:GBsxtpSfr}
}

\newcommand{\GBsxtdSfr}{
{\cal O}^{S,16,4}_{-\vec{D},\vec{A}\vec{B}\vec{C}}&=
 \frac{1}{\sqrt{6}}\left(
  \delta_{\vec{A},\vec{D}+\widehat{yz}}  \delta_{\vec{B},\vec{D}+\widehat{z}}   \delta_{\vec{C},\vec{D}+\widehat{y}}   
- \delta_{\vec{A},\vec{D}+\widehat{xz}}  \delta_{\vec{B},\vec{D}+\widehat{z}}   \delta_{\vec{C},\vec{D}+\widehat{x}}   
-2\delta_{\vec{A},\vec{D}+\widehat{xy}}  \delta_{\vec{B},\vec{D}+\widehat{y}}   \delta_{\vec{C},\vec{D}+\widehat{x}}   
\right)
\label{eq:GBsxtdSfr}
}

\newcommand{\GBsxtpSsx}{
{\cal O}^{S,16,6}_{+\vec{D},\vec{A}\vec{B}\vec{C}}&=
 \frac{1}{\sqrt{6}}\left(
  \delta_{\vec{A},\vec{D}+\widehat{xyz}} \delta_{\vec{B},\vec{D}+\widehat{yz}}  \delta_{\vec{C},\vec{D}+\widehat{x}}   
- \delta_{\vec{A},\vec{D}+\widehat{xyz}} \delta_{\vec{B},\vec{D}+\widehat{xz}}  \delta_{\vec{C},\vec{D}+\widehat{y}}   
-2\delta_{\vec{A},\vec{D}+\widehat{xyz}} \delta_{\vec{B},\vec{D}+\widehat{xy}}  \delta_{\vec{C},\vec{D}+\widehat{z}}   
\right)
\label{eq:GBsxtpSsx}
}

\newcommand{\GBsxtdSsx}{
{\cal O}^{S,16,6}_{-\vec{D},\vec{A}\vec{B}\vec{C}}&=
 \frac{1}{\sqrt{2}}\left(
  \delta_{\vec{A},\vec{D}+\widehat{xyz}} \delta_{\vec{B},\vec{D}+\widehat{yz}}  \delta_{\vec{C},\vec{D}+\widehat{x}}   
+ \delta_{\vec{A},\vec{D}+\widehat{xyz}} \delta_{\vec{B},\vec{D}+\widehat{xz}}  \delta_{\vec{C},\vec{D}+\widehat{y}}   
\right)
\label{eq:GBsxtdSsx}
}

\newcommand{\GBeigtMsxon}{
\,_{ijk}{\cal O}^{M,8,6^{1}}_{ \vec{D},\vec{A}\vec{B}\vec{C}}&=
 \frac{1}{\sqrt{3}}\delta_{iu}\delta_{jd}\delta_{ku}\left(
  \delta_{\vec{A},\vec{D}+\widehat{xyz}} \delta_{\vec{B},\vec{D}+\widehat{yz}}  \delta_{\vec{C},\vec{D}+\widehat{x}}   
- \delta_{\vec{A},\vec{D}+\widehat{xyz}} \delta_{\vec{B},\vec{D}+\widehat{xz}}  \delta_{\vec{C},\vec{D}+\widehat{y}}   
+ \delta_{\vec{A},\vec{D}+\widehat{xyz}} \delta_{\vec{B},\vec{D}+\widehat{xy}}  \delta_{\vec{C},\vec{D}+\widehat{z}}   
\right)
\label{eq:GBeigtMsxon}
}

\newcommand{\GBeigtMtw}{
\,_{ijk}{\cal O}^{M,8,2}_{ \vec{D},\vec{A}\vec{B}\vec{C}}&=
 \frac{1}{\sqrt{3}}\delta_{iu}\delta_{ju}\delta_{kd}\left(
  \delta_{\vec{A},\vec{D}}               \delta_{\vec{B},\vec{D}+\widehat{x}}   \delta_{\vec{C},\vec{D}+\widehat{x}}   
+ \delta_{\vec{A},\vec{D}}               \delta_{\vec{B},\vec{D}+\widehat{y}}   \delta_{\vec{C},\vec{D}+\widehat{y}}   
+ \delta_{\vec{A},\vec{D}}               \delta_{\vec{B},\vec{D}+\widehat{z}}   \delta_{\vec{C},\vec{D}+\widehat{z}}   
\right)
\label{eq:GBeigtMtw}
}

\newcommand{\GBeigtMsxtw}{
\,_{ijk}{\cal O}^{M,8,6^{2}}_{ \vec{D},\vec{A}\vec{B}\vec{C}}&=
 \frac{1}{\sqrt{3}}\delta_{iu}\delta_{ju}\delta_{kd}\left(
  \delta_{\vec{A},\vec{D}+\widehat{xyz}} \delta_{\vec{B},\vec{D}+\widehat{yz}}  \delta_{\vec{C},\vec{D}+\widehat{x}}   
- \delta_{\vec{A},\vec{D}+\widehat{xyz}} \delta_{\vec{B},\vec{D}+\widehat{xz}}  \delta_{\vec{C},\vec{D}+\widehat{y}}   
+ \delta_{\vec{A},\vec{D}+\widehat{xyz}} \delta_{\vec{B},\vec{D}+\widehat{xy}}  \delta_{\vec{C},\vec{D}+\widehat{z}}   
\right)
\label{eq:GBeigtMsxtw}
}

\newcommand{\GBeigtMth}{
\,_{ijk}{\cal O}^{M,8,3}_{ \vec{D},\vec{A}\vec{B}\vec{C}}&=
 \frac{1}{\sqrt{3}}\delta_{iu}\delta_{ju}\delta_{kd}\left(
  \delta_{\vec{A},\vec{D}}               \delta_{\vec{B},\vec{D}+\widehat{yz}}  \delta_{\vec{C},\vec{D}+\widehat{yz}}  
+ \delta_{\vec{A},\vec{D}}               \delta_{\vec{B},\vec{D}+\widehat{xz}}  \delta_{\vec{C},\vec{D}+\widehat{xz}}  
+ \delta_{\vec{A},\vec{D}}               \delta_{\vec{B},\vec{D}+\widehat{xy}}  \delta_{\vec{C},\vec{D}+\widehat{xy}}  
\right)
\label{eq:GBeigtMth}
}

\newcommand{\GBeigtMfr}{
\,_{ijk}{\cal O}^{M,8,4^{2}}_{ \vec{D},\vec{A}\vec{B}\vec{C}}&=
 \frac{1}{\sqrt{6}}(\delta_{iu}\delta_{ju}\delta_{kd}+\delta_{iu}\delta_{jd}\delta_{ku})\left(
  \delta_{\vec{A},\vec{D}+\widehat{yz}}  \delta_{\vec{B},\vec{D}+\widehat{z}}   \delta_{\vec{C},\vec{D}+\widehat{y}}   
+ \delta_{\vec{A},\vec{D}+\widehat{xz}}  \delta_{\vec{B},\vec{D}+\widehat{z}}   \delta_{\vec{C},\vec{D}+\widehat{x}}   
+ \delta_{\vec{A},\vec{D}+\widehat{xy}}  \delta_{\vec{B},\vec{D}+\widehat{y}}   \delta_{\vec{C},\vec{D}+\widehat{x}}   
\right)
\label{eq:GBeigtMfr}
}

\newcommand{\GBeigtMfv}{
\,_{ijk}{\cal O}^{M,8,5}_{ \vec{D},\vec{A}\vec{B}\vec{C}}&=
 \delta_{id}\delta_{ju}\delta_{ku}\left(
  \delta_{\vec{A},\vec{D}+\widehat{xyz}} \delta_{\vec{B},\vec{D}+\widehat{xyz}} \delta_{\vec{C},\vec{D}}               
\right)
\label{eq:GBeigtMfv}
}

\newcommand{\GBeprmMfr}{
\,_{ijk}{\cal O}^{M,8',4^{1}}_{ \vec{D},\vec{A}\vec{B}\vec{C}}&=
 \frac{1}{\sqrt{6}}(\delta_{iu}\delta_{ju}\delta_{kd}-\delta_{iu}\delta_{jd}\delta_{ku})\left(
  \delta_{\vec{A},\vec{D}+\widehat{yz}}  \delta_{\vec{B},\vec{D}+\widehat{z}}   \delta_{\vec{C},\vec{D}+\widehat{y}}   
- \delta_{\vec{A},\vec{D}+\widehat{xz}}  \delta_{\vec{B},\vec{D}+\widehat{z}}   \delta_{\vec{C},\vec{D}+\widehat{x}}   
+ \delta_{\vec{A},\vec{D}+\widehat{xy}}  \delta_{\vec{B},\vec{D}+\widehat{y}}   \delta_{\vec{C},\vec{D}+\widehat{x}}   
\right)
\label{eq:GBeprmMfr}
}

\newcommand{\GBsxtpMfron}{
\,_{ijk}{\cal O}^{M,16,4^{1}}_{+\vec{D},\vec{A}\vec{B}\vec{C}}&=
 \frac{1}{\sqrt{12}}(\delta_{iu}\delta_{ju}\delta_{kd}-\delta_{iu}\delta_{jd}\delta_{ku})\left(
  \delta_{\vec{A},\vec{D}+\widehat{yz}}  \delta_{\vec{B},\vec{D}+\widehat{z}}   \delta_{\vec{C},\vec{D}+\widehat{y}}   
+ \delta_{\vec{A},\vec{D}+\widehat{xz}}  \delta_{\vec{B},\vec{D}+\widehat{z}}   \delta_{\vec{C},\vec{D}+\widehat{x}}   
-2\delta_{\vec{A},\vec{D}+\widehat{xy}}  \delta_{\vec{B},\vec{D}+\widehat{y}}   \delta_{\vec{C},\vec{D}+\widehat{x}}   
\right)
\label{eq:GBsxtpMfron}
}

\newcommand{\GBsxtpMsxon}{
\,_{ijk}{\cal O}^{M,16,6^{1}}_{+\vec{D},\vec{A}\vec{B}\vec{C}}&=
 \frac{1}{\sqrt{6}}\delta_{iu}\delta_{jd}\delta_{ku}\left(
  \delta_{\vec{A},\vec{D}+\widehat{xyz}} \delta_{\vec{B},\vec{D}+\widehat{yz}}  \delta_{\vec{C},\vec{D}+\widehat{x}}   
- \delta_{\vec{A},\vec{D}+\widehat{xyz}} \delta_{\vec{B},\vec{D}+\widehat{xz}}  \delta_{\vec{C},\vec{D}+\widehat{y}}   
-2\delta_{\vec{A},\vec{D}+\widehat{xyz}} \delta_{\vec{B},\vec{D}+\widehat{xy}}  \delta_{\vec{C},\vec{D}+\widehat{z}}   
\right)
\label{eq:GBsxtpMsxon}
}

\newcommand{\GBsxtdMfron}{
\,_{ijk}{\cal O}^{M,16,4^{1}}_{-\vec{D},\vec{A}\vec{B}\vec{C}}&=
 \frac{1}{\sqrt{4}}(\delta_{iu}\delta_{ju}\delta_{kd}-\delta_{iu}\delta_{jd}\delta_{ku})\left(
  \delta_{\vec{A},\vec{D}+\widehat{yz}}  \delta_{\vec{B},\vec{D}+\widehat{z}}   \delta_{\vec{C},\vec{D}+\widehat{y}}   
- \delta_{\vec{A},\vec{D}+\widehat{xz}}  \delta_{\vec{B},\vec{D}+\widehat{z}}   \delta_{\vec{C},\vec{D}+\widehat{x}}   
\right)
\label{eq:GBsxtdMfron}
}

\newcommand{\GBsxtdMsxon}{
\,_{ijk}{\cal O}^{M,16,6^{1}}_{-\vec{D},\vec{A}\vec{B}\vec{C}}&=
 \frac{1}{\sqrt{2}}\delta_{iu}\delta_{jd}\delta_{ku}\left(
  \delta_{\vec{A},\vec{D}+\widehat{xyz}} \delta_{\vec{B},\vec{D}+\widehat{yz}}  \delta_{\vec{C},\vec{D}+\widehat{x}}   
+ \delta_{\vec{A},\vec{D}+\widehat{xyz}} \delta_{\vec{B},\vec{D}+\widehat{xz}}  \delta_{\vec{C},\vec{D}+\widehat{y}}   
\right)
\label{eq:GBsxtdMsxon}
}

\newcommand{\GBsxtpMtw}{
\,_{ijk}{\cal O}^{M,16,2}_{+\vec{D},\vec{A}\vec{B}\vec{C}}&=
 \frac{1}{\sqrt{6}}\delta_{iu}\delta_{ju}\delta_{kd}\left(
  \delta_{\vec{A},\vec{D}}               \delta_{\vec{B},\vec{D}+\widehat{x}}   \delta_{\vec{C},\vec{D}+\widehat{x}}   
+ \delta_{\vec{A},\vec{D}}               \delta_{\vec{B},\vec{D}+\widehat{y}}   \delta_{\vec{C},\vec{D}+\widehat{y}}   
-2\delta_{\vec{A},\vec{D}}               \delta_{\vec{B},\vec{D}+\widehat{z}}   \delta_{\vec{C},\vec{D}+\widehat{z}}   
\right)
\label{eq:GBsxtpMtw}
}

\newcommand{\GBsxtpMfrtw}{
\,_{ijk}{\cal O}^{M,16,4^{2}}_{+\vec{D},\vec{A}\vec{B}\vec{C}}&=
 \frac{1}{\sqrt{4}}(\delta_{iu}\delta_{ju}\delta_{kd}+\delta_{iu}\delta_{jd}\delta_{ku})\left(
 -\delta_{\vec{A},\vec{D}+\widehat{yz}}  \delta_{\vec{B},\vec{D}+\widehat{z}}   \delta_{\vec{C},\vec{D}+\widehat{y}}   
- \delta_{\vec{A},\vec{D}+\widehat{xz}}  \delta_{\vec{B},\vec{D}+\widehat{z}}   \delta_{\vec{C},\vec{D}+\widehat{x}}   
\right)
\label{eq:GBsxtpMfrtw}
}

\newcommand{\GBsxtpMsxtw}{
\,_{ijk}{\cal O}^{M,16,6^{2}}_{+\vec{D},\vec{A}\vec{B}\vec{C}}&=
 \frac{1}{\sqrt{6}}\delta_{iu}\delta_{ju}\delta_{kd}\left(
  \delta_{\vec{A},\vec{D}+\widehat{xyz}} \delta_{\vec{B},\vec{D}+\widehat{yz}}  \delta_{\vec{C},\vec{D}+\widehat{x}}   
- \delta_{\vec{A},\vec{D}+\widehat{xyz}} \delta_{\vec{B},\vec{D}+\widehat{xz}}  \delta_{\vec{C},\vec{D}+\widehat{y}}   
-2\delta_{\vec{A},\vec{D}+\widehat{xyz}} \delta_{\vec{B},\vec{D}+\widehat{xy}}  \delta_{\vec{C},\vec{D}+\widehat{z}}   
\right)
\label{eq:GBsxtpMsxtw}
}

\newcommand{\GBsxtdMtw}{
\,_{ijk}{\cal O}^{M,16,2}_{-\vec{D},\vec{A}\vec{B}\vec{C}}&=
 \frac{1}{\sqrt{2}}\delta_{iu}\delta_{ju}\delta_{kd}\left(
  \delta_{\vec{A},\vec{D}}               \delta_{\vec{B},\vec{D}+\widehat{x}}   \delta_{\vec{C},\vec{D}+\widehat{x}}   
- \delta_{\vec{A},\vec{D}}               \delta_{\vec{B},\vec{D}+\widehat{y}}   \delta_{\vec{C},\vec{D}+\widehat{y}}   
\right)
\label{eq:GBsxtdMtw}
}

\newcommand{\GBsxtdMfrtw}{
\,_{ijk}{\cal O}^{M,16,4^{2}}_{-\vec{D},\vec{A}\vec{B}\vec{C}}&=
 \frac{1}{\sqrt{12}}(\delta_{iu}\delta_{ju}\delta_{kd}+\delta_{iu}\delta_{jd}\delta_{ku})\left(
  \delta_{\vec{A},\vec{D}+\widehat{yz}}  \delta_{\vec{B},\vec{D}+\widehat{z}}   \delta_{\vec{C},\vec{D}+\widehat{y}}   
- \delta_{\vec{A},\vec{D}+\widehat{xz}}  \delta_{\vec{B},\vec{D}+\widehat{z}}   \delta_{\vec{C},\vec{D}+\widehat{x}}   
-2\delta_{\vec{A},\vec{D}+\widehat{xy}}  \delta_{\vec{B},\vec{D}+\widehat{y}}   \delta_{\vec{C},\vec{D}+\widehat{x}}   
\right)
\label{eq:GBsxtdMfrtw}
}

\newcommand{\GBsxtdMsxtw}{
\,_{ijk}{\cal O}^{M,16,6^{2}}_{-\vec{D},\vec{A}\vec{B}\vec{C}}&=
 \frac{1}{\sqrt{2}}\delta_{iu}\delta_{ju}\delta_{kd}\left(
  \delta_{\vec{A},\vec{D}+\widehat{xyz}} \delta_{\vec{B},\vec{D}+\widehat{yz}}  \delta_{\vec{C},\vec{D}+\widehat{x}}   
+ \delta_{\vec{A},\vec{D}+\widehat{xyz}} \delta_{\vec{B},\vec{D}+\widehat{xz}}  \delta_{\vec{C},\vec{D}+\widehat{y}}   
\right)
\label{eq:GBsxtdMsxtw}
}

\newcommand{\GBsxtpMth}{
\,_{ijk}{\cal O}^{M,16,3}_{+\vec{D},\vec{A}\vec{B}\vec{C}}&=
 \frac{1}{\sqrt{6}}\delta_{iu}\delta_{ju}\delta_{kd}\left(
  \delta_{\vec{A},\vec{D}}               \delta_{\vec{B},\vec{D}+\widehat{yz}}  \delta_{\vec{C},\vec{D}+\widehat{yz}}  
+ \delta_{\vec{A},\vec{D}}               \delta_{\vec{B},\vec{D}+\widehat{xz}}  \delta_{\vec{C},\vec{D}+\widehat{xz}}  
-2\delta_{\vec{A},\vec{D}}               \delta_{\vec{B},\vec{D}+\widehat{xy}}  \delta_{\vec{C},\vec{D}+\widehat{xy}}  
\right)
\label{eq:GBsxtpMth}
}

\newcommand{\GBsxtdMth}{
\,_{ijk}{\cal O}^{M,16,3}_{-\vec{D},\vec{A}\vec{B}\vec{C}}&=
 \frac{1}{\sqrt{2}}\delta_{iu}\delta_{ju}\delta_{kd}\left(
  \delta_{\vec{A},\vec{D}}               \delta_{\vec{B},\vec{D}+\widehat{yz}}  \delta_{\vec{C},\vec{D}+\widehat{yz}}  
- \delta_{\vec{A},\vec{D}}               \delta_{\vec{B},\vec{D}+\widehat{xz}}  \delta_{\vec{C},\vec{D}+\widehat{xz}}  
\right)
\label{eq:GBsxtdMth}
}

\newcommand{\GBsxtpMsv}{
\,_{ijk}{\cal O}^{M,16,7}_{+\vec{D},\vec{A}\vec{B}\vec{C}}&=
 \frac{1}{\sqrt{2}}\delta_{id}\delta_{ju}\delta_{ku}\left(
  \delta_{\vec{A},\vec{D}+\widehat{yz}}  \delta_{\vec{B},\vec{D}+\widehat{xz}}  \delta_{\vec{C},\vec{D}+\widehat{xy}}  
- \delta_{\vec{A},\vec{D}+\widehat{xz}}  \delta_{\vec{B},\vec{D}+\widehat{yz}}  \delta_{\vec{C},\vec{D}+\widehat{xy}}  
\right)
\label{eq:GBsxtpMsv}
}

\newcommand{\GBsxtdMsv}{
\,_{ijk}{\cal O}^{M,16,7}_{-\vec{D},\vec{A}\vec{B}\vec{C}}&=
 \frac{1}{\sqrt{2}}\delta_{id}\delta_{ju}\delta_{ku}\left(
  \delta_{\vec{A},\vec{D}+\widehat{yz}}  \delta_{\vec{B},\vec{D}+\widehat{xz}}  \delta_{\vec{C},\vec{D}+\widehat{xy}}  
+ \delta_{\vec{A},\vec{D}+\widehat{xz}}  \delta_{\vec{B},\vec{D}+\widehat{yz}}  \delta_{\vec{C},\vec{D}+\widehat{xy}}  
\right)
\label{eq:GBsxtdMsv}
}

\section{Introduction}

Lattice-QCD calculations have entered a precision era, with total uncertainties below one percent for some simple properties of
mesons and Standard Model parameters that can be determined from them~\cite{Dowdall:2013rya,Bazavov:2014wgs,Bazavov:2017lyh,%
Bazavov:2018omf,Lytle:2018evc,Bazavov:2018kjg}.
It is important for interpreting experiments in nuclear and particle physics to extend such precision calculations to nucleon
properties.
For example, various nucleon expectation values (with no momentum transfer) are needed for precision nucleon beta decay (scalar and
tensor charges), direct dark matter detection (sigma terms), and high-energy scattering (moments of parton distribution
functions)~\cite{Detmold:2018qcd,Davoudi:2018qcd}.
With nonzero momentum transfer, there are form factors pertinent to lepton-nucleon scattering~\cite{Kronfeld:2018qcd}.
In particular, lattice-QCD calculations of vector-current form factors can be compared to measurements in electron-nucleon
scattering, while very similar calculations of (nucleon) axial-current form factors are needed as inputs to the analysis of
neutrino-nucleus scattering.

To carry out a lattice QCD calculation one must first choose a discretization for the quarks and gluons.
Because of the doubling problem of lattice fermion fields, the quarks are the more complicated consideration.
The precise meson-sector calculations referred to above employ the ``highly improved staggered quark'' (HISQ)
action~\cite{HISQAction}.
Not only are the discretization effects small (by design and, it turns out, in
practice~\cite{Donald:2012ga,McNeile:2011ng,Davies:2010ip}), but also the MILC collaboration has generated two dozen ensembles of
$\text{SU}(3)$ gauge fields with 2+1+1 flavors of sea quarks (where ``2'' implies the up and down quarks are chosen to have equal
mass, and the strange- and charm-quark masses are tuned close to their physical value).
The MILC HISQ ensembles~\cite{MILC:Configs,*tastesplitting,Bazavov:2017lyh} have four lattice spacings ($a\approx0.15$, 0.12, 0.09,
0.06~fm) with pion masses near 135~MeV, 210~MeV, and 300~MeV, a fifth ($a\approx0.042$~fm) at 135 and 300~MeV, plus a sixth
($a\approx0.03$~fm) at 300~MeV only.
It is worth investigating how useful these ensembles are for nucleon physics.
Here we present our first step in this direction: a calculation of the nucleon mass employing the HISQ action for the valence quarks
and using the MILC HISQ ensembles with physical pion masses, which have lattice spacings ranging from $a\approx0.15$, 0.12, and
0.09~fm.
The valence masses are chosen equal to the equal-mass light pair in the sea.
An advantage of using only the physical-pion ensembles is that we do not need to extrapolate unphysical-pion-mass data to the
physical limit.

These ensembles have already been used for nucleon physics
 using different fermion formulations for the valence quarks.
One set of papers uses Wilson fermions with the clover action for the valence quarks~\cite{\pndme};
 another set uses gradient-flowed M\"obius domain-wall fermions for the valence quarks~\cite{\callat}.

Any simulation with rooted staggered fermions violates unitarity at order~$a^2$
 \cite{Bernard:1993sv,Bernard:2006zw},
 but the mixed-action simulations of Refs.~\cite{\allmixed} introduce further
 violations~\cite{Bar:2002nr,Bowler:2004hs,Bar:2005tu,Chen:2005ab}.
On the other hand, if the same staggered action is used for valence quarks as is used in
 the sea-quark determinant, no new unitarity
 violations arise~\cite{Bernard:1993sv,Bernard:2006zw,Bernard:2013eya}.
These properties can be seen in chiral perturbation theory, where mixed-action
 simulations require more low-energy constants than the HISQ-on-HISQ setup used here.
Thus, our setup is simpler than those of Refs.~\cite{\allmixed}.

The challenge for an all-staggered calculation stems from the remaining doubling: one staggered fermion field yields four Dirac
fermions.
The quantum number labeling the four species is known as ``taste''.
In the continuum, infinite-volume limit, $\text{SU}(4)$ taste and $\text{SO}(4)$ spacetime symmetries are expected to become
separately exact.
At nonzero lattice spacing, however, the taste-rotation symmetry group (of the transfer matrix) is a finite group lying in a
diagonal subgroup of
$\text{SU}(4)\times\text{SO}(4)$~\cite{vandenDoel:1983mf,*Golterman:1984cy,*Golterman:1985dz,Golterman:1984dn,Kilcup:1986dg}.
Consequently, it is complicated to construct staggered-baryon creation and annihilation operators~\cite{Golterman:1984dn},
especially when isospin and strangeness are incorporated~\cite{Bailey:2006zn}.
These complications need to be confronted only once for each correlation function, after which one can study whether the all-HISQ
formulation is promising for simple quantities (e.g., masses and form factors).
If successful, increasingly more complicated quantities can be determined.
Here, we start with the nucleon mass.

This paper is organized as follows: in Sec.~\ref{sec:StaggeredBaryons} the construction of the staggered baryon irreducible
interpolating operators is discussed, in Sec.~\ref{sec:SimDetails} the simulation details are given, and in
Sec.~\ref{sec:FittingMethods} our different fitting methodologies are described.
Specific details about fitting the staggered nucleon two-point correlators is given in Sec.~\ref{sec:DataAnalysis}, and these
results are combined with all sources of systematic errors in Sec.~\ref{sec:NucleonDetermination} to produce a final estimate of the
nucleon mass.
We then discuss our conclusions in Sec.~\ref{sec:conclusions}.
Appendices~\ref{app:Group}, \ref{app:irrepident}, and~\ref{appendix:staggeredop} spell out the group-theoretic construction of
staggered baryon operators in detail.
Appendix~\ref{appendix:moredata} contains additional information about the Bayesian fits discussed in Sec.~\ref{sec:DataAnalysis},
namely the priors and posteriors from the Bayesian fits in Sec.~\ref{sec:DataAnalysis} and additional plots.

\section{Baryons Built with Staggered Fermions}
\label{sec:StaggeredBaryons}

Here we outline the construction of our baryon operators using staggered quarks, and refer the reader to Appendices~\ref{app:Group},
\ref{app:irrepident}, and~\ref{appendix:staggeredop} for more technical details.

With staggered fermions~\cite{Susskind:1976jm}, the doubling problem is partly solved.
The simplest discretization contains a set of ``doubling symmetries'' that, in the end, imply that a single lattice fermion field
corresponds to 16 Dirac fermions in the continuum limit.
A subset of doubling symmetries can be simultaneously diagonalized~\cite{Kawamoto:1981hw,Sharatchandra:1981si}, leaving four
identical, decoupled one-component fields.
Three of these four copies can simply be removed, leaving four tastes instead of 16.
This procedure retains one exact axial symmetry, which is a nonsinglet with respect to taste.
This remnant is enough to ensure several important consequences of chiral symmetry: if the bare mass vanishes, the pion mass
vanishes; renormalization constants related by chiral symmetry are equal, etc.

On the other hand, the translational and doubling symmetries are not separate after this diagonalization.
The lattice action is invariant under a composition of the two known as ``shifts'', which multiply the fermion field with a sign
that depends on the originating site and direction of the translation.

\subsection{Staggered Baryon Quantum Numbers}

The diagonalization of the doubling symmetries leads to an intimate relationship between the spin-taste quantum numbers of a hadron
creation/annihilation operator and the spatial distribution of the constituent quark fields.
For computing masses and matrix elements, the relevant symmetry group is the subgroup of lattice transformations restricted to a
single timeslice.
This group is called the geometric timeslice group~\cite{Golterman:1984dn}, and denoted ``GTS''.
For a meson bilinear, operators that transform irreducibly under GTS can be constructed by fixing both the relative displacement
between the quark and antiquark in the bilinear, in combination with fixing the relative signs between the bilinear from one lattice
site to its neighbors.
Rotation symmetries interchange the staggered phases identified with the sites, and shift symmetries induce phase changes in meson
operators.

For baryons, we need the fermionic irreducible representations (irreps), which are much more complicated.
GTS has three fermionic irreducible representations, labeled $8$, $8'$, and $16$, which are simply the dimensions of the irreps.
The staggered quark field with zero momentum transforms irreducibly under the~$8$, where the 8 elements of the representation map
onto the 8 vertices of a spatial unit cube.
Shift transformations and rotations interchange quark fields at the unit-cube sites, and also change the
phases of the sites relative to one another.

The characters of the $8'$~irrep are mostly the same as those of the 8~irrep, \emph{except} that the characters of the rotations by
$\pi/2$ about a lattice axis have the opposite sign.
The 16~irrep is less simple to describe.
Under shifts, the sixteen elements split into two disjoint subsets of eight elements each.
Under rotations, the elements map onto linear combinations of elements from both
subsets, albeit in a convention-dependent way. 
One can choose a convention such that rotations about a single lattice axis only permute and sign-flip elements
within each separate subset.
The characters of 16~irrep are obtained from those of the 8~irrep via a tensor product with the irrep of
$\mathds{Q}_8\rtimes\text{SW}_3$ induced from the $E$~irrep of~$\text{SW}_3$~\cite{Kilcup:1986dg}.

The mapping of the continuum spin representations to the lattice octahedral representations is given in
Table~\ref{tab:octspincontent}.
\begin{table}
    \centering
    \caption{Continuum spin irrep decompositions into lattice octahedral irreps when excluding and including the taste symmetry
        group.}
    \label{tab:octspincontent}
    \begin{tabular}{cc@{\qquad}cc@{\quad}c}
        \hline\hline
        Spin & Lattice                   &  Spin & Lattice    &   Lattice    \\
             &                           &       & w/o shifts & with shifts \\
        \hline
        0    & $A_1$                     & $1/2$ & $G_1$                   & 8 \\
        1    & $T_1$                     & $3/2$ & $H$                     & 16 \\
        2    & $E\oplus T_2$             & $5/2$ & $G_2\oplus H$           & $8'\oplus 16$\\
        3    & $A_2\oplus T_1\oplus T_2$ & $7/2$ & $G_1\oplus G_2\oplus H$ & $8 \oplus 8'\oplus 16$ \\
        \hline\hline
\end{tabular}
\end{table}
There is a one-to-one correspondence between the 8, $8'$, and 16 representations of the GTS group to the conventionally named $G_1$,
$G_2$, and $H$ representations of the double cover of the cubic rotation group~\cite{JOHNSON1982147}.
This mapping is a consequence of the reduction of the continuum taste symmetry group $\text{SU}(4)$ to the Clifford group
$\Gamma_4=(\mathds{Q}_8\times \text{D}_4)/\mathds{Z}_2$~\cite{Kilcup:1986dg,Meyer:2017ddy}.
Here, $\mathds{Q}_8$ is the order-8 quaternion group and $\text{D}_4$ is the order-8 dihedral group.
Both $\mathds{Q}_8$ and $\text{D}_4$ only have one fermionic representation, which in both cases is 2-dimensional.
The method of induced representations tells us that these fermionic representations are the only representations that can appear
when lifting a fermionic representation from the octahedral group to the GTS group.
In this way, the only modification of the $G_1/G_2/H$ representations when including the taste symmetry is to increase the dimension
of these representations from $2/2/4$ to $8/8'/16$.

In phenomenological models, the structure of baryonic wave functions is typically understood by embedding the $\text{SU}(2)$ spin
and $\text{SU}(3)$ flavor groups into an $\text{SU}(6)$ symmetry group.
As baryonic wave functions need to be overall antisymmetric, and the color component is antisymmetric, it is necessary to isolate
the representations of $\text{SU}(6)$ that are symmetric combinations of spin and flavor.

This embedding procedure may also be performed for the staggered fermions by combining the $\text{SU}(4)$ taste symmetry group with
the aforementioned spin and flavor groups.
These three groups are thus embedded into $\text{SU}(24)$~\cite{Bailey:2006zn}.
The representations of $\text{SU}(24)$ are decomposed back into $\text{SU}(2)\times\text{SU}(3)\times\text{SU}(4)$ in order to
understand the symmetry structure of the resulting operators.
This procedure yields the usual baryon octet and decuplet (paired with symmetric taste representations) as well as several other
representations that are mixed-symmetric or antisymmetric in taste.

Notably, it is possible to construct baryon operators that are nonsymmetric in taste.
These operators are also nonsymmetric in spin and flavor, in such a way that, when combined with the antisymmetric color component,
the overall baryon wave function is antisymmetric.
Such operators are not part of the physical world without taste.
For example, in the physical world, symmetrizing over spin and isospin only allows for combinations of spin \onehalf\ (\threehalves)
with isospin \onehalf\ (\threehalves), which correspond to the nucleon ($\Delta$).
With the additional taste symmetry, it is also possible to build operators with nontrivial symmetrization over tastes, which
create states that have isospin \threehalves\ with spin \onehalf, and vice versa.
Further, Bailey~\cite{Bailey:2006zn} shows that in the continuum limit each of these additional states lies in a multiplet that is
related by a $\text{SU}(12)$ flavor-taste symmetry transformation to the physical nucleon or $\Delta$ states.
Consequently, these taste nonsymmetric (iso)spin (\threehalves) \onehalf\ representations must give identical physics in the
continuum limit to the physical nucleon or $\Delta$ baryon.
We refer to these states as ``$N$-like'' or ``$\Delta$-like''.

It should be emphasized that the $N$-like and $\Delta$-like multiplets are distinct in
the continuum limit.
The continuum taste symmetry transformations are local operations insensitive to the finite spatial extent, even though the finite
volume breaks spin symmetry.
As a consequence, the mixing of the baryon spin-taste spectra on the lattice is entirely due to taste-breaking effects that enter at
$\order(\alpha_sa^2)$ for the improved staggered action used here.
The same separation applies to the excited-state spectra as well, meaning that the $N$-like and $\Delta$-like states may be
considered to have their own sets of excited states that mix only at $\order(\alpha_sa^2)$.

In each irrep of GTS, multiple taste partners of the same baryon can contribute, which we call multiplicity of tastes, e.g., three
$N$-like tastes lie in the~8.
The expected multiplicity of the lowest-lying multiplet (i.e., not orbital or radial excitations) of $N$-like and $\Delta$-like
states are given in Table~\ref{tab:gtsmultiplicities}.
\begin{table}
    \centering
    \caption{Multiplicities of the $N$-like and $\Delta$-like states in each GTS irrep for a given isospin.
        Refer to the text for an explanation of how $I=\threehalves$ combines with taste to give a nucleon-like state.}
    \label{tab:gtsmultiplicities}
    \setlength{\tabcolsep}{6pt}
    \begin{tabular}{ccc}
        \hline\hline
        Irrep & $I={3}/{2}$ & $I={1}/{2}$ \\
        \hline
         $8$  & $3N+2\Delta$ & $5N+1\Delta$ \\
         $8'$ & $0N+2\Delta$ & $0N+1\Delta$ \\
         $16$ & $1N+3\Delta$ & $3N+4\Delta$ \\
        \hline\hline
    \end{tabular}
\end{table}
Excited multiplets are expected to have the same taste multiplicities if they share the same particle content and $J^P$ quantum
numbers.
In the numerical work presented below, we only use the 16 irrep of the isospin-\threehalves\ constructions of
Table~\ref{tab:gtsmultiplicities}, because the $16$ irrep only has a single $N$-like state, whereas the 8 has three and the $8'$ has
none.

\subsection{Interpolating Operator Construction}
\label{subsec:OpConstruct}

Here we give an overview of our interpolating operator construction and refer the reader to Ref.~\cite{Bailey:2006zn} and
Appendix~\ref{appendix:staggeredop} for more technical details.
When constructing staggered-baryon interpolating operators, it is easier to work on one timeslice with combinations of quark fields
that are defined by their displacement from the origin modulo~2.
There are eight sets of these quark combinations which can be labeled by a unit-cube corner $\vec{A}$ with $A_\ell \in \{0,1\}$ and
$\ell \in \{1,2,3\}$.
We refer to these objects as ``corner walls'', which are conventional wall sources projected to a single corner in every unit
cube, and write them as
\begin{equation}
    \chi^a_{\vec{A}} = \sum_{\vec{x}@\vec{A}} \chi^a(\vec{x}) .
    \label{eq:sumquark}
\end{equation}
The sum over ``$\vec{x}@\vec{A}$'' is defined by summing $\vec{x}$ over all sites on a timeslice that are displaced modulo~2 from
the origin by the vector $\vec{A}$:
\begin{equation}
    \sum_{\vec{x}@\vec{A}} f(\vec{x}) \equiv
    \sum_{\vec{y}}^{(N_s/2)^3} \sum_{\vec{x}}^{N_s^3} \delta_{2\vec{y}+\vec{A},\vec{x}} f(\vec{x})
\end{equation}
for some general function $f(\vec{x})$.
Here, $N_s$ is the (even) number of sites in a spatial direction.
These corner wall sources must be connected by Wilson lines to construct gauge-invariant operators, which may be accomplished with
explicit insertions or implicitly with gauge fixing.

Interpolating operators with nontrivial taste quantum numbers have two or more quarks at different spatial sites.
To make these operators gauge invariant, parallel transporters must be inserted to connect the quarks at different spatial sites.
Our parallel transporters are defined as (with color indices suppressed)
\begin{equation}
    \overset{\leftrightarrow}{U}_i (\vec{x},\vec{x}')
    = \frac{1}{2} \left[ U_i (\vec{x}) \delta_{\vec{x},\vec{x}'-\hat{i}}
    + U_i^{\dagger} (\vec{x}-\hat{i}) \delta_{\vec{x},\vec{x}'+\hat{i}} \right] ,
\end{equation}
which have an equal sum of links in both the positive and negative directions away from a lattice site, ensuring the operators have
simple transformations under discrete rotations.

The parallel transporters from perpendicular directions are chained together to build operators that obey the full set of spin-taste
symmetries allowed by the staggered lattice symmetry group.
The dressing with parallel transporters may be denoted with a vector $\vec{B}$, with $B_\ell \in \{0,1\}$ and $\ell \in \{1,2,3\}$,
indicating the displacement away from the starting site.
We write these eight parallel transport dressings with the condensed notation,
\begin{widetext}
\begin{align}
    V_{\vec{B}}(\vec{x},\vec{y})
        = \left\{ \begin{array}{c@{\;\text{when } \displaystyle\sum_\ell B_\ell =\;}l}
        \delta_{\vec{x},\vec{y}}, &  0 \\
        \overset{\leftrightarrow}{U}_i (\vec{x},\vec{y}), &  1, B_i = 1 \\
        \frac{1}{2} \displaystyle\sum_{\vec{x}'} \displaystyle\sum_{i\neq j}
           \overset{\leftrightarrow}{U}_i (\vec{x},\vec{x}')
           \overset{\leftrightarrow}{U}_j (\vec{x}',\vec{y}), & 2 , B_i = B_j = 1 \\
        \frac{1}{6} \displaystyle\sum_{\vec{x}',\vec{x}''} \displaystyle\sum_{i\neq j\neq k}
           \overset{\leftrightarrow}{U}_i (\vec{x},\vec{x}')
           \overset{\leftrightarrow}{U}_j (\vec{x}',\vec{x}'')
           \overset{\leftrightarrow}{U}_k (\vec{x}'',\vec{y}), & 3 \end{array} \right.  .
    \label{eq:paralleltransport}
\end{align}
\end{widetext}
To construct gauge invariant operators, the quark fields are dressed with these parallel transporters:
\begin{equation}
    \widetilde{\chi}^a_{\vec{A},\vec{B}}(\vec{x})
    = \sum_{\vec{y}@\vec{A}}\sum_{b} V^{ab}_{\vec{B}}(\vec{x},\vec{y}) \chi^b(\vec{y}),
    \label{eq:dressedquark}
\end{equation}
 where $a$ and $b$ are color indices.

A baryon operator, $\mathcal{B}$, is a quark trilinear that is overall antisymmetric when considering color, flavor, spin, and
taste.
Because color is antisymmetric, $\mathcal{B}$ must be symmetric under simultaneous interchange of the flavor and spin-taste of any
two quarks, namely,
\begin{equation}
    \mathcal{B}_{\vec{D},\vec{A}\vec{B}\vec{C}}^{ijk} = \mathcal{B}_{\vec{D},\vec{B}\vec{A}\vec{C}}^{jik}
        = \mathcal{B}_{\vec{D},\vec{A}\vec{C}\vec{B}}^{ikj} ,
\end{equation}
where the flavor indices $i$, $j$, $k$, and the spin-taste unit-cube indices $\vec{A}$, $\vec{B}$, $\vec{C}$ are for the three
quarks in the baryon.
The remaining unit-cube index $\vec{D}$ is discussed below.

As mentioned above, we restrict ourselves to the isospin $I={3}/{2}$ representations of a baryon in this paper.
As the $I=\threehalves$ irrep built from three $I=\onehalf$ irreps is completely symmetric, in the following we drop the flavor
indices for clarity.
Then baryon operators are constructed as
\begin{align}
    \mathcal{B}_{\vec{D},\vec{A}\vec{B}\vec{C}}(\vec{x}) &= \frac{1}{6} \displaystyle\sum_{abc} \epsilon^{abc}
    \nonumber \\
    &\times
    \widetilde{\chi}^a_{\vec{D}+\vec{A},\vec{A}}(\vec{x})
    \widetilde{\chi}^b_{\vec{D}+\vec{B},\vec{B}}(\vec{x})
    \widetilde{\chi}^c_{\vec{D}+\vec{C},\vec{C}}(\vec{x}) .
    \label{eq:sinkop}
\end{align}
Here, the index $\vec{D}$ is the site where all parallel transporters meet.
Next, the baryon operators must be symmetrized over the flavor and spin-taste unit-cube site indices.
The unit-cube sites of the quarks are symmetrized via
\begin{align}
    \mathcal{S}_{\vec{D},\vec{A}\vec{B}\vec{C}} = \frac{1}{6} \left(
        \mathcal{B}_{\vec{D},\vec{A}\vec{B}\vec{C}} + \mathcal{B}_{\vec{D},\vec{B}\vec{A}\vec{C}} +
        \mathcal{B}_{\vec{D},\vec{C}\vec{A}\vec{B}} \right.
    \nonumber\\ \left.\quad {}+
        \mathcal{B}_{\vec{D},\vec{C}\vec{A}\vec{B}} + \mathcal{B}_{\vec{D},\vec{B}\vec{C}\vec{A}} +
        \mathcal{B}_{\vec{D},\vec{C}\vec{B}\vec{A}} \right) .
 \label{eq:symmetrizedbaryon}
\end{align}

To build interpolating operators that transform in a desired irreducible representation of GTS, one must define appropriate tensors
and contract them with the symmetrized baryon operators $\mathcal{S}_{\vec{D},\vec{A}\vec{B}\vec{C}}$ in
Eq.~(\ref{eq:symmetrizedbaryon}).
For the fermionic irreps of the GTS symmetry group, these tensors can be written as
$\mathcal{O}^{R}_{s\vec{D},\vec{A}\vec{B}\vec{C}}$, where $R$ is an index distinguishing different irreducible representations and
spatial combinations, and $s$ is an additional index for the 16~irrep.
It trivially takes one value for the two 8-dimensional irreps, and in the 16-dimensional irrep $s=\pm1$.
The spin-taste unit-cube index $\vec{D}$ serves as the component of the irrep.
In Appendix~\ref{appendix:staggeredop}, we give explicit formulas for the $\mathcal{O}^{R}_{s\vec{D},\vec{A}\vec{B}\vec{C}}$ that we
use in this paper.

Finally, the baryon interpolating operator that transforms within a definite GTS irrep is 
\begin{equation}
    B^{R}_{s\vec{D}}(\vec{x}, t) = \sum_{\vec{A}\vec{B}\vec{C}}
        \mathcal{O}^{R}_{s\vec{D},\vec{A}\vec{B}\vec{C}}
        \mathcal{S}_{\vec{D},\vec{A}\vec{B}\vec{C}} (\vec{x}, t) ,
    \label{eq:baryop}
\end{equation}
where $\vec{D}$ and $s\vec{D}$ denote the representation index for irrep 8, $8'$, or 16,
 so $\vec{D}$ is not summed over.

The antibaryon operator is a similarly symmetrized trilinear, but with antiquarks rather than quarks.
We use the conjugate of the corner-wall construction in Eq.~(\ref{eq:sumquark}), without parallel transporters, rather than
Eq.~(\ref{eq:dressedquark}).
We denote this object as $\overline{\mathcal{S}}$ after replacing $\tilde{\chi}_{\vec{A},\vec{B}}(\vec{x})\to\bar{\chi}_{\vec{A}}$
(with no $\vec{B}$ or $\vec{x}$ dependence).
These operators are
\begin{equation}
    \overline{B}^{\bar{R}}_{s\vec{D}}(t) = \sum_{\vec{A}\vec{B}\vec{C}} \mathcal{O}^{\bar{R}}_{s\vec{D},\vec{A}\vec{B}\vec{C}}
        \overline{\mathcal{S}}_{\vec{A}\vec{B}\vec{C}}(t) .
    \label{eq:srcop}
\end{equation}
We retain the spatial dependence in $B^{R}_{s\vec{D}}$, but not in $\overline{B}^{\bar{R}}_{s\vec{D}}$, because in
Sec.~\ref{sec:SimDetails} we use them as sink and source, respectively.

As described in Appendix~\ref{appendix:staggeredop}, operators in each irrep can be obtained from distinct ``classes'' of the three
quark fields.
Here, ``class'' is shorthand~\cite{Golterman:1984dn,Bailey:2006zn} for distinct spatial distributions of the three quark fields
within the unit cube.
As operators with identical quantum numbers, these classes of operators all excite the same $N$-like or $\Delta$-like states.%
\footnote{Classes do not arise for staggered mesons, because the decomposition of $8\otimes8$ contains at most one copy of any
irrep.
On the other hand, the decomposition of $8\otimes8\otimes8$ contains multiplicity 20, 4, and 20 for the 8, $8'$, and 16,
respectively.} %
As described in Appendix~\ref{appendix:staggeredop}, in this work we use the four classes of the $I=\threehalves$, GTS~16 irrep,
which are labeled as class~2, 3, 4, and 6~\cite{Golterman:1984dn,Bailey:2006zn}.
With any of these operators at the source or sink, we consequently have a $4\times 4$ matrix correlation function.

\section{Simulation Details}
\label{sec:SimDetails}

We use gauge-field configurations generated by the MILC collaboration~\cite{MILC:Configs,*tastesplitting,Bazavov:2017lyh}.
For the gauge fields, they employed the one-loop tadpole-improved L\"{u}scher-Weisz gauge action improved through
$\order(\alpha_sa^2)$ \cite{Hart:GluonImprovement} and included $2+1+1$ flavors in the sea, the up and down quarks (with equal
mass $m_l$), the strange quark ($m_s$), and the charm quark ($m_c$).
For the sea quarks, MILC employed the HISQ action~\cite{HISQAction}, also improved to $\text{O}(\alpha_s a^2)$ by removing one-loop
taste-changing processes.
For valence quarks, we use the HISQ action with the same bare masses as their sea counterparts.
This choice introduces no additional unitarity violations from a mixed action~\cite{Bernard:1993sv,Bernard:2006zw,Bernard:2013eya}.
Further, the remnant chiral symmetry ensures there are no unwanted near-zero modes in the propagators at nonzero quark
mass~\cite{Bardeen:1997gv}.

To enable a continuum extrapolation, we choose three ensembles, with lattice spacings in the range $a\approx 0.09$--0.15~fm.
Details of these ensembles are given in Table~\ref{tab:ensembledetails}.
\begin{table}
    \centering
    \caption{Details of the gauge-field ensembles used in this study.
        $\beta$ is the gauge coupling; 
        $a$~(fm) is the lattice spacing in a mass-independent $f_{p4s}$ scheme~\cite{Bazavov:2017lyh};
        $am_l$, $am_s$, and $am_c$ are the sea quark masses;
        $N_s \times N_T$ gives the spatial and temporal extent of the lattices; and
        $n_{\text{cfg}}$ is the number of configurations used for each ensemble.
    }
    \label{tab:ensembledetails}
    \begin{tabular}{cccccccc}
        \hline \hline 
        Set&  $\beta$ & $a$ (fm) &  $am_l$ &$am_s$ & $am_c$ & $N_s\times N_T$ & $n_{\text{cfg}}$ \\
        \hline
        $1$ & $5.8$ & $0.1529(4)$ & $0.002426$ & $0.06730$ & $0.8447$ & $32 \times 48$ & $3500$ \\
        $2$ & $6.0$ & $0.1222(3)$ & $0.001907$ & $0.05252$ & $0.6382$ & $48 \times 64$ & $1000$ \\
        $3$ & $6.3$ & $0.0879(3)$ & $0.001200$ & $0.03630$ & $0.4320$ & $64 \times 96$ & $1047$ \\
        \hline \hline
  \end{tabular}
\end{table}
The spatial volumes of the lattices are large enough to ensure single particle finite volume effects are exponentially
small~\cite{luscher1986}.
Each ensemble has $m_l$ tuned to reproduce the physical pion mass.
They differ from those listed in Refs.~\cite{MILC:Configs,*tastesplitting,Bazavov:2017lyh} by retuning the light sea-quark masses to
reproduce the pion masses more accurately.
The taste-Goldstone pion mass $M_{\pi_5}$ on sets~1 and~2 is 135~MeV, while the mass on set~3 is 128~MeV \cite{Davies:2019efs}.
The retuning does not alter the lattice spacing values, which are determined from the mass-independent scheme.

Although MILC has generated many ensembles with unphysically large~$m_l$, we do not use those ensembles here.
In this way, we can circumvent using baryon chiral perturbation theory (or some other physically motivated function) to guide the
unphysical data to the physical value.
There is some evidence~\cite{WalkerLoud:2008pj,Hansen:2016qoz} that high-order functional forms are necessary.
To gain control over the chiral extrapolation it might, in this case, be more costly, because numerous ensembles could be needed.

To compute baryon masses, we construct the two-point correlation function
\begin{equation}
    C^{R\bar{R}}(t) = \sum_{s,\vec{D}} \sum_{\vec{x}}
    \left\langle B^{R}_{s\vec{D}}(\vec{x}, t) \overline{B}^{\bar{R}}_{s\vec{D}}(0) \right\rangle ,
    \label{eqn:corrfcn}
\end{equation}
where the source is defined in Eq.~(\ref{eq:srcop}) and the sink is defined in Eq.~(\ref{eq:sinkop}).
In order to increase statistics for ensembles~$1$ and~$2$, we choose two well separated timeslices for $t=0$.
The locations of these two timeslices are chosen randomly for each configuration.
Successive configurations generated within each ensemble are expected to be correlated.
These autocorrelations were studied in Ref.~\cite{MILC:Configs,*tastesplitting} and were, however, found not to be appreciable, so
that these configurations can be treated as statistically independent.
Even so, we reduce the autocorrelations by blocking two consecutive configurations to obtain each sample.

By expressing hadron correlators in terms of quark fields and then using Wick's theorem in the Feynman path integral,
one can write the two-point correlation functions in Eq.~(\ref{eqn:corrfcn}) in terms of quark propagators.
Quark propagators emanating from timeslice~$t$ are found by solving
\begin{equation}
    \sum_{b, y} \slashed{D}^{ab}_{xy} G^{bc}_{\vec{A}} (y,t) = \sum_{\vec{z}@\vec{A}} \delta^{ac}
        \delta_{\vec{x}, \vec{z}} \delta_{x_4,t}
\end{equation}
for the Green function $G^{ac}_{\vec{A}}(x,t)$, where $\slashed{D}^{ab}_{xy}$ is the kernel of the HISQ action~\cite{HISQAction}.

A straightforward way to construct the full set of correlation functions would require 64 different quark propagators, one for every
source corresponding to the parallel-transported field, $\widetilde{\chi}^a_{\vec{A},\vec{B}}(\vec{x})$ in
Eq.~(\ref{eq:dressedquark}).
To reduce the number of propagators, we have fixed the gauge fields to Coulomb gauge.
Then the links connecting the quark fields in the source interpolating operators are no longer necessary.
Instead, only the eight propagators with the corner-wall sources specified in Eq.~(\ref{eq:sumquark}) must be computed.
It turns out that the gauge fixing improves the signal significantly.
Without gauge fixing, it would be necessary to introduce the parallel transporters within each unit cube.
Contributions from different cubes would average to zero, albeit introducing some gauge noise.
With gauge fixing, however, every part of the corner walls is linked to the others, providing a helpful volume factor in the signal.

The parallel transporters at the sink are applied after all quark propagators have been calculated.
The only nonzero correlation functions are those where the quantum numbers are conserved, e.g., where $R$ and $\bar{R}$ belong to
the same irreducible representation.
The correlation functions are also nonzero when all unit-cube sites $\vec{D}$ are summed without any staggered phase factor, as
reflected in Eq.~(\ref{eqn:corrfcn}), which increases statistics eightfold.

\section{Fitting Methodologies}
\label{sec:FittingMethods}

After generating data for the two-point correlation function, Eq.~(\ref{eqn:corrfcn}), the next step is to extract the baryon
masses.
In this section, we discuss general aspects of the problem; in Sec.~\ref{sec:DataAnalysis}, we apply these considerations to the
data at hand.

\begin{widetext}
Inserting a complete set of eigenstates of the QCD Hamiltonian into Eq.~(\ref{eqn:corrfcn}) yields many exponential contributions,
including those associated with $N$-like, $\Delta$-like, and $N\pi$ scattering states in the positive-parity sector as well as many
higher excitations and other negative-parity states.
To capture the important physics of this problem, we model the correlation function as a sum of the exponential contributions
\begin{subequations}
    \label{eq:corr_raw}
    \begin{align}   
    C^{(r_1,r_2)}(t) &= C^{(r_1,r_2)}_N(t) + C^{(r_1,r_2)}_\Delta(t) + C^{(r_1,r_2)}_r(t) + C^{(r_1,r_2)}_{-}(t) , \\[0.7em]
    C^{(r_1,r_2)}_N(t) &\equiv a^{(r_1)}_Nb^{(r_2)}_N\Big(e^{-M_Nt}-(-1)^te^{-M_N(T-t)}\Big), \\
    C^{(r_1,r_2)}_\Delta(t) &\equiv \sum_{i=1}^{3} a^{(r_1)}_{\Delta_i}b^{(r_2)}_{\Delta_i}
        \Big(e^{-M_{\Delta_i}t}-(-1)^te^{-M_{\Delta_i}(T-t)}\Big) ,  \\
    C^{(r_1,r_2)}_r(t) &\equiv \sum_{i=1}^{n}a^{(r_1)}_{r,i}b^{(r_2)}_{r,i}\Big(e^{-M_{r,i}t}-(-1)^te^{-M_{r,i}(T-t)}\Big) , \\
    C^{(r_1,r_2)}_{-}(t) &\equiv \sum_{i=1}^{m}a^{(r_1)}_{-,i}b^{(r_2)}_{-,i}
        \Big(e^{-M_{-,i}(T-t)}-(-1)^te^{-M_{-,i}t} \Big),
    \end{align}
\end{subequations}
\end{widetext}
where $a^{(r_1)}$ and $b^{(r_2)}$ are the source and sink overlap amplitudes, $M$ is the mass of the corresponding state, $t$ is the
propagation time, and $T=N_Ta$ is the time extent of the lattice.
The superscripts $r_1$ and $r_2$ indicate the classes of the source and sink operators.
For clarity, we separate the two-point correlator into $C_N(t)$, $C_\Delta(t)$, $C_r(t)$, and $C_-(t)$, which describe,
respectively, the lowest-lying $N$-like state, the lowest three $\Delta$-like states, all other positive-parity excited states, and
all negative-parity states.
Contributions from $N\pi$ scattering states are discussed below and are argued to be largely incorporated into
$C^{(r_1,r_2)}_\Delta$, with a small, residual contamination that is included in the excited state systematic uncertainty.

Due to the antiperiodic boundary conditions of the lattice, two-point correlator data at time $t$ and $T-t$ should converge to the
same result (up to a sign) at infinite statistics.
Therefore, it is convenient to average the correlator data around $T/2$, substituting
\begin{equation}
	C^{(r_1,r_2)}(t) \rightarrow \frac{C^{(r_1,r_2)}(t) - (-1)^{N_T-t/a}C^{(r_1,r_2)}(T-t)}{2}
    \label{eq:fold}
\end{equation} 
and then fitting in $t$ only up to~$T/2$.

Although an infinite number of states contribute to the spectral decomposition, the exponential suppression in
Eq.~(\ref{eq:corr_raw}) of excited states effectively reduces the sums to a small number of states.
Even so, a few still contribute to the correlator after a few time steps.
Thus, one obstacle in extracting accurate information about a particular state is correctly disentangling its contribution from the
other states' contributions.

In this work, we are only interested in the ground state nucleon, so the contamination just mentioned comes from excited states.
We treat all excited states, for both positive and negative parities, as nuisance parameters.
Three kinds of excited state contributions must be addressed:
1)~the three tastes of $\Delta$-like states in $C_\Delta^{(r_1,r2)}$,
2)~the other parity-even excited states in $C_r^{(r_1,r2)}$, and
3)~the negative parity partners in $C_-^{(r_1,r2)}$.
The latter two, $C_r^{(r_1,r2)}$ and $C_-^{(r_1,r2)}$, include finite-volume scattering states, the lowest of which are $N\pi$ 
combinations.
As discussed below,
$N\pi$ states near the $\Delta$ mass are not cleanly disentangled from the $\Delta$-like states modeled
by~$C_\Delta^{(r_1,r2)}$.

Concerning 1), the nucleon and $\Delta$ baryons have distinct quantum numbers, so in other fermion formulations the $\Delta$ baryon
does not contribute to their nucleon correlation functions~\cite{Gupta:2018qil}.
Here, however, the $16$~irrep of the staggered GTS contains both $N$-like and $\Delta$-like
states, as shown in Table~\ref{tab:gtsmultiplicities}.
From experiment~\cite{Tanabashi:2018oca}, the $\Delta$ mass ($\approx 1232$~MeV) is closer to the nucleon mass ($\approx 940$~MeV)
than any other $J^P=\onehalf^+$ or $\threehalves^+$ single-particle excitation, so it is the most important excited state in
staggered-baryon correlator data.
Further, the 16~irrep contains three tastes of $\Delta$-like states, which are separated by a splitting of order~$\alpha_s a^2$.

Regarding 2), after the $N$-like and $\Delta$-like states, the next state in the positive-parity spectrum is (in infinite volume)
the $N\pi$ state in a $P$~wave and, thus, energy near $1250$~MeV on our ensembles.
The next single-particle state in the spectrum is expected to be the so-called Roper resonance $N(1440)$, which has
$J^P=\onehalf^+$.
Disentangling these excited states from the $\Delta$-like states is, however, unlikely within our statistical precision.
Of course, in a finite volume resonances and two-particle states cannot be cleanly separated.
Even so, we aim to include these states as nuisance parameters with suitable prior widths when fitting correlator data to fit away
any excited-state contamination.
For a detailed account of the effects of these states in finite volume, we refer the reader to Ref.~\cite{Hansen:2016qoz}.

Last, concerning 3), all our correlation functions show oscillatory behavior as a consequence of the negative-parity states.
The lowest single-particle contribution to this channel is expected to be the $N(1520)$ with $J^P=\threehalves^-$.
The lowest finite-volume two-body state in this channel is expected to be the $N\pi$ in an $S$~wave with zero momentum and, thus,
energy around $1080$~MeV.
We expect finite-volume corrections from scattering between the two states \cite{Luscher1986II}, with potentially large effects near
the $J^P=\onehalf^-$ $N(1535)$ resonance.
In the meson sector, extracting two-body eigenstates from correlation functions built from single particle interpolating operators
has not been possible \cite{Dudek:2012xn}.
With staggered baryons, even though we only use single-baryon interpolating operators, we may still be able to resolve the lowest
negative-parity two-body eigenstates, because the next-lowest state has an appreciable splitting.
Moreover, the operator constructions for the $16$~irrep classes are somewhat nonlocal, being spread out over the whole unit cube.
In fact, evidence of negative-parity $N\pi$ states has been found using only three-quark operators with Wilson
fermions~\cite{Mahbub:2013bba}.

These contributions all must be dealt with carefully in order to extract nucleon physics.
Because we compute four different classes of staggered-baryon operators, as described in Sec.~\ref{subsec:OpConstruct}, we obtain a
$4\times4$ matrix correlation function.
Further, we adopt two distinct fitting strategies to ensure that we have removed excited-state contamination reliably.
In Sec.~\ref{sec:BayesApproach}, we apply multistate Bayesian curve fitting~\cite{Lepage:2001ym} to the matrix correlation
function, using all information in the spectral decomposition, Eq.~(\ref{eq:corr_raw}).
In Sec.~\ref{sec:GEVP}, we solve the generalized eigenvalue problem
(GEVP)~\cite{Michael:1982gb,Kronfeld:1989tb,Luscher:1990ck,Blossier:2009kd}, adapted to correlators with oscillating
states~\cite{DeTar:2014gla}.
This is particularly suited to staggered baryons because the Golterman-Smit-Bailey construction naturally provides several distinct
operator classes.
We have found that the two analyses yield consistent results for the nucleon mass.

\subsection{Bayesian Approach}
\label{sec:BayesApproach}

In the Bayesian fitting approach, we simultaneously fit multiple different correlators using Lepage's corrfitter~\cite{corrfitter}
and related packages~\cite{lsqfit,gvar}.
We omit correlator data built from the class-3 interpolating operator, which we have empirically observed to have a poor
signal-to-noise ratio due to a smaller overlap with the lowest $N$-like state, corroborating earlier
findings~\cite{Ishizuka:1993mt}.
As such, we fit a $3\times3$ matrix of correlation functions built from the $r_i=2$, $4$, and $6$ operator classes residing at
either the source or sink.

Within the Bayesian methodology, every fit parameter is assigned a prior distribution.
The fit function can incorporate, in principle, an arbitrarily large number of states.
Any state insufficiently constrained by the data will return a posterior distribution close to the prior, and
so will have negligible effect upon the fit results.
With such an approach, in contrast to plateau fitting (of the effective mass), we can include as many states as needed to
successfully fit the correlation functions at small $t$ without compromising fit quality.
As baryon correlation functions suffer from an exponential signal-to-noise degradation at large times, fitting to small~$t$
increases the amount of available data, leading to a more precise nucleon mass.

Suitably wide priors have to be chosen for each fit parameter, based on available information.
In practice, one has knowledge only about the first few excited-state mass splittings, while one has very little knowledge about the
remaining spectrum or overlap amplitudes.
For the nucleon fit parameters, we want to assume no significant prior knowledge, so the prior widths are chosen wide enough to
leave them effectively unconstrained.
As is standard, we shall demonstrate that our fitted nucleon mass is stable against reasonable variations of the prior widths.
To ensure the correct ordering of states, we choose a log-normal distribution for the mass differences between adjacent states and a
normal distribution for all other priors, as implemented in corrfitter~\cite{corrfitter}.

We have observed that fits including four exponentials in the positive parity channel---corresponding to the $N$-like state
and all three $\Delta$-like tastes---in addition to a high-mass positive parity exponential and four negative-parity
exponentials are unstable under a variation of fit choices, such as the number of states in the fit function and prior choices.
This is caused by the presence of three states with masses separated by $\order(\alpha_s a^2)$.
Resolving three states with such small splittings is not possible within our statistical errors, and including these three states in
a fit function gives rise to a flat direction in the $\chi^2$ landscape.

Knowing the cause of the issue, it is easy to overcome it by removing the flat directions systematically.
For simplicity, we can examine a single correlation function.
Let $\delta m_{\text{taste}}$ denote the typical taste splitting between the $\Delta$-like masses.
The taste splittings between HISQ pions are $\order(\alpha_s a^2)$~\cite{HISQAction}, the largest of which is between the
taste-scalar pion and Goldstone pion and is around $200$~MeV on our ensembles~\cite{tastesplitting}.
We order the three taste-split states as $M_{\Delta_1}<M_{\Delta_2}<M_{\Delta_3}$ and take $M_{\Delta_3}-M_{\Delta_1}\sim\delta
m_{\text{taste}}$ as the largest taste splitting.

To marginalize the $\Delta$-like contribution, we replace $C_\Delta(t)$ in Eq.~(\ref{eq:corr_raw}) with a functional
form containing two exponentials instead of three,
\begin{equation}
    C_{\Delta'}(t) \equiv \sum_{i=1}^{2} a_{\Delta'_i}b_{\Delta'_i}
        e^{-M_{\Delta'_i}t},
    \label{eq:twoexp}
\end{equation}
suppressing the backward propagating terms for clarity.
It should be safe to use $C_{\Delta'}(t)$ in place of $C_\Delta(t)$ for times $t$ such that
\begin{equation}
    \left |\frac{C_{\Delta'}(t) - C_\Delta(t)}{C(t)}\right | <  \text{statistical error},
    \label{eq:inequality}
\end{equation}
as has been used successfully before~\cite{Hughes:2017xie}.

To explore the systematic error introduced by Eq.~(\ref{eq:twoexp}), let us focus on a single 
channel and Taylor expand both $C_{\Delta'}(t)$ and $C_{\Delta}(t)$ around $M_{\Delta'_1}$, finding
\begin{align}
    C_\Delta(t) &= \sum_{i=1}^3 a_{\Delta_i}b_{\Delta_i}e^{-M_{\Delta'_1}t} \nonumber\\
            &\hspace{1em}+ \sum_{n=1}^\infty\sum_{i=1}^3 \frac{a_{\Delta_i}b_{\Delta_i}}{n!}(-\delta m_{i} t)^n e^{-M_{\Delta'_1}t},
    \label{eq:delta} \\
    C_{\Delta'}(t) &= (a_{\Delta'_1}b_{\Delta'_1}+a_{\Delta'_2}b_{\Delta'_2}) e^{-M_{\Delta'_1}t} \nonumber\\
            &\hspace{1em}+ \sum_{n=1}^\infty \frac{a_{\Delta'_2}b_{\Delta'_2}}{n!} (-\delta M t)^n e^{-M_{\Delta'_1}t} ,
    \label{eq:deltaprime}
\end{align}
where $\delta m_{i}\equiv M_{\Delta_i} - M_{\Delta'_1} $ and $\delta M \equiv M_{\Delta'_2} - M_{\Delta'_1}$.
Treating $a_{\Delta_i}$, $b_{\Delta_i}$, and $M_{\Delta_i}$ as fixed, one can solve for $M_{\Delta'_i}$ and the products of
overlaps $a_{\Delta'_i}b_{\Delta'_i}$ order-by-order:
\begin{align}
    \delta M &\approx \frac{\sum_{i=1}^3 a_{\Delta_i}b_{\Delta_i}(\delta m_i)^2}{\sum_{i=1}^3a_{\Delta_i}b_{\Delta_i}\delta m_i},
    \label{eq:dMp1} \\
    a_{\Delta'_2}b_{\Delta'_2} &\approx
        \frac{(\sum_{i=1}^3a_{\Delta_i}b_{\Delta_i}\delta m_i)^2}{\sum_{i=1}^3a_{\Delta_i}b_{\Delta_i}(\delta m_i)^2},
    \label{eq:overlap1} \\
    a_{\Delta'_1}b_{\Delta'_1} &\approx -a_{\Delta'_2}b_{\Delta'_2} + \sum_{i=1}^3 a_{\Delta_i}b_{\Delta_i},
    \label{eq:overlap2}
\end{align}
to $\order((\delta m_it)^2)\sim\order((\delta m_{\text{taste}}t)^2)$.
In this way, the two- and three-state fits are equivalent if the statistical error is larger than $\order((\delta
m_{\text{taste}}t)^2)$.
Note that the analysis with a $4\times4$ matrix including the class~3 entries would be more complicated~\cite{Kronfeld:1989tb}.

\begin{figure}[b]
    \centering
    \includegraphics[width=\columnwidth]{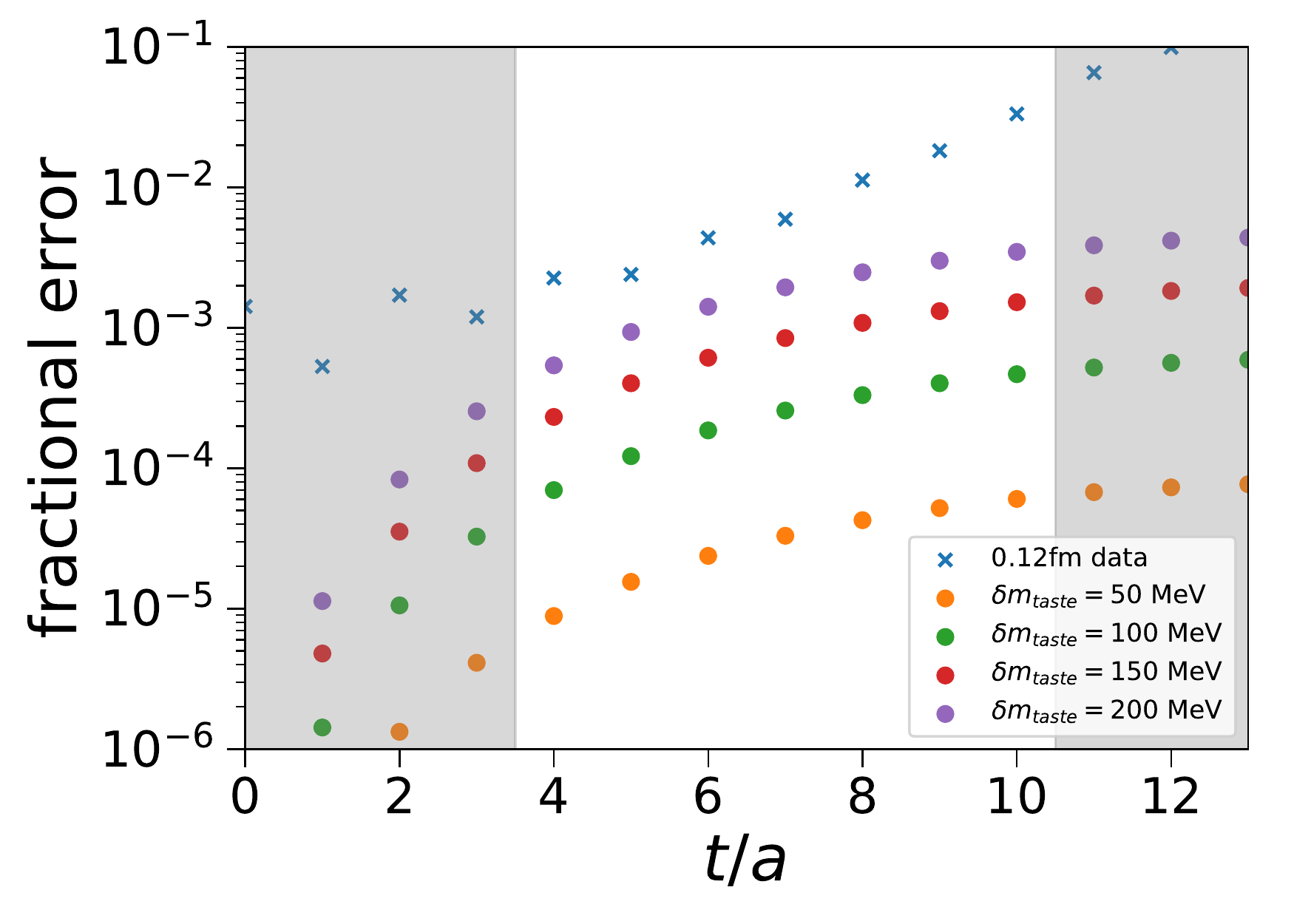}
    \caption{(color online) Comparison between the statistical fractional (relative) error in the correlator data---shown as
        crosses---and the systematic error---shown as circles---induced by replacing the fit function as defined by the left hand
        side of Eq.~(\ref{eq:inequality}).
        The test parameter values used to generate these points are listed in Table~\ref{tab:testvalue}.
        Parameters for $C_{\Delta'}(t)$ are calculated using Eq.~(\ref{eq:dMp1})--(\ref{eq:overlap2}).
        The $t_{\text{fit}}$ within the white range are those used in the later analysis.
        The correlator data used to generate the crosses are our most statistically precise, coming from the $a\approx 0.12$~fm
        ensemble with a class-2 operator at both source and sink.}
		\label{fig:new_model}
\end{figure}

In Fig.~\ref{fig:new_model}, we compare our statistical precision to the quantity defined in Eq.~(\ref{eq:inequality}) for various
$\delta m_\text{taste}$.
\begin{table}
    \centering
    \caption{Input values used to generate Fig.~\ref{fig:new_model}.}
    \label{tab:testvalue}
    \begin{tabular}{cc}
        \hline\hline
        Parameter & Test input\\
        \hline
        $M_N$ & $940$~MeV\\
        $M_{\Delta_1}$ & $M_{\Delta_2}-\delta m_{\text{taste}}/2$\\
        $M_{\Delta_2}$ & $1230$~MeV\\
        $M_{\Delta_3}$ & $M_{\Delta_2}+\delta m_{\text{taste}}/2$\\
        $a_Nb_N$&1\\
        $a_{\Delta_i}b_{\Delta_i}$&1\\
        \hline\hline
    \end{tabular}
\end{table}%
These data are from the $a\approx 0.12$~fm ensemble with the class 2 operator residing at both the source and sink.
This correlator has the smallest statistical error of all our data, giving it the best opportunity to resolve the various
taste-split states.
As can be seen, the systematic error introduced by using $C_{\Delta'}(t)$ is smaller than our statistical precision, and so we are
unable to distinguish between $C_{\Delta}(t)$ and $C_{\Delta'}(t)$.
If we do try to fit to $C_{\Delta}(t)$, we observe the expected flat direction in the $\Delta$ parameters.

In practice, a $3\times3$ matrix of correlators is used to disentangle the $\Delta$-like states.
Given high enough statistics, a fit to the correlator matrix could resolve the small mass splittings and the
two-state model would not suffice.
(Even at infinite statistics, we would not be able to eliminate all excited-state contamination,
 simply because we have a finite basis and finite number of timeslices,
 but an infinite tower of states.)
At finite precision, however, the two-state model should be good enough to reduce
 the excited-state contamination of the three $\Delta$-like states,
 \emph{if} we find stable posteriors for $a(M_{\Delta'_1} - M_N)$
 and $a\delta M=a(M_{\Delta'_2}-M_{\Delta'_1})$.
We discuss the outcome for our data in Sec.~\ref{sec:BayeSys}. 

The two-state model can be further validated with the data by the following exercise.
Assuming the $\Delta$-like states are degenerate, we can rewrite Eq.~(\ref{eq:corr_raw}) as
\begin{align}
    \mathbf{C}(t) &= \mathbf{C}_N(t) + \mathbf{C}_\Delta(t) + \cdots,
    \label{eq:Cmx} \\
    \mathbf{C}_N(t)      &\equiv \mathbf{N}      \left(e^{-M_Nt}     - (-1)^te^{-M_N(T-t)}\right),
    \label{eq:Nmx}\\
    \mathbf{C}_\Delta(t) &\equiv \mathbf{\Delta} \left(e^{-M_\Delta} - (-1)^te^{-M_\Delta(T-t)}\right),
    \label{eq:Dmx}
\end{align}
where $\mathbf{N}$ and $\mathbf{\Delta}$ are the time-independent $3\times3$ coefficient matrices for the $N$-like and $\Delta$-like states.
We now fit the data to Eqs~(\ref{eq:Cmx})--(\ref{eq:Dmx}) with independent
entries in $\mathbf{N}$ and $\mathbf{\Delta}$ but common masses. We then find the number of
nonvanishing singular values of $\mathbf{N}$ and $\mathbf{\Delta}$, thereby determining the rank of these
matrices. On all three ensembles, we find one
nonvanishing singular value for $\mathbf{N}$ and two for $\mathbf{\Delta}$, and further, that this outcome is stable under
bootstrap resampling.
This test validates using the two-state model so to marginalize the $\Delta$-like contribution.

Future analyses with higher statistics should revisit the omission of the class-3 correlators.
In our data, their exponential falloff and signal-to-noise behavior suggest a small coupling of the class-3 operator to the
lowest-lying $N$-like state.
Because the $N$-like and $\Delta$-like states end up in different multiplets in the continuum limit, the class-3 operators are
mostly $\Delta$-like, as explained further in Appendix~\ref{appendix:staggeredop}.
We have tried some fits including it: it is then possible, but not compelling to use $C_\Delta^{(r_1,r_2)}(t)$, \emph{i.e.}, to fit
three states.
For a complete description of the states in this region of the spectrum (\emph{e.g.}, in calculation of the $\Delta$ mass or
$N\to\Delta$ form factors), two-body $N\pi$ operators are expected to be essential too, because, as mentioned above, the lowest
$N\pi$ states have an energy close to the $\Delta$ mass.

The $N\pi$ states in the spectrum warrant some further discussion.
As mentioned above, the $J=\onehalf$ states are in an S~wave starting around 1080~MeV, while the the $J=\threehalves$ states are in
a P~wave that, for our volumes, are near the $\Delta$ mass.
Although the overlap of these states with our one-body operators can be expected to be small, Hansen and
H.~Meyer~\cite{Hansen:2016qoz} caution that an accumulation could cause a false plateau.
Because our source and sink operators are not identical, our matrix of correlators need not be even approximately symmetric
and, thus, our data are especially susceptible.
For our data to exhibit false plateaux, negative exponential contributions in the diagonal correlation
functions $C^{(r_1,r_1)}$ would have to be present, which would indicate very different behavior for the wall and point
operators in the same operator class.
We have studied our data, looking for these negative contributions in the diagonal
correlation functions.
We find that i) any negative fit contributions appear only in the two highest excited states and ii) the negative fit contributions
are rare and suppressed by at least one, if not several, orders of magnitude relative to the $N$-like fit exponential.
The mass gap between these states and the nucleon mass is large, making this unlikely to be cause for concern.

Note also that Ref.~\cite{Hansen:2016qoz} observes problems---a ``false plateau''---in a three-point function yielding the nucleon 
axial charge~$g_A$.
Tracing the analysis given there, it turns out that the main culprit for the false plateau is an excited-state matrix element of 
the axial current, which is immaterial here.
Moreover, Ref.~\cite{Hansen:2016qoz} has in mind a strategy of fitting only to a plateau, whereas we deliberately aim to fit the 
region in which the excited states are not especially small, in order to fit them away.

In summary, the final function that we perform a Bayesian fit to is
\begin{align}
    C^{(r_1,r_2)}(t) &= C^{(r_1,r_2)}_N(t) + C^{(r_1,r_2)}_{\Delta'}(t) \nonumber\\
        &\hspace{1em}+ C^{(r_1,r_2)}_r(t) + C^{(r_1,r_2)}_{-}(t) .
	\label{eq:corr_final}
\end{align}
We have found empirically that $C_r(t)$ and $C_-(t)$ contribute very little to the correlator during our fit range.
Note that one should be careful about assigning physical meaning to the fit parameters of
$C^{(r_1,r_2)}_{\Delta'}(t)$, $C^{(r_1,r_2)}_r(t)$, and $ C^{(r_1,r_2)}_{-}(t)$.
Their role is to describe the contributions other than the lowest-lying $N$-like state while yielding stable results for the 
corresponding parameters.

\subsection{GEVP Approach}
\label{sec:GEVP}

The construction of staggered baryons naturally gives rise to different classes of interpolating operators, which can form a basis
for the GEVP.
With this GEVP approach we can extract the mass of the lowest $N$-like state.

After setting aside the class-3 correlator data, we have a $3\times 3$
matrix $\mathbf{C}(t)\equiv C^{(r_1,r_2)}(t)$, where $r_1, r_2 \in \{2, 4, 6\}$ again denotes the class of the operator at the
source or sink.
The GEVP~\cite{Michael:1982gb,Kronfeld:1989tb,Luscher:1990ck,Blossier:2009kd} is defined via
\begin{align}
	\mathbf{C}(t)v^R_i(t,t_0) &= \lambda_i(t, t_0)\mathbf{C}(t_0)v^R_i(t,t_0),
    \label{eq:gevpright}\\
	[v_i^L(t,t_0)]^T\mathbf{C}(t) &= \lambda_i(t, t_0)[v^L_i(t,t_0)]^T\mathbf{C}(t_0) ,
    \label{eq:gevpleft}
\end{align}
where $v^L_i(t,t_0)$ and $v^R_i(t,t_0)$ are the generalized left and right eigenvectors, $i\in\{1,2,3\}$.
They share the same set of generalized eigenvalues $\lambda_i(t, t_0)$.

For staggered fermions, the oscillating parity-partner states must also be addressed.
We fit the ground state eigenvalues, $\lambda_0(t, t_0)$, to a reduced form of Eq.~(2.16) in Ref.~\cite{DeTar:2014gla},
\begin{equation}
    \lambda_0(t,t_0=\text{fixed}) = Ae^{-M_Nt} - (-1)^{t/a} Be^{-M_-t}
    \label{eq:gevpt0}
\end{equation}
where $M_N$ is the nucleon mass, $M_-$ accounts for an oscillating state, and $A$ and $B$ are coefficients
satisfying $A_i+B_i\approx1$.
Equation~(\ref{eq:gevpt0}) sets $b_n=d_n=0$ in Eq.~(2.16) of Ref.~\cite{DeTar:2014gla}.
The further excited state contamination can be controlled by varying $t_0$ and $t$, as shown below.

Based on the discussion in Sec.~\ref{sec:BayesApproach}, we can marginalize over a $\Delta$-like state and constrain the $\Delta'$
contributions.
As a consequence, we expect the lowest-lying $N$-like state and the two $\Delta^{\prime}$ exponentials to be the three
distinguishable states from the asymptotic time GEVP with a $3\times3$ basis.
The initial time $t_0$ is chosen to suppress excited-state contributions to the eigenvalues, which in this work come from the higher
nonoscillating states and all of the oscillating states.
The traditional approach employed to suppress excited state contamination in the GEVP is to choose a large enough $t_0$ in either
Eq.~(\ref{eq:gevpright}) or (\ref{eq:gevpleft}).
Below we show control over excited contributions by varying the choice of $t_0$ in Sec.~\ref{sec:GEVPAnalysis}, where we observe
such excited states have a negligible impact on the quality of fits.

The presence of the staggered phase $(-1)^{t/a}$ in Eq.~(\ref{eq:gevpt0}) can be mitigated via a combination that becomes an exact
cancellation in the asymptotic large-time limit.
To achieve such an affect, we take the symmetrized combination
\begin{align}
	\frac{1}{4} \bigg\{ &[\mathbf{C}(t_0-1)]^{-1} \mathbf{C}(t-1) \nonumber + 2[\mathbf{C}(t_0)]^{-1} \mathbf{C}(t) \nonumber\\
	&+ [\mathbf{C}(t_0+1)]^{-1} \mathbf{C}(t+1) \bigg\} \widetilde{v}^R_i(t) = \widetilde{\lambda}_i(t, t_0) \widetilde{v}^R_i(t),
	\label{eq:gevpwindow}
\end{align}
which cancels most of the opposite-parity contributions.
A similar expression holds for the left eigenvectors $\widetilde{v}_i^L(t)$.
At large $t$ and $t_0$, $\lambda_i(t,t_0) \to \widetilde{\lambda}_i(t, t_0)$, but for intermediate $t$ and $t_0$ the explicit
cancellation of oscillations makes $\widetilde{\lambda}_i(t, t_0)$ much better behaved.
In the scenario with a $1\times 1$ correlator matrix, this approach is equivalent to the smoothed effective mass
\cite{Bailey:2008wp,DeTar:2014gla}.

After performing the GEVP analysis and determining the eigenvalues, we perform unconstrained plateau fits to 
\begin{equation}
	-\frac{\ln \widetilde{\lambda}_i(t,t_0)}{\tau} = M_i + C_ie^{-\widetilde{\delta M_i} t_0},
	\label{eq:gevpwindowfit}
\end{equation}
where $C_i$ and $\widetilde{\delta M_i}$ are fit parameters used to account for any residual excited-state contribution, and
$\tau=t-t_0$.
In the subsequent analysis, we fix $\tau/a = 2$ and vary~$t_0$.
In this way we can obtain the nucleon mass $M_N=M_1$.

\section{Fitting Nucleon Correlator Data}
\label{sec:DataAnalysis}

In this section, we present specific details about fitting the 16~irrep two-point correlators.
The $N$-like mass on each ensemble is determined as a fit parameter from this process.
The Bayesian and GEVP analyses are discussed in turn.

\subsection{Bayesian Analysis}

Table~\ref{tab:summaryprior} summarizes the priors used in the Bayesian fits for the three ensembles used in this paper.
\begin{table*}
    \centering
    \caption{Summary of priors used in nominal Bayesian fits.
        Note that the notation 4E+4O means that we are fitting to four even and four odd parity states.
        Refer to Table~\ref{tab:detailpos} for more detailed priors and posteriors.
        Entries for masses and mass splittings in~MeV.}
    \label{tab:summaryprior}
    \setlength{\tabcolsep}{6pt}
    \begin{tabular}{ccccccc}
        \hline\hline
        Ensemble&$t_\text{min}/a$& No. of States& $M_N$ Prior &
        $M_{\Delta'_1}-M_N$ Prior & $M_{\Delta'_2}-M_{\Delta'_1}$ Prior & $M_{-,1}$ Prior\\
        \hline
        1 & 4 &4E+4O& $940(50)$&$290(100)$&$150(50)$&$1400(200)$\\
        2 & 5 &4E+4O& $940(50)$&$290(100)$&$100(50)$&$1400(200)$\\
        3 & 6 &4E+4O& $940(50)$&$290(100)$& $50(50)$&$1400(200)$\\
        \hline\hline
    \end{tabular}
\end{table*}
\begin{figure*}
    \centering
    \includegraphics[width=\textwidth]{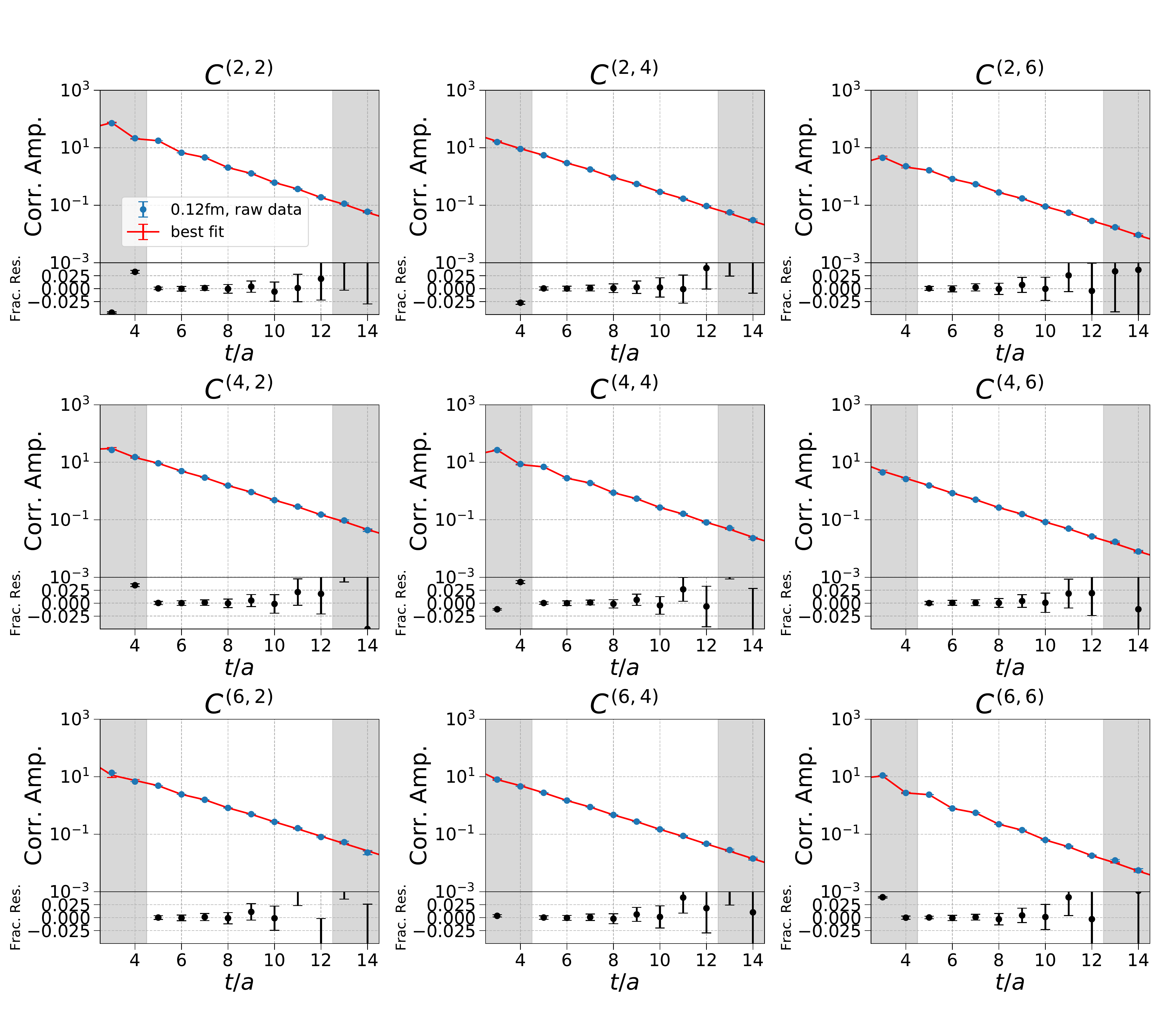}
    \caption{Joint nominal fit of the $0.12$~fm ensemble correlators $C^{(r_1,r_2)}$, with $r_1,r_2=2,4,6$ denoting the different
        classes of source and sink operators (defined in Appendix \ref{appendix:staggeredop}).
        The nominal fit posteriors can be found in Table~\ref{tab:detailpos}.
        The solid blue circles are the raw two-point data; and the red lines with error bars are the nominal fit result; the solid
        black circles in the bottom panels display the fractional residues defined in Eq.~(\ref{eq:fracerr}).
        The white regions are the fit ranges for each correlator.}
    \label{fig:corr012}
\end{figure*}
Detailed prior and posterior values are given in Table~\ref{tab:detailpos} of Appendix~\ref{appendix:moredata}.
For each two-point correlator, $t_\text{min}$ is fixed in physical units across all three ensembles to be $\approx 0.6$~fm and
$t_\text{max}$ is chosen to be the timeslice where the noise-to-signal ratio exceeds $5\%$.

As mentioned in Sec.~\ref{sec:FittingMethods}, we take prior central values for the excited-state contributions based on
infinite-volume and noninteracting energy splittings.
We choose prior widths for the excited states in the region of the $\Delta$ mass that are, however, large enough to accommodate the
fact that finite-volume states could receive corrections from avoided level crossings with $N\pi$-like scattering states, among
others.
The mass difference between the lowest $N$-like state and the $\Delta'_1$ state is chosen so that $M_{\Delta'_1}\approx1230$~MeV,
which is the physical $\Delta$ resonance mass.
We choose a prior width of $100$~MeV.
To roughly match the observed taste-split masses between staggered mesons, the mass difference between $\Delta'_1$ and $\Delta'_2$
states is given a prior of $150\pm 50$~MeV, $100\pm 50$~MeV and $50\pm 50$~MeV on the $0.15$~fm, $0.12$~fm and $0.09$~fm ensembles.
The other mass differences, for both even- and odd-parity states, are $400\pm 200$~MeV.
    
\subsubsection{Fit Analysis}
\label{sec:bayesianfitanalysis}

\begin{figure*}
    \centering
    \includegraphics[width=\textwidth]{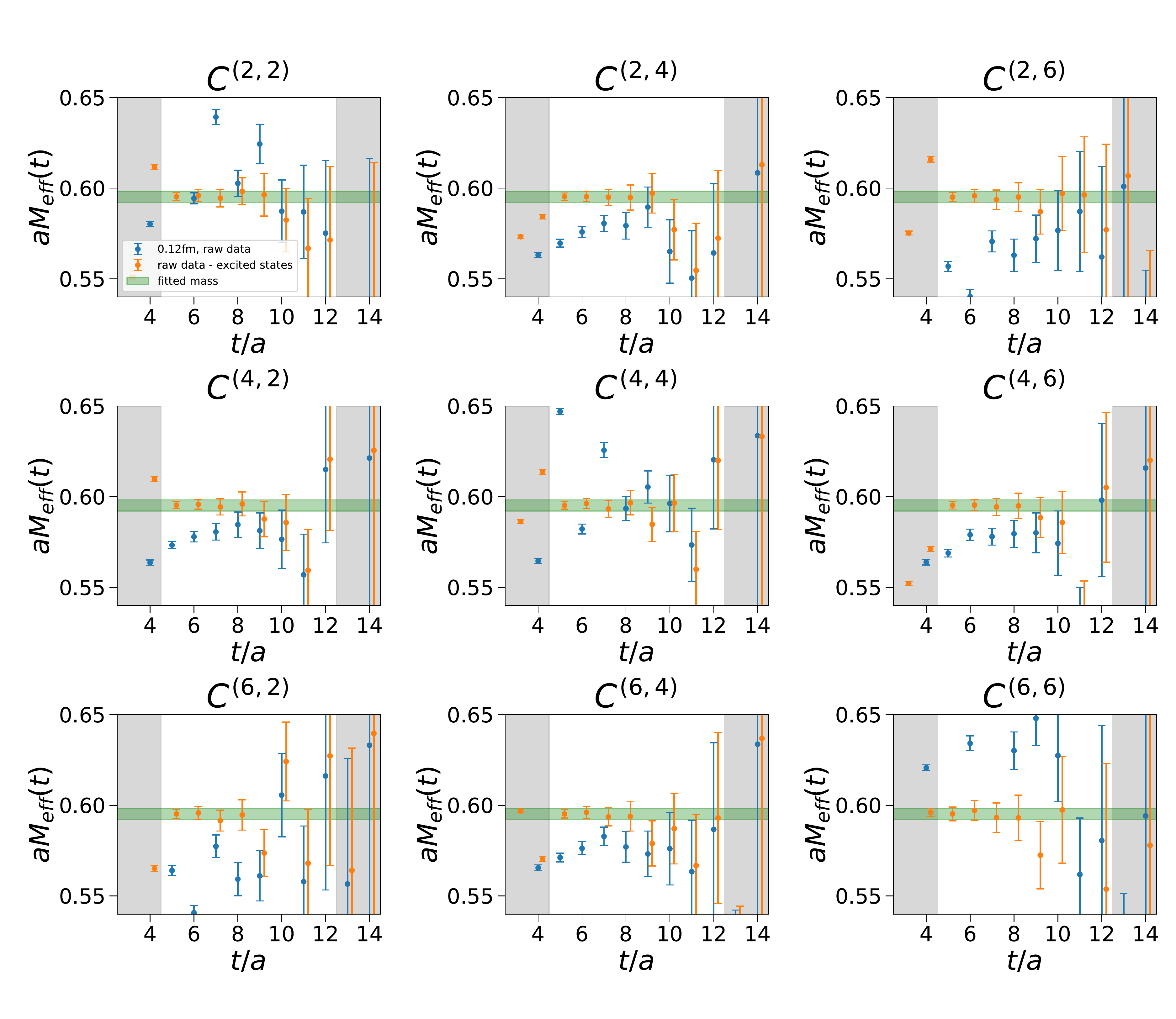}
    \caption{Effective mass [as defined in Eq.~(\ref{eq:meffdef})] of the $0.12$~fm ensemble correlators $C^{(r_1, r_2)}$,
        with~$r_1$, $r_2=2,4,6$ denoting the different classes of source and sink operators (defined in
        Appendix~\ref{appendix:staggeredop}).
        The solid blue circles are the effective masses of raw two-point correlators, whereas the solid orange circles are the
        effective masses of the two-point correlators after the nominal fitted values for the higher excited states have been
        subtracted out.
        The green bands show the posterior mass from the joint Bayesian fit to the matrix correlator.}
    \label{fig:meff012}
\end{figure*}
For brevity, we focus on the ensemble with $a\approx0.12$~fm, with the corresponding information for $a\approx0.15$, $0.09$~fm in
Appendix~\ref{appendix:moredata}.
We plot the raw correlator data overlaid with the fit result in Fig.~\ref{fig:corr012}, showing the results for the correlator
matrix element, $C^{(r_1,r_2)}$, where $r_i=2,4,6$ labels the source and sink operator classes.
The fractional residues in the bottom panels of these figures are defined by
\begin{equation}
    \text{frac. res.} \equiv \frac{\text{data} - \text{nominal fit}}{\text{data}}
    \label{eq:fracerr}
\end{equation} 
where the correlations between the data and the nominal fit are not included and the exhibited error bars are statistical.

To further examine the quality of the nominal fits, we show effective masses in Fig.~\ref{fig:meff012}.
Again, we plot results for each correlator matrix element $C^{(r_1,r_2)}$.
The effective mass is defined by
\begin{equation}
    M_\text{eff}(t) \equiv \frac{1}{\tau}\ln\left(\frac{C(t)}{C(t+\tau)}\right),
    \label{eq:meffdef}
\end{equation}	
where $C(t)$ is the two-point correlator at timeslice $t$, and $\tau=2$ is chosen to reduce the effects of oscillations.
For some two-point correlators (solid blue circles), effective mass plateaus are not visible due to the excited state contributions
combined with the oscillating terms [cf., Eq.~(\ref{eq:corr_raw})].
Traditional plateau fits to these effective masses would clearly fail.
After carefully fitting away the excited states with the Bayesian method, however, and subtracting the excited-state contributions
from the raw data, a much better plateau (solid orange circles) appears across the fit range.
There is also consistency between plateaus for different choices of source and sink operators.
Without excited states, the solid orange circles would only contain a single exponential contribution from the ground state, and
produce a flat plateau.
Consequently, we have successfully eliminated most of the excited states contamination.
The final posterior estimates of the $N$-like masses are done with $1000$ bootstrap samples for each ensemble.
They are listed in the Bayesian fit column of Table~\ref{tab:allnucleon}.
    
\subsubsection{Systematic Checks}
\label{sec:BayeSys}

\begin{figure}
    \centering
    \includegraphics[width=\columnwidth]{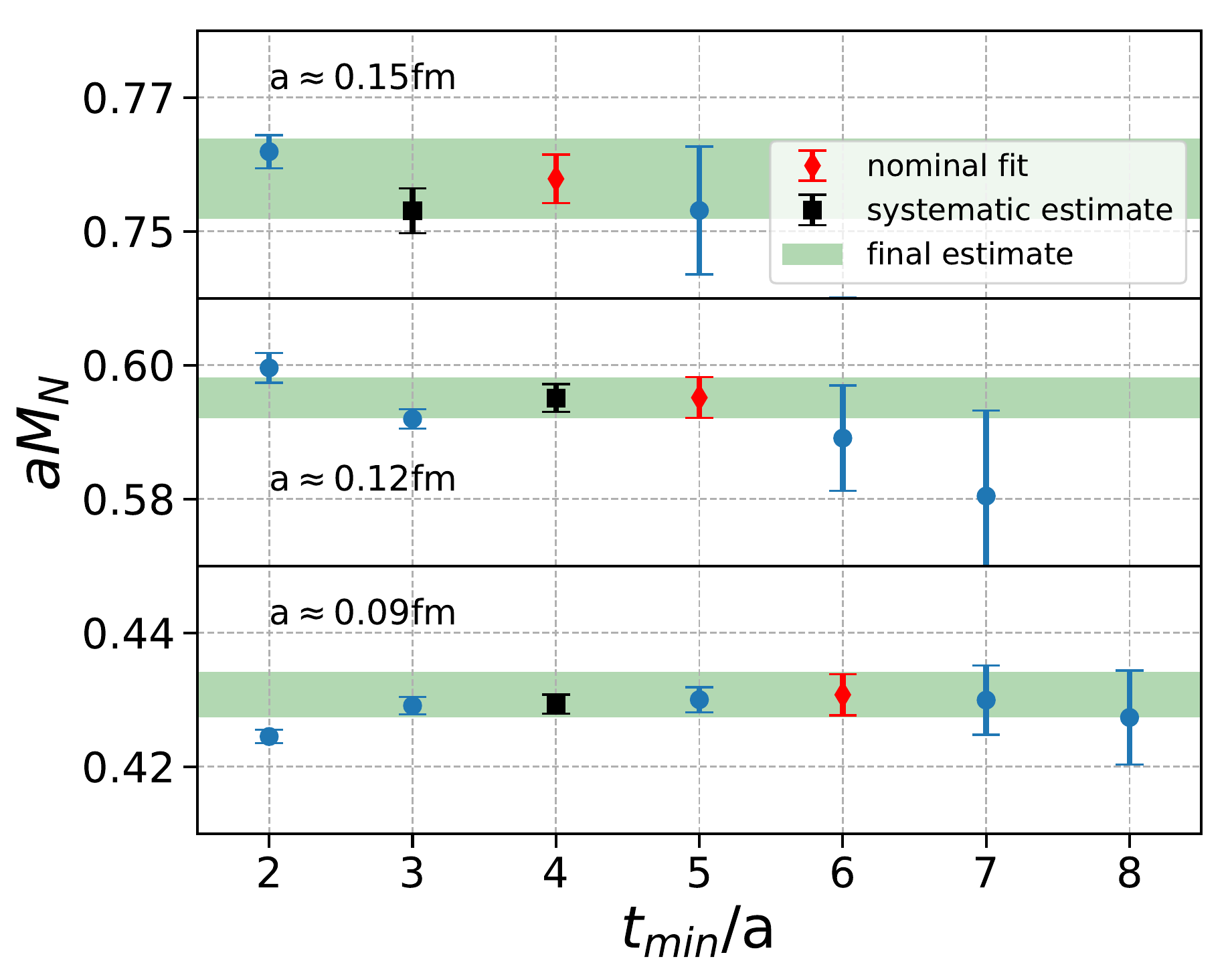}
    \caption{Stability of the nucleon mass posterior as a function of $t_\text{min}/a$.
        The red diamonds denote the nominal fits, the black squares denote the timeslices used to estimate the fitting systematics
        (see text for further details), and the green bands show the final estimates, which include both systematic and statistical 
        errors from fitting combined in quadrature.}
    \label{fig:tminstab}
\end{figure}

\begin{figure}
    \centering
    \includegraphics[width=\columnwidth]{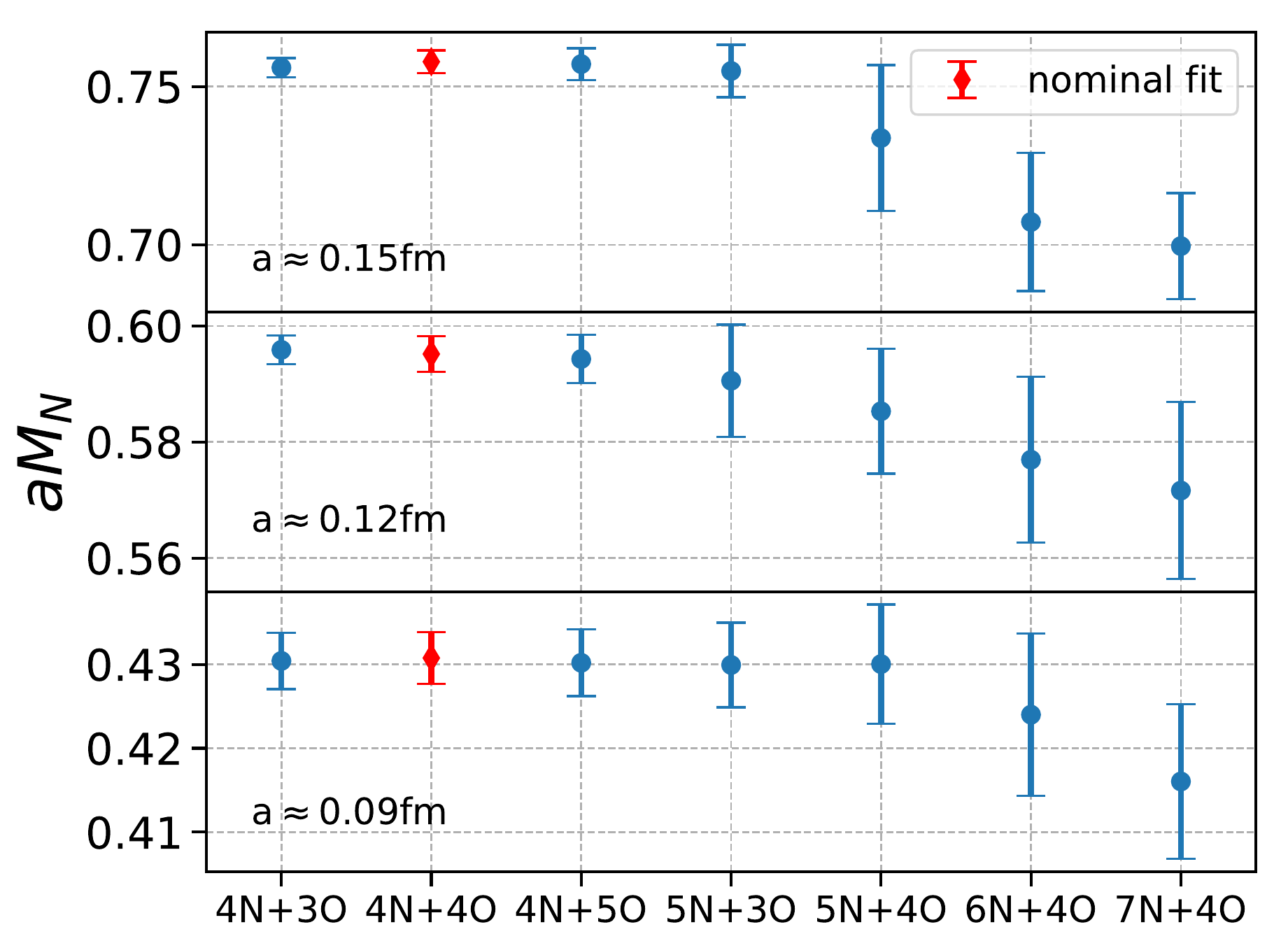}
    \caption{Stability of the nucleon mass as a function of the number of additional radial excitations included in the fit 
        function.
        The red diamonds denote the nominal fits.}
    \label{fig:statestab}
\end{figure}

We have plotted several stability plots in Figs.~\ref{fig:tminstab}~and~\ref{fig:statestab} to demonstrate control over various types
of systematic effects when extracting the $N$-like mass.
\begin{figure}
    \centering
    \includegraphics[width=\columnwidth]{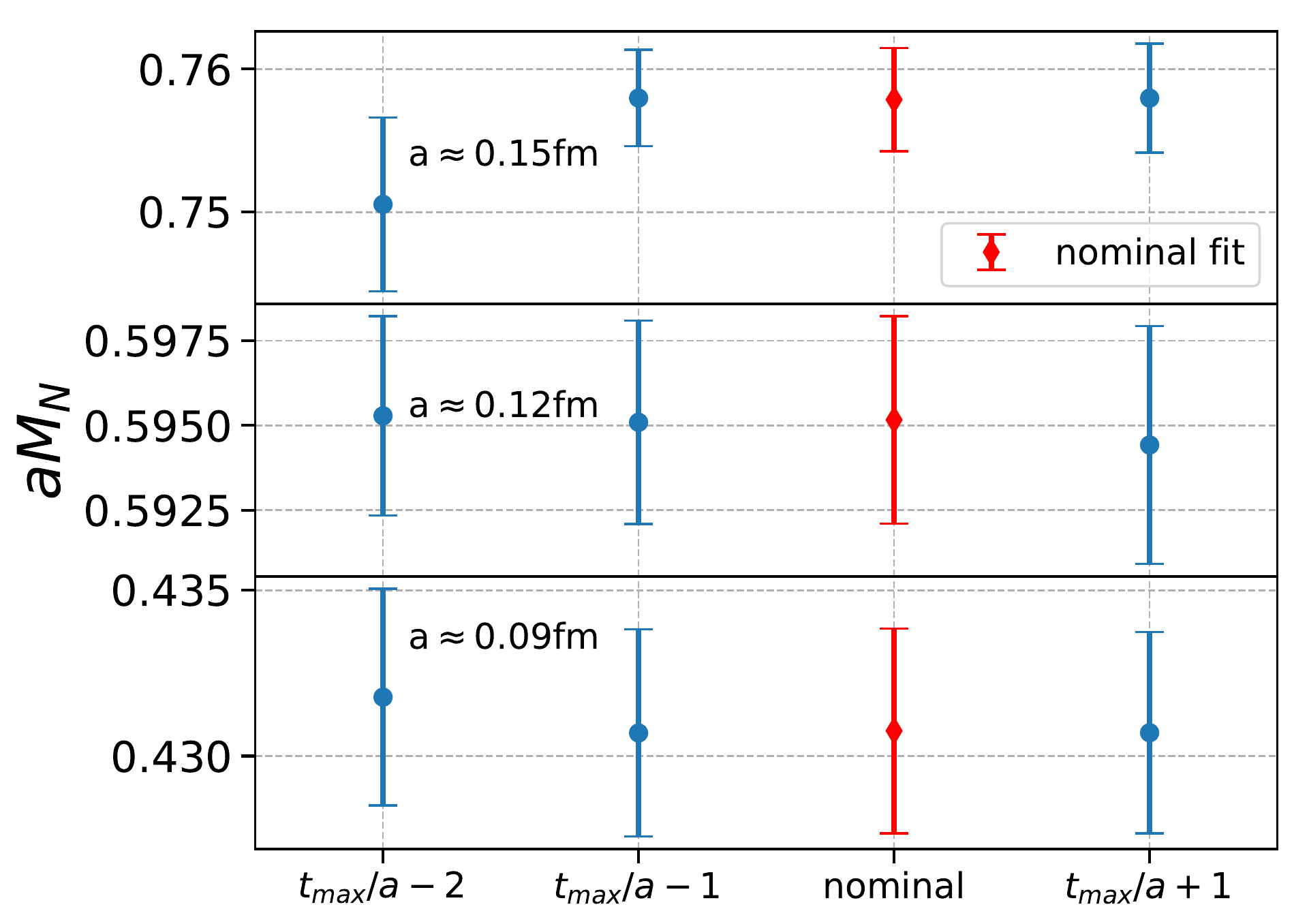}
    \caption{Stability of the nucleon mass posterior as a function of $t_\text{max}/a$.
        The red diamonds denote the nominal fits.}
    \label{fig:tmaxstab}
\end{figure}
Excited state contamination of any kind would manifest as a variation of the ground state $N$-like mass posterior as the value of
$t_\text{min}/a$ varies.
Figure~\ref{fig:tminstab} shows the stability of our Bayesian results under such a variation.
Note that as $t_\text{min}$ is increased, fewer data points are included in the fit, and, thus, the errors increase.
We observe that the ground state $N$-like mass is stable under a change of $t_\text{min}$ for all three ensembles, indicating that
excited-state contamination is under control in these Bayesian fits.

To estimate any small residual excited state contamination, let $t_\text{nom}/a$ be the nominal value of $t_\text{min}/a$, which is
shown as the red diamonds in Fig.~\ref{fig:tminstab}.
We choose $t_\text{sys}\approx 0.15$~fm to be a fixed physical length across all ensembles and examine the posteriors
for $t_\text{min}=t_\text{nom}-t_\text{sys}$, which are shown as black
squares in Fig.~\ref{fig:tminstab}.
The central-value difference between the red diamonds and black squares is an estimate of the systematic error from fitting.
We expect the source of this difference to be stem from any residual excited-state contamination and/or the choice of fit range.
Although these two results are not independent, we conservatively combine the statistical error and the above fitting systematic
estimate in quadrature to obtain the total $N$-like mass posterior width.

We also study how the ground-state mass posterior changes as a function of the number of states included in the Bayesian fit
function, as shown in Fig.~\ref{fig:statestab}.
Based on the stability of the ground state posterior, our nominal fit contains four even and four odd parity states, which we denote
by 4E+4O.
As can be seen, including too many higher excited states and, thus, many more poorly constrained priors, can cause noticeable
changes in the low-energy posteriors.
This behavior could be avoided if we had some prior knowledge of the overlap factors, which could then be used to impose prior
widths of natural size.

A third systematic test demonstrates that the variation of posteriors from changing $t_{\text{max}}/a$ is not significant, as shown in
Fig.~\ref{fig:tmaxstab}.
The breakdown of the final uncertainties can be found in Table~\ref{tab:allnucleon}.

\begin{figure}
    \centering
    \includegraphics[width=\columnwidth]{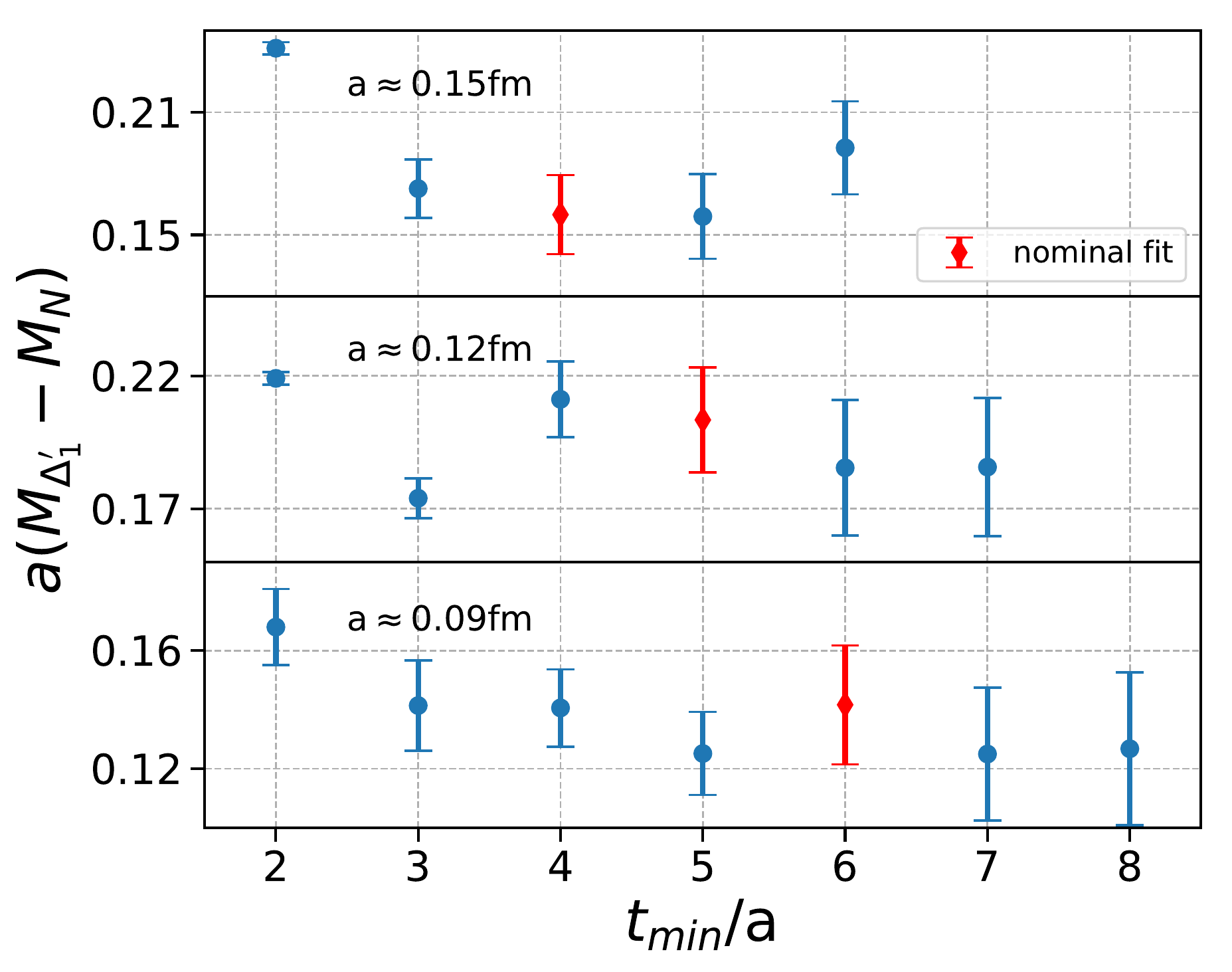}
    \caption{Stability of $a(M_N - M_{\Delta'_1})$ posterior as a function of $t_\text{min}/a$.
        The red diamonds denote the nominal fits.}
    \label{fig:tmindelta1stab}
\end{figure}

\begin{figure}
    \centering
    \includegraphics[width=\columnwidth]{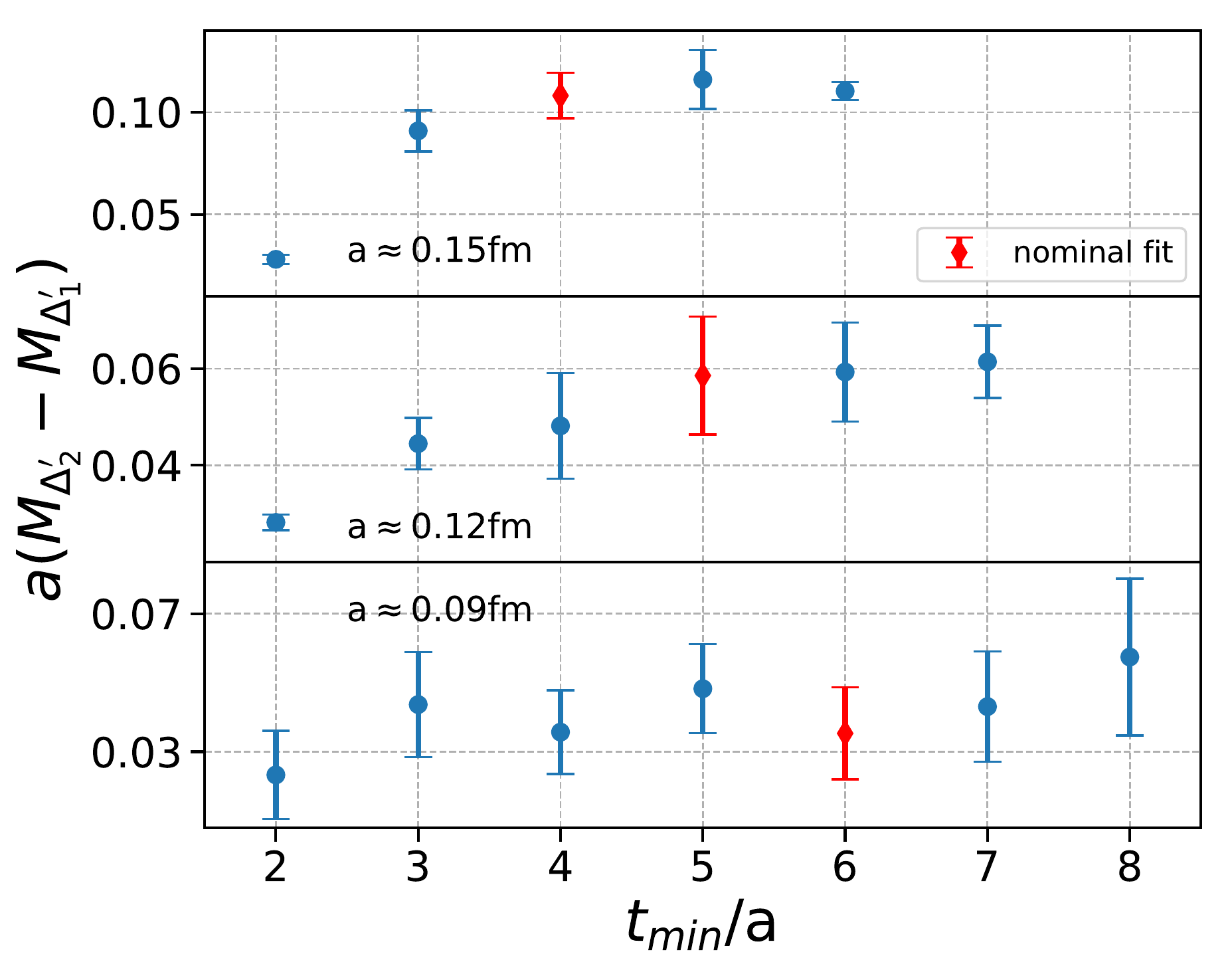}
    \caption{Stability of $a(M_{\Delta'_2}-M_{\Delta'_1})$ posterior as a function of $t_\text{min}/a$.
        The red diamonds denote the nominal fits.}
    \label{fig:tmindelta2stab}
\end{figure}

As discussed in Sec.~\ref{sec:BayesApproach}, we must also test the stability of $a(M_{\Delta'_1}-M_N)$ and $a\delta
M=a(M_{\Delta'_2}-M_{\Delta'_1})$ as functions of $t_\text{min}/a$.
Figures~\ref{fig:tmindelta1stab} and~\ref{fig:tmindelta2stab}
 demonstrate stability in both quantities similar to that seen for the nucleon mass
 in Fig.~\ref{fig:tminstab}.
Similar stability is observed in $a(M_{\Delta'_1} - M_N)$ and $a(M_{\Delta'_2}-M_{\Delta'_1})$ as additional radial excitations are
added.
These observations confirm our expectation that the two-state model is enough to describe the
$\Delta$-like contributions, especially since the corresponding fit quantities
 are not target observables, but instead nuisance parameters
 employed to reduce the bias in the nucleon mass from these states below
 the statistical precision of the correlator data.

\subsubsection{Negative-parity states}

As shown in Eq.~(\ref{eq:corr_raw}), our correlation functions contain contributions from negative-parity states with a
characteristic $(-1)^{t/a}$ in the time evolution.
The lowest-lying single-particle negative-parity state should be the $N(1520)$, while the lowest-lying two-particle threshold should
consist of $S$-wave $N\pi$ states with energy around and above $M_N + M_\pi=1100$~MeV.
With staggered quarks, the $N\pi$ states spread out over several levels, corresponding to the different pions tastes.
Experience from the meson sector and studies of nucleon correlators in chiral perturbation theory~\cite{Bar:2015zwa} provide no
reason to expect that $N\pi$ states contribute enough to single-particle-correlator data to be determined reliably.
That said, our operators differ from those in other formulations, being spread over a unit cube.
Particularly in light of the results of Ref.~\cite{Mahbub:2013bba}, we should keep an open mind.

We have studied these states with the Bayesian methodology.
The default prior for the lowest-lying negative-parity energy is $1400(200)$~MeV (cf., Table~\ref{tab:summaryprior}), which yields a
posterior centered around $1250(50)$~MeV for the $0.15$~fm and $0.12$~fm ensembles (cf., $aM_{-,1}$ in Table~\ref{tab:detailpos}).
The $0.09$~fm ensemble does not exhibit this behavior, instead returning a posterior of $1400(50)$ in agreement with the prior.
We have tried further Bayesian fits on the coarser ensembles with the prior centered within the range 1250--1500~MeV and a similar
width.
Such fits always return a posterior centered around 1250~MeV, and the same holds for any prior with significant probability at
1250~MeV.
To pull the posterior away from 1250~MeV, it is necessary to choose a prior with center separated from 1250~MeV by at least a few
multiples of the prior width.
In such cases, it is possible to obtain a posterior centered somewhere in the range 1250--1500~MeV.
Note also that even though the GEVP should not be expected to isolate any state besides the positive-parity nucleon and two
$\Delta'$s, it also returns $1400\pm40$~MeV in the negative-parity channel.

These findings suggest that the data contain some information about $N\pi$ states, but it is too weak to pull a ``bad prior'' down
to the expected threshold energy.
It is certainly more likely that the 1250~MeV
 signal consists of multiparticle states than the $N(1520)$.
While this study is interesting, a definitive work would require multibody interpolating operators.

In the context of our determination of the nucleon mass, this study of the negative-parity channel is relevant for the simple 
reason that the $N$-like posterior is stable under the changes mentioned here.

\subsection{GEVP Analysis}
\label{sec:GEVPAnalysis}

We have also performed a GEVP analysis in order to extract both $\lambda_1(t,t_0)$ and $\widetilde{\lambda}_1(t,t_0)$.
From the solutions of the GEVP in Eqs.~(\ref{eq:gevpright}) and (\ref{eq:gevpwindow}), we use Eqs.~(\ref{eq:gevpt0}) and
(\ref{eq:gevpwindowfit}) respectively to fit the ground state nucleon mass.
Table~\ref{tab:gevpfitparam} summarizes the fitting parameters used in the GEVP analysis.
\begin{table*}
    \centering
    \caption{The fit parameters used in the GEVP analyses to extract the ground state nucleon eigenvalues $\lambda_1$ and
        $\widetilde{\lambda}_1$, defined in Eqs.~(\ref{eq:gevpleft}) and~(\ref{eq:gevpwindow}).}
    \label{tab:gevpfitparam}
    \begin{tabular}{ccccc}
        \hline\hline
        Ensemble & $t_0/a$ ($\lambda_1$ fit)& $t_\text{min}/a$  ($\lambda_1$ fit) &$(t-t_0)/a $ ($\widetilde{\lambda}_1$ fit)&
            $t_\text{min}/a$  ($\widetilde{\lambda}_1$ fit) \\
        \hline
        1 & $3$&$2$&$2$&$3$\\
        2 & $5$&$1$&$2$&$4$\\
        3 & $5$&$2$&$2$&$5$\\
        \hline\hline
    \end{tabular}
\end{table*}
To compare consistently between both fitting strategies, we impose that $t_0+t_\text{min}$ in the $\lambda_1$ fits is equal to
$t-t_0 + t_\text{min}$ in the $\widetilde{\lambda}_1$ fits.

\begin{figure}
    \centering
    \includegraphics[width=\columnwidth]{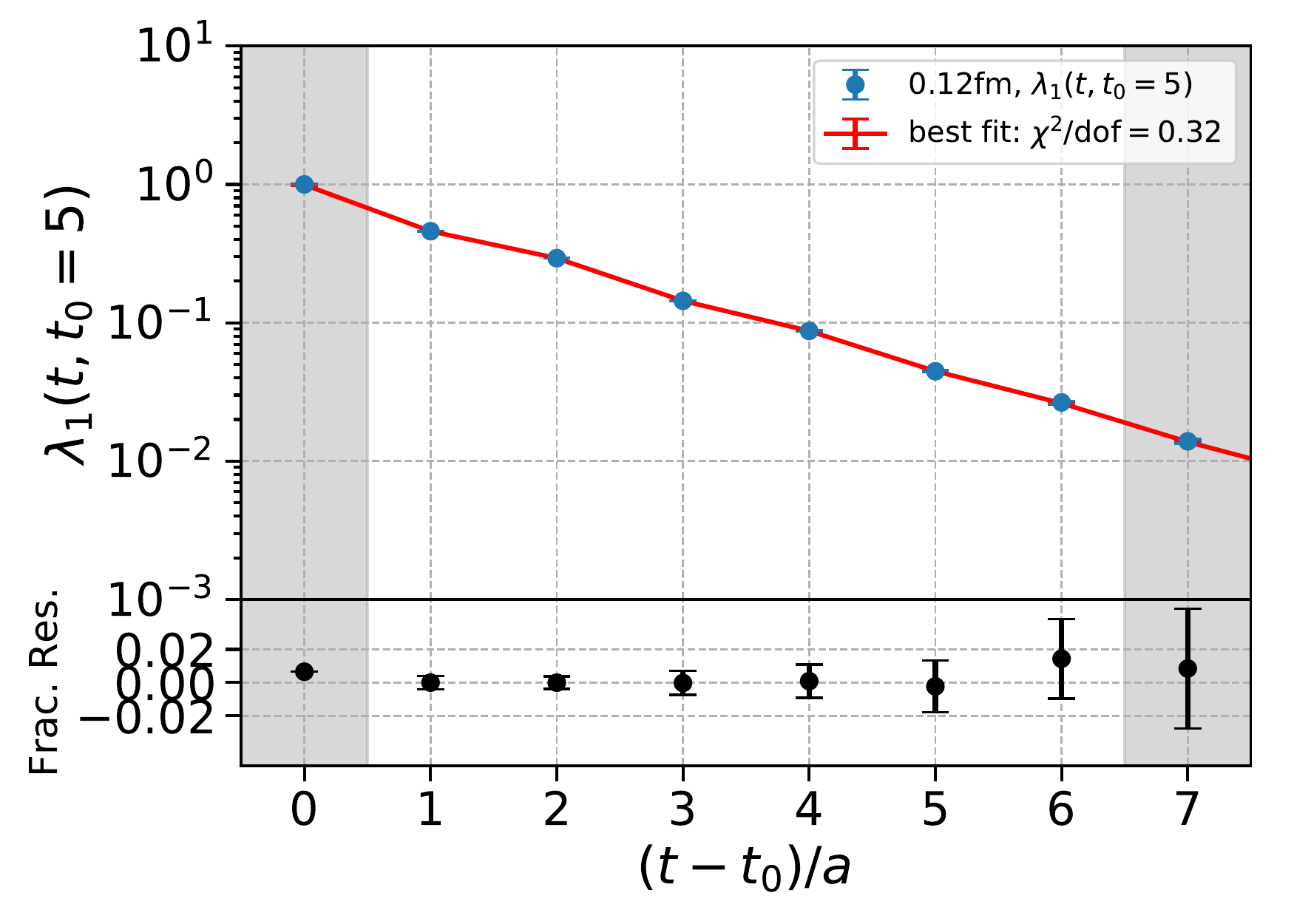}
    \caption{The GEVP eigenvalue $\lambda_1$ (defined in Eq.~(\ref{eq:gevpright})) on the $0.12$~fm ensemble.
        The fit function is given in Eq.~(\ref{eq:gevpt0}). The white region is the fitted time range.
        The fractional residues of the nominal fit (defined in Eq.~(\ref{eq:fracerr})) are shown in the bottom panel.}
    \label{fig:gevp012}
\end{figure}

For the $\lambda_1(t,t_0)$ analyses, we fix $t_0/a=3,5,5$ for the $0.15$~fm, $0.12$~fm, and $0.09$~fm ensembles to minimize the
effects of oscillating states.
In Fig.~\ref{fig:gevp012}, we have plotted the results for $\lambda_1(t,t_0=5)$ overlaid with the fit curve for the $0.12$~fm
ensemble.
Similar plots for the other ensembles are shown in Fig.~\ref{fig:gevp-other} in Appendix~\ref{appendix:moredata}.

\begin{figure*}
    \centering
    \includegraphics[width=0.98\columnwidth]{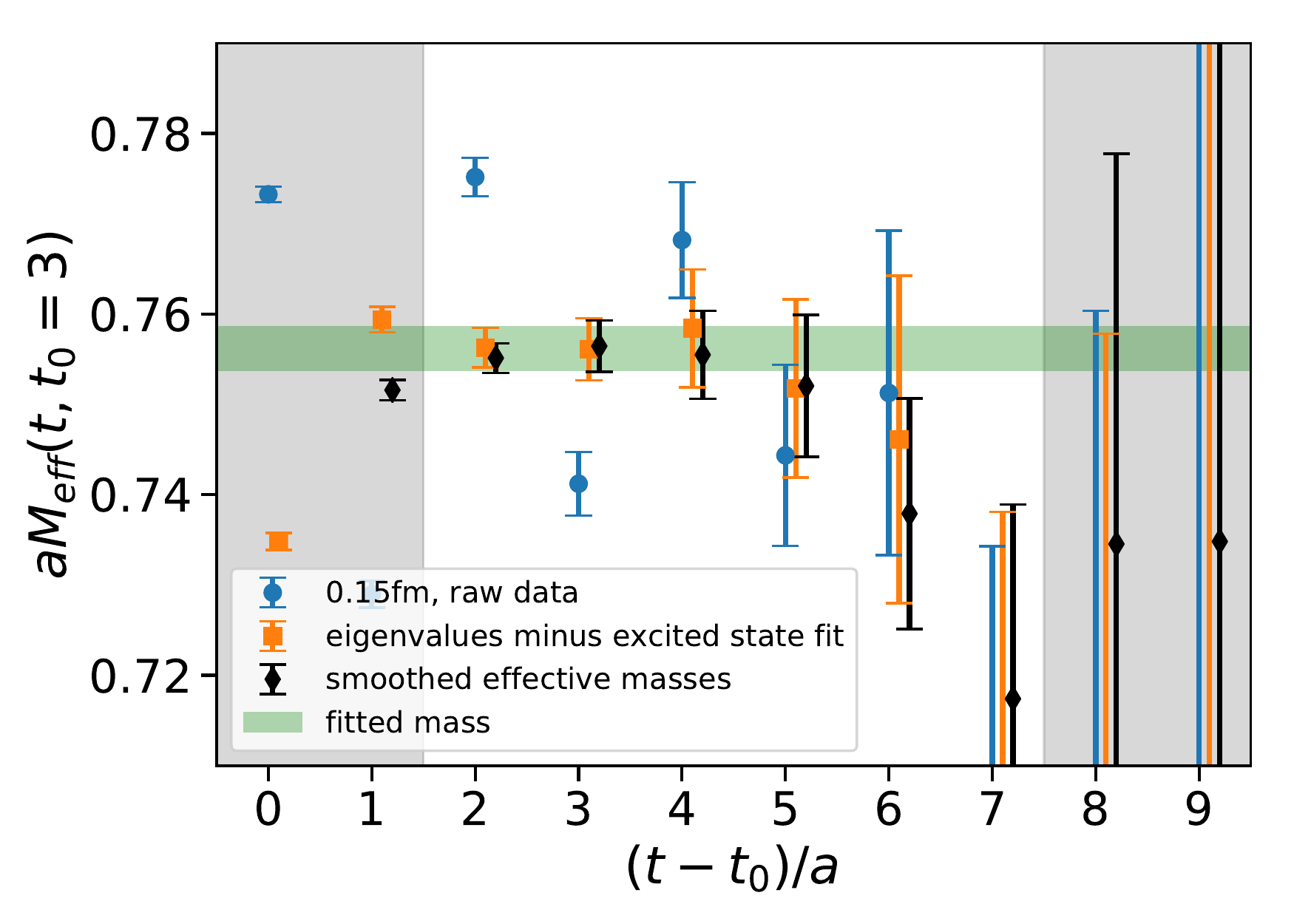} \hfill
    \includegraphics[width=0.98\columnwidth]{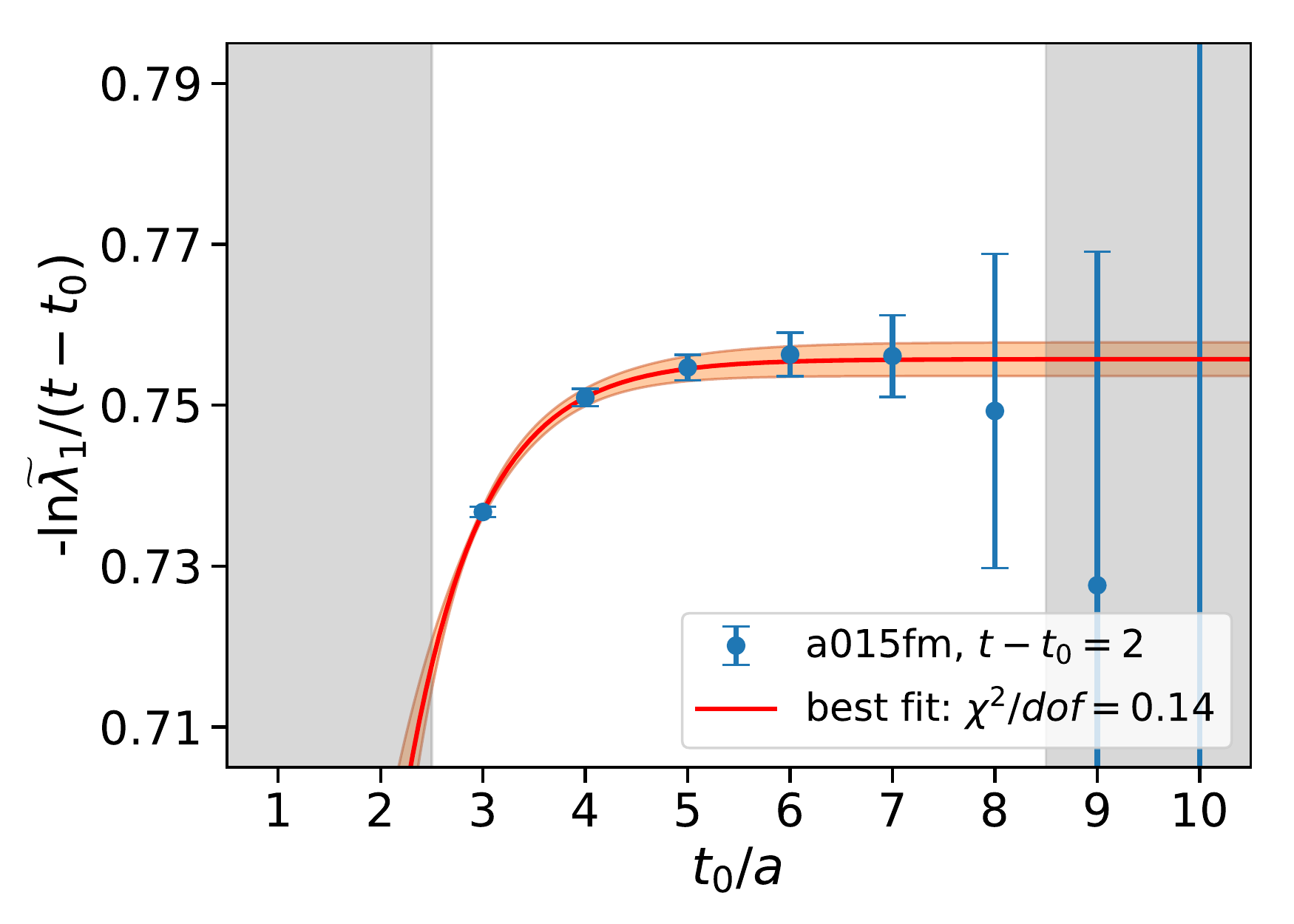}\\
    \includegraphics[width=0.98\columnwidth]{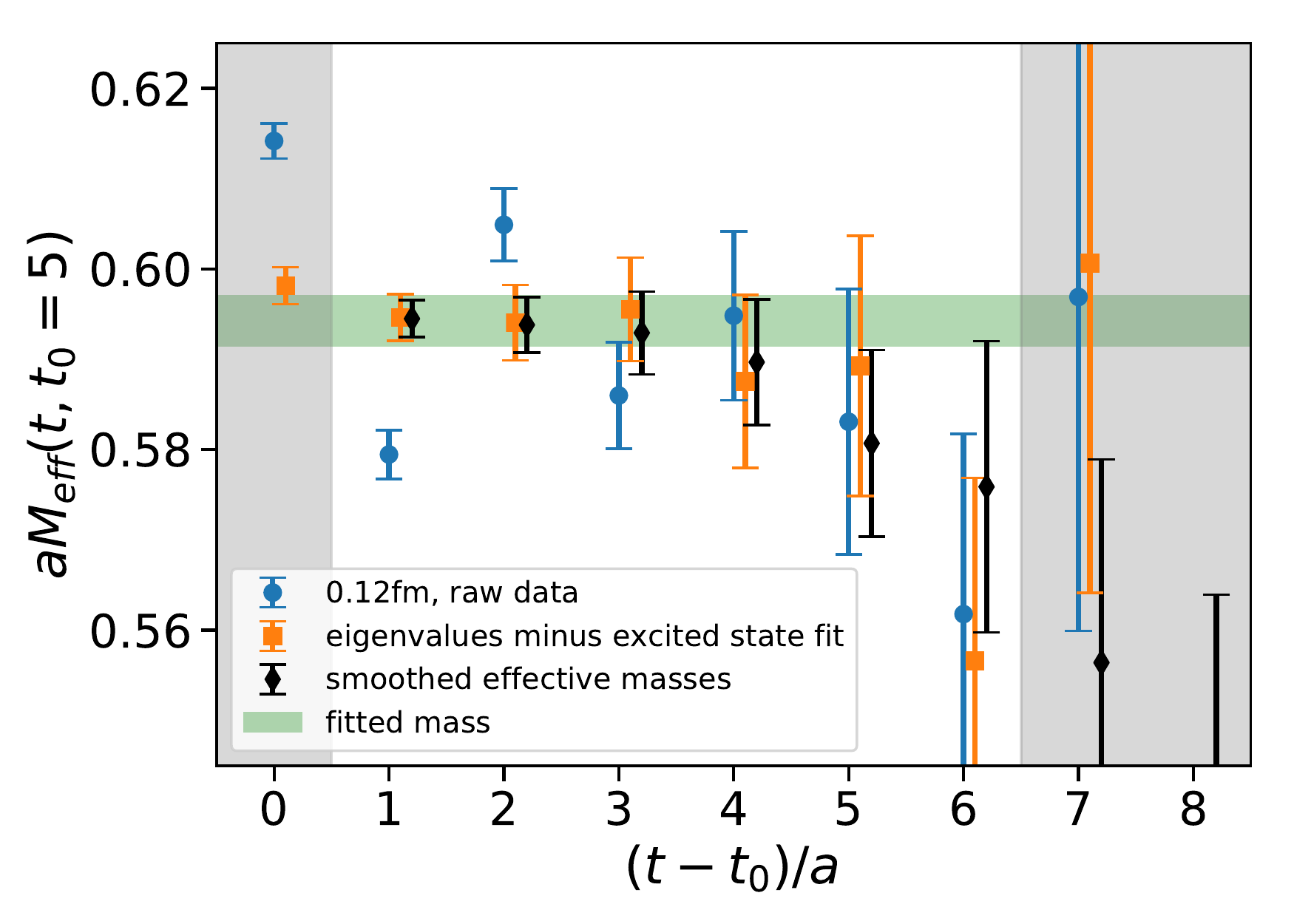} \hfill
    \includegraphics[width=0.98\columnwidth]{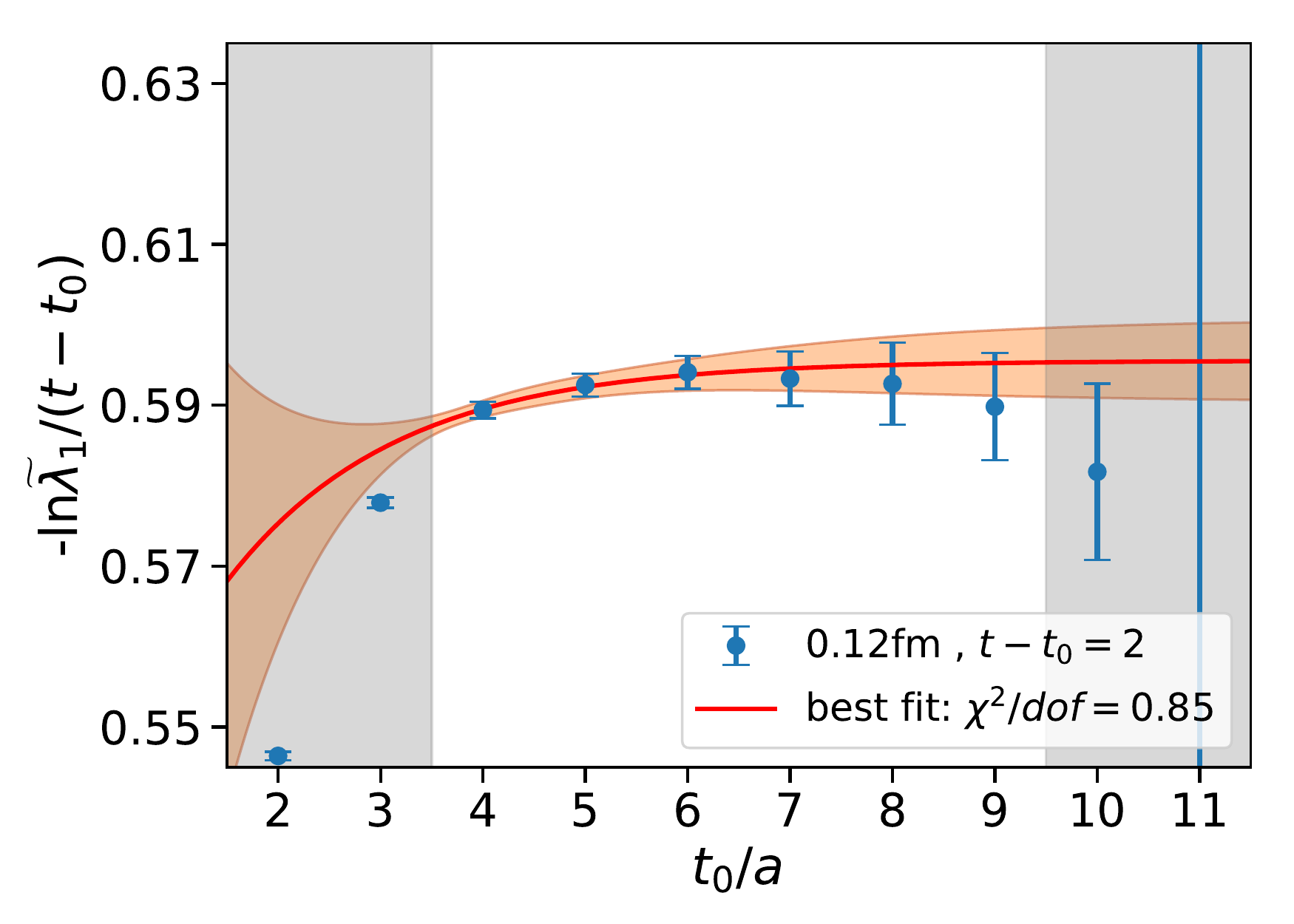}\\
    \includegraphics[width=0.98\columnwidth]{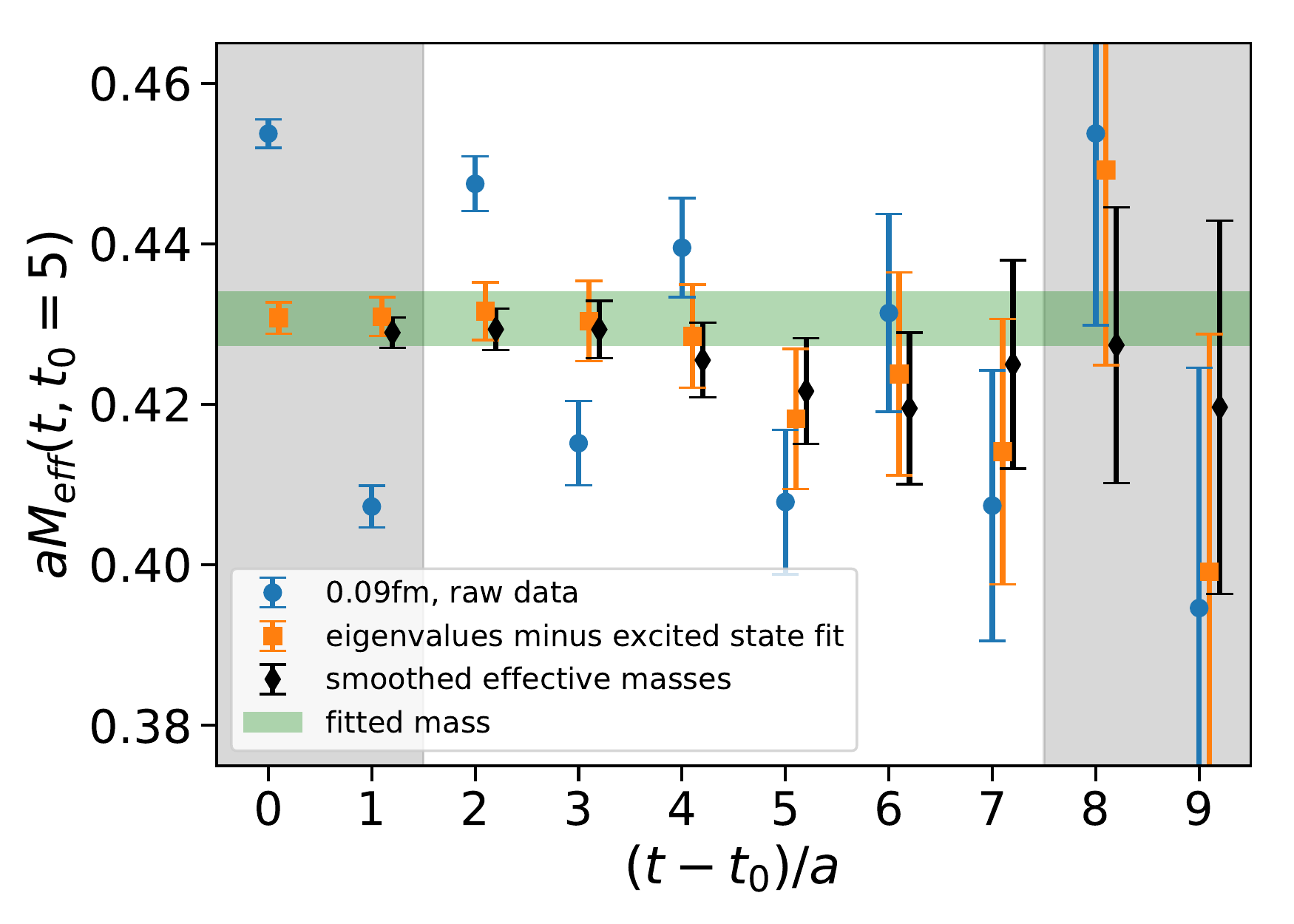} \hfill
    \includegraphics[width=0.98\columnwidth]{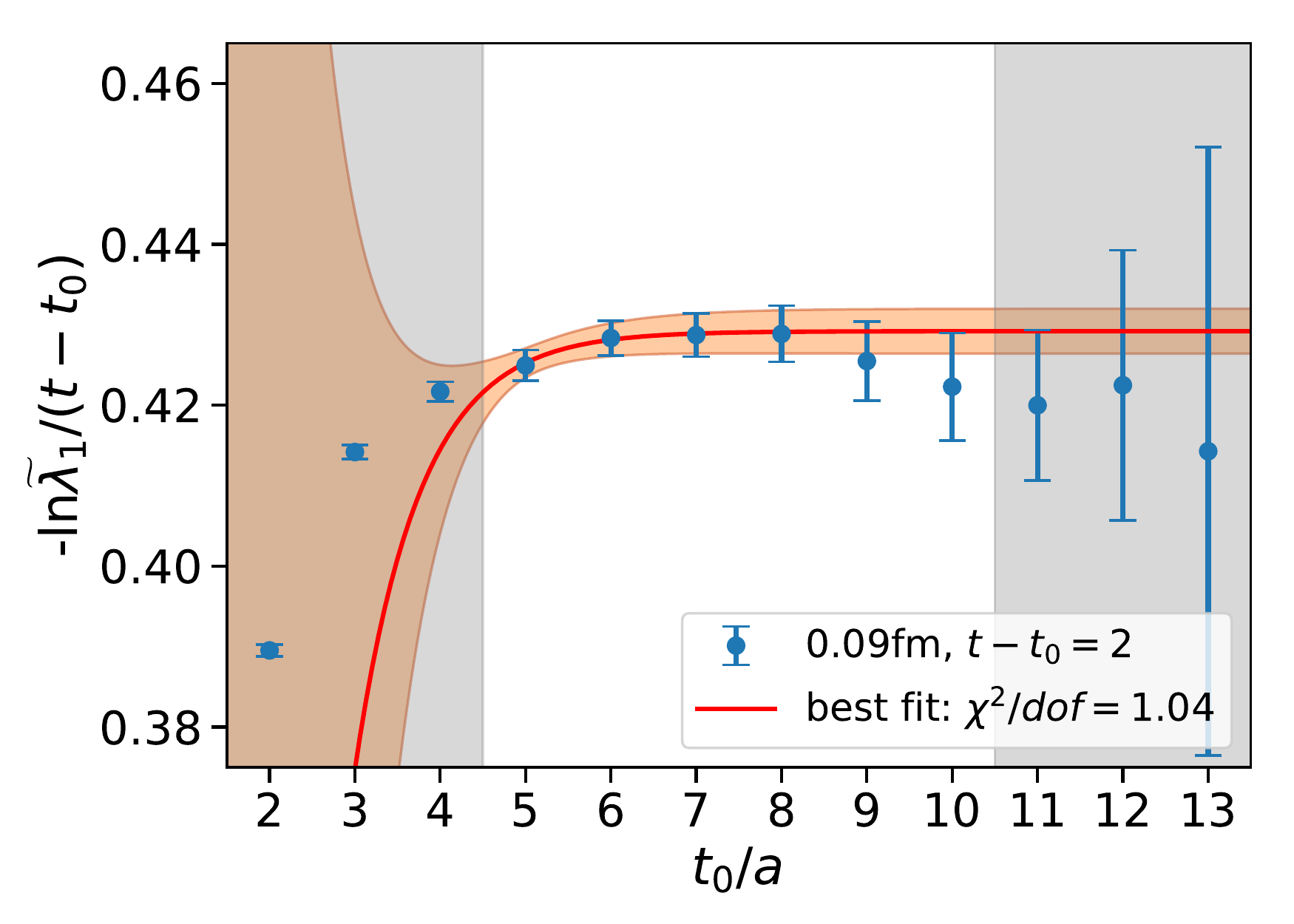}\\
    \caption{Results from the GEVP analyses.
        The left column shows the effective masses of the eigenvalue $\lambda_1$.
        The blue circles are the raw effective masses of $\lambda_1$, and the orange squares are the effective masses of the raw
        data which has the excited state central values from the nominal fit subtracted out.
        The black diamonds are the smoothed effective masses defined in Eq.~(\ref{eqn:smootheff}).
        The green band is the result of the Bayesian fit described in Sec.~\ref{sec:bayesianfitanalysis}.
        The right column shows the data for $\widetilde{\lambda}_1$ (defined in Eq.~(\ref{eq:gevpwindow})) and the corresponding
        plateau fits (brown bands) to $-\ln\widetilde{\lambda}_1/\tau$ for $\tau/a \equiv (t-t_0)/a = 2$ using
        Eq.~(\ref{eq:gevpwindowfit}).}
    \label{fig:allgevp}
\end{figure*}

Next, in the left column of Fig.~\ref{fig:allgevp}, we show three different definitions of the effective masses for the
$\lambda_1(t,t_0)$ data just described.
The first effective mass (solid blue circles) is the usual effective masses as defined in Eq.~(\ref{eq:meffdef}) using
$\lambda_1(t,t_0)$ instead of~$C(t)$.
The second effective mass (solid orange squares) is obtained by subtracting the central values of the excited states from the
nominal fit to $\lambda_1(t,t_0)$ and then using Eq.~(\ref{eq:meffdef}).
This third effective mass (black diamonds) is obtained by reducing the oscillations in the effective masses via the smoothed
effective mass~\cite{Bailey:2008wp,DeTar:2014gla}
\begin{equation}
	\overline{M}_{\text{eff}}(t) = \frac{1}{4}\left(2M_{\text{eff}}(t) + M_{\text{eff}}(t-1) + M_{\text{eff}}(t+1)\right).
	\label{eqn:smootheff}
\end{equation}
We can see that $\overline{M}_{\text{eff}}(t)$ shows no perceptible oscillations and agrees well with both
Bayesian estimates and the direct fitting to $\lambda_1(t,t_0)$, giving confidence in the reliability of the extracted nucleon mass.

The results of solving for the eigenvalues of the smoothed GEVP in Eq.~(\ref{eq:gevpwindow}) are shown in the right column of
Fig.~\ref{fig:allgevp}.
Without any post-processing, the plateaus of $\widetilde{\lambda}_1$ are clearly identifiable, with no sign of oscillations.
Thus, with no issues associated with negative-parity states, we simply perform unconstrained fits to Eq.~(\ref{eq:gevpwindow}) and
extract the $N$-like mass.

\begin{figure}[b]
    \centering
    \includegraphics[width=0.9\columnwidth]{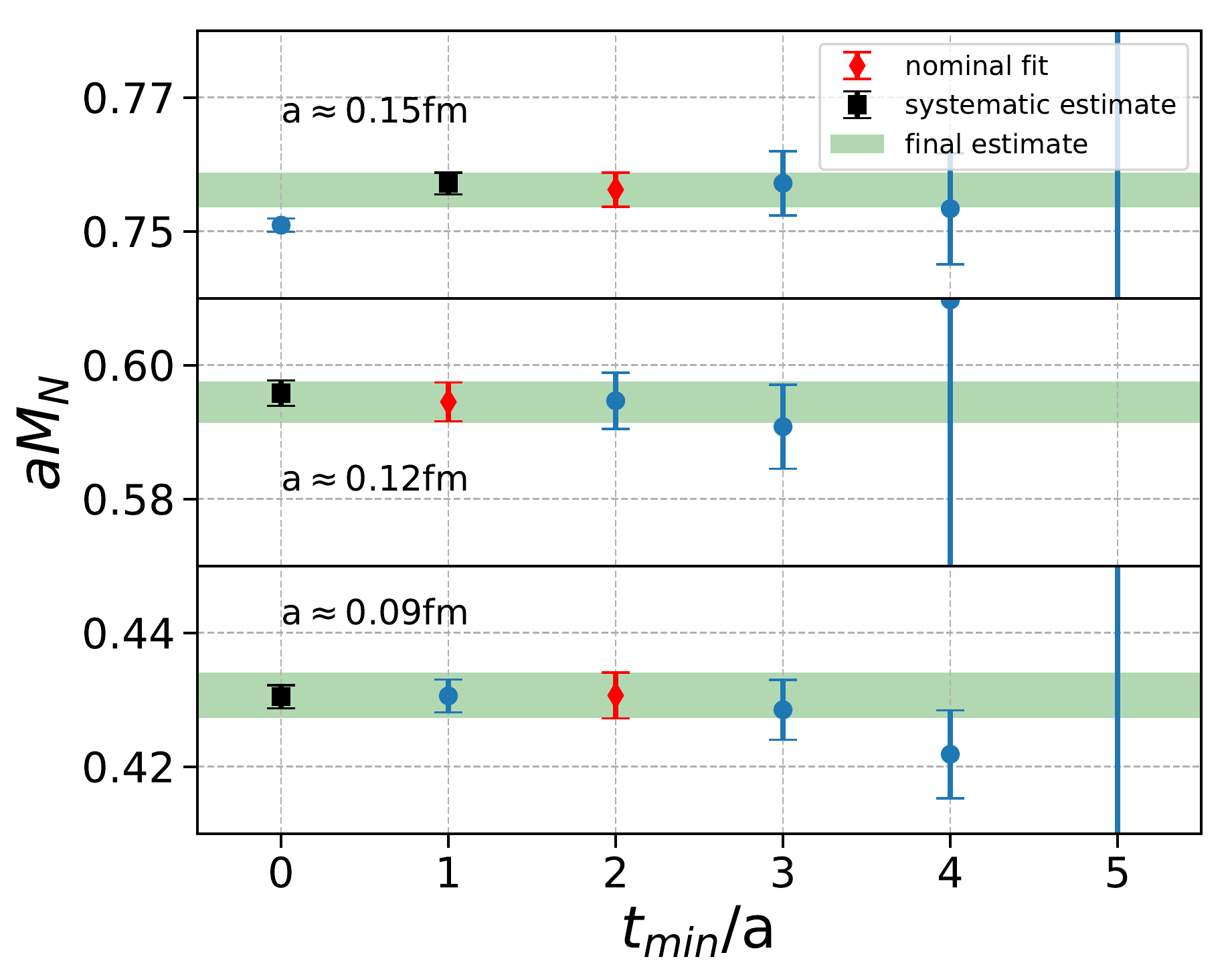} \\
    \includegraphics[width=0.9\columnwidth]{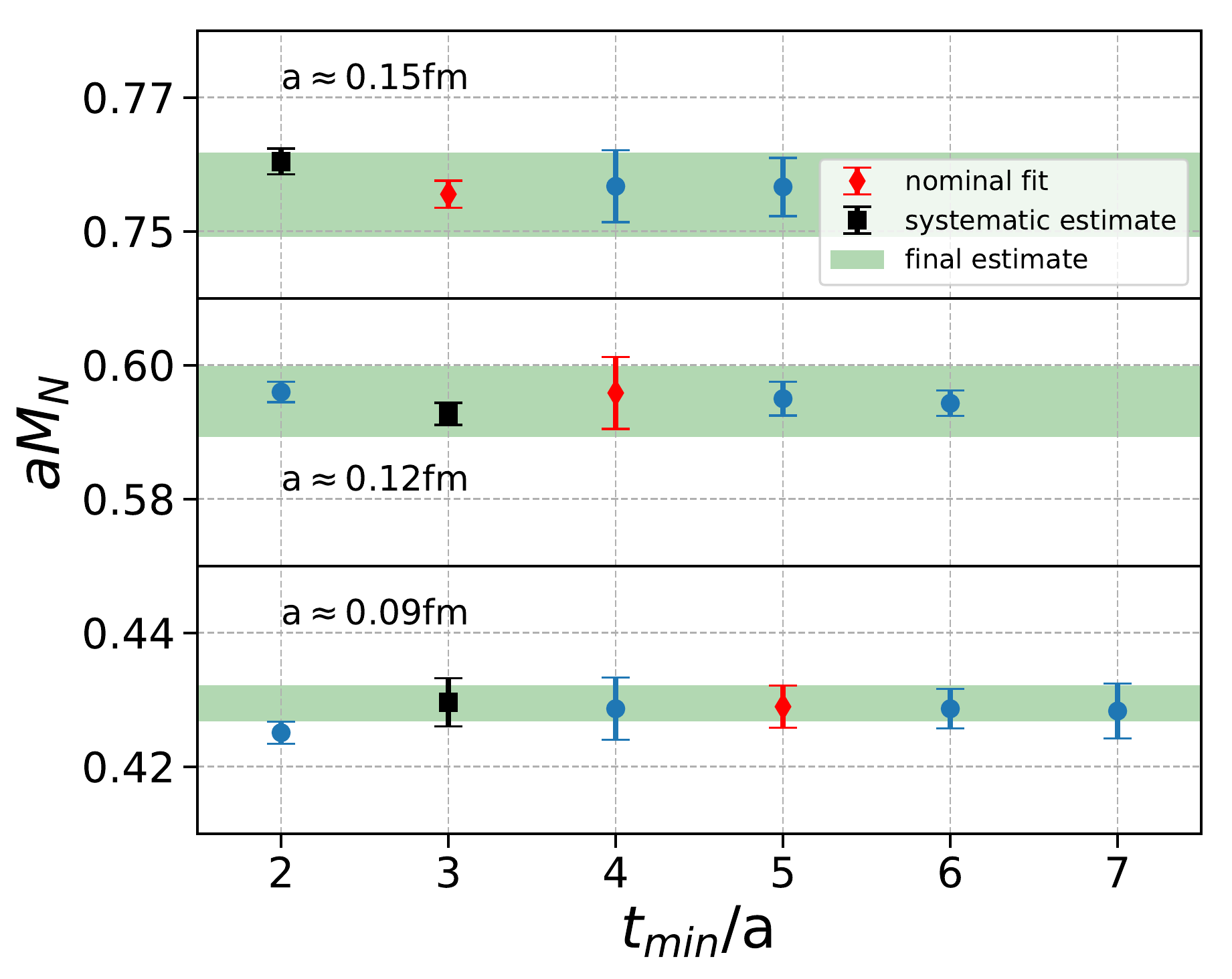}
    \caption{Stability of the nucleon mass as a function of $t_\text{min}/a$ for $\lambda_1$ (top) and
      $\widetilde{\lambda}_1$ (bottom).
      Note that the same physical distance of approximately $0.7$~fm is kept fixed
      on all ensembles for both $\lambda_1$ and $\widetilde{\lambda}_1$.
      Plotting symbols as in Fig.~\ref{fig:tminstab}.}
    \label{fig:stabgevp}
\end{figure}

We estimate the fitting systematics of both $\lambda_1$ and $\widetilde{\lambda}_1$ using the same procedures as in
Sec.~\ref{sec:BayeSys}.
These fitting systematics are listed in Table~\ref{tab:allnucleon}.
All aspects are qualitatively similar to those in Sec.~\ref{sec:BayeSys}.
Still, to emphasize this point, in Fig.~\ref{fig:stabgevp} we show the stability of the GEVP extracted nucleon mass as a function of
$t_\text{min}/a$.

Table~\ref{tab:allnucleon} lists the $N$-like mass estimates for all ensembles from the three analyses.
The extracted $N$-like mass values from the three Bayesian and GEVP analyses all agree within their (uncorrelated) $1\sigma$
uncertainties.
As the smoothed effective mass with the GEVP removes opposite parity contamination and shows a visible plateau, we take the $N$-like
mass values from $\widetilde{\lambda}_1(t,t_0)$ for use in a continuum extrapolation and all further discussion.
Note that since all three analyses agree with one another, the value of the continuum nucleon mass does not depend on which fitting
methodology we use.

\begin{table*}
    \centering
    \caption{The $N$-like mass, in lattice units, extracted from the three different fitting methodologies described in the text.
        $\widetilde{\lambda}_1$ are the fit results from Eq.~(\ref{eq:gevpwindow}), whereas $\lambda_1$ are from (\ref{eq:gevpt0}).
        The first uncertainties are statistical and the second are fitting systematics (described in Sec.~\ref{sec:BayeSys}).} 
    \label{tab:allnucleon}
    \begin{tabular}{cccc}
        \hline\hline
        Ensemble   & $\widetilde{\lambda}_1$ fit   & $\lambda_1 $ fit   & Bayesian fit   \\
        \hline
        1 & $0.7555(22)(59)$ & $0.7562(25)(9)$ & $0.7579(36)(48)$    \\
        2 & $0.5946(48)(22)$ & $0.5945(29)(13)$ & $0.5952(31)(1)$    \\
        3 & $0.4295(26)(8)$ & $0.4307(34)(2)$ & $0.4308(31)(14)$   \\
        \hline\hline
    \end{tabular}
\end{table*}

\section{Nucleon Mass Determination}
\label{sec:NucleonDetermination}

In the previous section we have extracted the $N$-like masses from physical-mass ensembles at three lattice spacings.
In this section, we use these values to extract a physical value of the nucleon mass that can be compared to experiment.

\subsection{Sources of systematic error}

In our calculation, the sources of systematic uncertainty include excited-state contamination, a slightly unphysical quark mass on
one ensemble, finite-volume effects, isospin-breaking effects, and scale-setting errors.
Errors arising from excited-state contamination have already been addressed, estimated, and controlled in
Sec.~\ref{sec:DataAnalysis}.

As mentioned in Table~\ref{tab:ensembledetails}, all three of our ensembles have nearly-physical pion masses.
As such, we avoid the potentially large chiral extrapolation errors (compounded by using a slowly converging chiral fit function
\cite{WalkerLoud:2008pj}) and do not need to include an error from a chiral extrapolation.

Nevertheless, the mistuned light-quark mass on the $0.09$~fm ensemble is an important effect.
Although the other two ensembles have negligible mistuning, the task at hand is to combine the three results.
The taste-Goldstone pion on the $0.09$~fm ensemble has a mass of $128.3(7)$~MeV, as mentioned in Sec.~\ref{tab:ensembledetails},
which is about $7$~MeV smaller than the taste-Goldstone pion mass from the other two ensembles.
References~\cite{Yang:2018nqn, Walker-Loud:2013yua} study the nucleon mass vs.\ pion mass and observe that
$M_N\approx800~\text{MeV}+M_\pi$ (within uncertainties) over a wide range of pion mass and with different actions.
Since our mistuning is small, this observation suggests that the nucleon mass on the $0.09$~fm ensemble is approximately 7~MeV too
small.
We therefore take account of the $0.09$~fm ensemble mistuning by applying a shift of $+7(7)$~MeV before performing the continuum
extrapolation.

We can check this estimate with baryon chiral perturbation theory.
A~detailed NNLO SU(2) analysis leads to the conclusion that the perturbative expansion for the nucleon mass converges for
$M_{\pi}<350$~MeV~\cite{Beane:2004ks}, which easily covers the pion masses used in this work.
This SU(2) analysis gives a $+5$~MeV shift to our $N$-like mass on the $0.09$~fm ensemble from mistuning which is well within the
range $7(7)$~MeV from the linear fit.

Another check is to use a Taylor expansion of $M_N$ that is linear in $M_{\pi}^2$.
The slope is the nucleon sigma term $\sigma_{N\pi}$.
It can be determined from lattice QCD: the $N_F = 2+1+1$ FLAG average \cite{Aoki:2019cca,Alexandrou:2014sha} is
$64.9(1.5)(13.2)$~MeV, while an estimate from the
Roy-Steiner equations~\cite{Hoferichter:2015dsa} yields $\sigma_{N\pi} = 59.1(3.5)$~MeV.
Using either value of $\sigma_{N\pi}$ yields a $6$~MeV shift to the $N$-like mass, which is also within the $7(7)$~MeV range from
our first estimate.

Single-particle finite-volume errors are exponentially small in $M_{\pi}L$~\cite{luscher1986}.
For the $a\approx0.15$~fm, $0.12$~fm, and $0.09$~fm ensembles, the corresponding values of $M_{\pi}L$ are $3.4$, $4.0$, and $3.7$
respectively.
The lattice data of Ref.~\cite{Durr:2008zz} supports a model based on a resummation of the L\"uscher
formula~\cite{Colangelo:2010ba}, but the statistical errors on the nucleon masses are too large---around $5\%$---to be conclusive.
Applying this model with our ensemble parameters, only the $a=0.15$~fm nucleon mass would receive an appreciable correction, namely
$-5$~MeV.
However, applicability of this model is still unclear, as Ref.~\cite{Bali:2014nma} observes no change in the nucleon mass with a
variation of $M_{\pi}L$ between 3.4 and~6.7, with $\sim 1\%$ statistical and fitting uncertainties.
Moreover, Ref.~\cite{Jang:2019jkn} has ensembles which have $M_{\pi}L$ ranging from 3.3--5.5 and finds a positive $4$~MeV shift
between $M_{\pi}L=3.3$ and~$4$, beyond which any further change is negligible.
Due to this conflicting information, even about the sign of the correction, we apply a $0(5)$~MeV error on the $a=0.15$~fm nucleon
mass arising from finite-volume corrections.
Our final result is insensitive to this finite-volume correction and we leave an in-depth study of potential finite-volume
corrections of the nucleon mass on the MILC HISQ ensembles to a future investigation.

Our lattice simulation is isospin symmetric, i.e., the up- and down-quark masses have the same value and quantum electrodynamics is
omitted.
Both of these effects effects give rise to the proton-neutron mass difference, which is less than $1$~MeV.
As $1$~MeV is small compared to our statistical error, we apply no additional uncertainty from these effects.

\subsection{Continuum extrapolation}
\label{sec:nucmassfinal}

\begin{figure}
    \centering
    \includegraphics[width=\columnwidth]{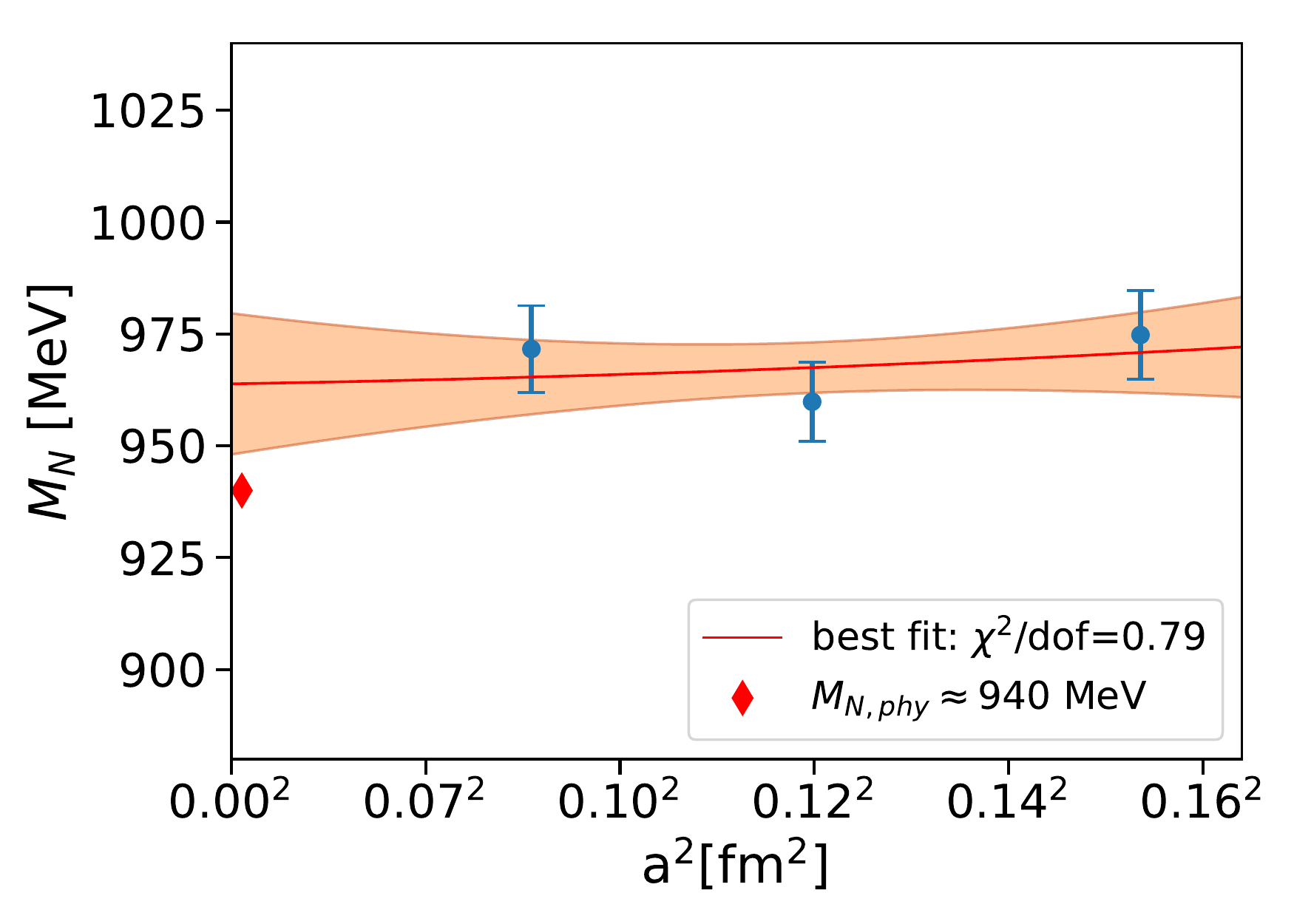}
    \caption{The continuum extrapolation of the nucleon mass using Eq.~(\ref{eq:coneq}).}
    \label{fig:continuum}
\end{figure}

Using the ensemble-by-ensemble nucleon masses given in Sec.~\ref{sec:GEVPAnalysis}, we can include all sources of systematic error
and perform a continuum extrapolation to produce a nucleon mass which can be compared to experiment.
To do so, we perform a Bayesian fit to the functional form
\begin{equation}
	M_{N}(a) = M_{N,\text{phy}}\left\{1 + o_2 \left(\Lambda_{\text{QCD}}a\right)^2
	+ o_4\left(\Lambda_{\text{QCD}}a\right)^4
	\right\},
	\label{eq:coneq}
\end{equation}
where $M_{N,\text{phy}}$ is the physical nucleon mass, $\Lambda_{\text{QCD}}$ is taken to be 500~MeV, and $o_2$ and $o_4$ are fit
coefficients.
Equation~(\ref{eq:coneq}) is fit to the $\tilde{\lambda}_1$ results in Table~\ref{tab:allnucleon} after converting to~MeV.
The lattice spacings for this conversion are taken from Table~\ref{tab:ensembledetails} and are assumed to be uncorrelated with
other errors.
We do not constrain the prior on $o_2$ and choose an order-one prior for $o_4=0(1)$.
The list of priors and posteriors are given in Table \ref{tab:connfit}, and the continuum extrapolation is shown in
Fig.~\ref{fig:continuum}.
\begin{table}
	\centering
	\caption{The results from the nucleon mass continuum extrapolation.}
	\label{tab:connfit}
	\begin{tabular}{ccc}
	    \hline\hline
		Parameter & Prior & Posterior \\
		\hline
		 $M_{N,\text{phy}}$ [MeV]  &  $940 (50)$ & $964 (16)$\\
		 $o_2$ & unconstrained  & $0.02 (23)$\\
		 $o_4$ & $0.0 (1.0)$    & $0.17 (97)$ \\
	    \hline\hline
	\end{tabular}
\end{table}
If we constrain $o_2=0.0(3)$, in line with the HISQ action improvement, we obtain the same posterior as with no constraint on~$o_2$.
We choose the fit with unconstrained $o_2$ for our central result.

Our final estimate for the continuum nucleon mass, including all sources of systematic errors, is 
\begin{align}
    M_{N, \text{phy}}
    &= 964(8)_\text{stat}(5)_\text{fit}(4)_a(3)_\text{FV}(8)_\text{mis}~\text{MeV}, 
    \label{eqn:finalMNbreakdown}\\
    &= 964(16)~\text{MeV},
    \label{eqn:finalMN}
\end{align}
where ``stat'', ``fit'', ``$a$'', ``FV'', and ``mis'' represent the statistical, fitting, scale-setting, finite-volume, and the
0.09~fm ensemble quark mistuning errors contribution to the final continuum nucleon mass uncertainties.
With no prior constraint on $M_{N,\text{phy}}$, we find the posterior $M_{N,\text{phy}}=966(8)_\text{stat}$ and the same systematic
uncertainties as in Eq.~(\ref{eqn:finalMNbreakdown}).

Our determination of the nucleon mass is $1.6\sigma$ above the experimental value, which arises from the high nucleon mass
on the $0.09$~fm ensemble, as can be seen clearly in Fig.~\ref{fig:continuum}.
Either higher statistics at $a\approx0.09$~fm or additional ensembles with smaller lattice spacings could be employed to see whether
this is a statistical fluctuation.
Such work is planned.

It is interesting to see what happens if we do not apply a $+7(7)$~MeV correction on this ensemble for the quark-mistuning.
If instead we apply a $0(7)$~MeV correction, the final value of our nucleon mass is $955(16)$~MeV which is within the $1\sigma$
error of our final result in Eq.~(\ref{eqn:finalMN}), as expected.
Even though this result is closer to experiment, we do not prefer it, because the size and direction of the shift is on solid
footing.
The only robust way to reconcile this issue is to generated an ensemble with a more precisely tuned light-quark mass.

\subsection{Comparison with Other Calculations}
\label{sec:comparison}

The average of the proton and neutron masses found in experiment~\cite{Tanabashi:2018oca} is $939$~MeV, and the uncertainties on
these masses are about the level of one part per million.
In this work, we determine a nucleon mass of $964(16)$~MeV.
Although this work is the first to determine the nucleon mass from first principles using staggered baryons, other first-principles
results exist in the literature.
Those which quote a full error budget, and hence are comparable to the present work in scope, can be found in
Refs.~\cite{Durr:2008zz,Jang:2019jkn}.%
\footnote{There are other studies~\cite{Bali:2012qs,Bali:2014nma} that are comparable to ours in scope but which use the nucleon
mass to set the lattice scale, and hence cannot be compared to our final result.
A very recent paper~\cite{Abramczyk:2019fnf} reports a result on one lattice spacing, using it as a cross check and, therefore, 
does not provide a full error budget.}

\begin{figure}
    \centering
    \includegraphics[width=\columnwidth]{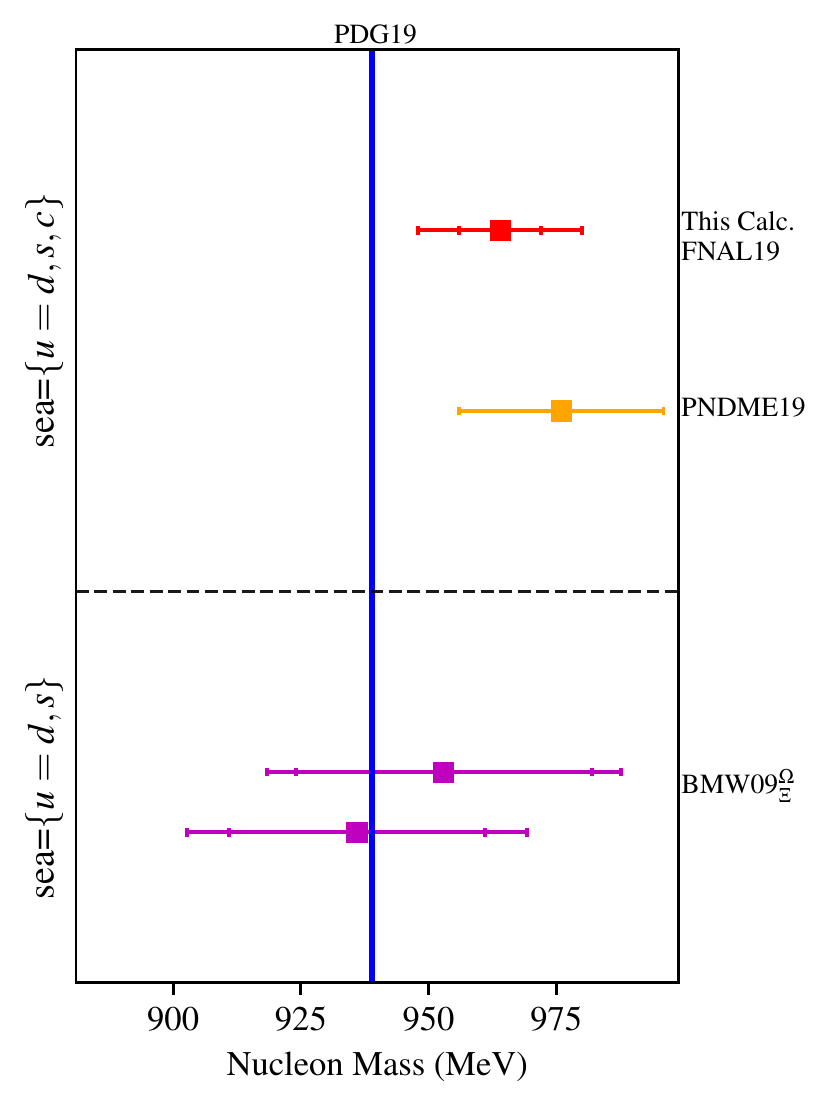}
    \caption{A comparison of our result against others found in the literature, as discussed in Sec.~\ref{sec:comparison}.
        The outer bars denote the $1\sigma$ error from both statistics and systematics, while the inner bar consists of the
        statistical error only.
        The labels on the y-axis denote the PNDME collaborations result \cite{Jang:2019jkn} and the BMW result \cite{Durr:2008zz}.
        The upper BMW result sets the scale using the $\Omega$ baryon and the lower sets the scale with the $\Xi$ baryon.}
    \label{fig:Comparison}
\end{figure}

Reference~\cite{Durr:2008zz} uses a tree-level $\order(a^2)$-improved Symanzik gauge action, $(2+1)$ tree-level improved Wilson
fermions, and includes 20 different ensembles covering three lattice spacings ($0.13$~fm, $0.09$~fm, and $0.07$~fm), 4--5 different
light quark masses (giving pion masses ranging from approximately $190$--$650$~MeV), three ensembles with different physical
volumes, and eight ensembles with different strange quark masses.
Reference~\cite{Durr:2008zz} gives two determinations of the nucleon mass: $936(25)(22)$~MeV and $953(29)(19)$~MeV, where the first
error is statistical and second is systematic.
Here, the two values differ by the quantity used to set the scale: the first uses the $\Xi$ baryon mass and the second uses
the~$\Omega$.
Reference~\cite{Jang:2019jkn} uses 11 ensembles generated by the MILC collaboration (general details of which are in
Sec.~\ref{sec:SimDetails}).
The ensembles have $(2+1+1)$-HISQ sea-quarks with pion masses ranging from $128$--$320$~MeV and four lattice spacings covering
$0.06$~fm, $0.09$~fm, $0.12$~fm, and $0.15$~fm.
Their valence quarks have the improved Wilson-clover action.
With a combined chiral--continuum--finite-volume ansatz for the systematic extrapolation, they find a nucleon mass of $976(20)$~MeV.
As such, our result is the most precise first-principles determination of the nucleon mass in the literature, and is relatively low
cost calculation.
A comparison of the results from the three collaborations is shown in Fig.~\ref{fig:Comparison}.

\section{Discussion and Conclusions}
\label{sec:conclusions}

In this work, we have extracted a precise and accurate value for the nucleon mass, including a full error budget, using lattice QCD
with the HISQ action for both valence and sea quarks.
We find $M_N=964\pm16$~MeV [Eq.~(\ref{eqn:finalMN})].
Some notable details of our simulations are three lattice spacings ranging from $a = 0.09$--0.15~fm, all of which are tuned to a
nearly-physical pion mass.
All ensembles have $u, d, s$ and $c$ quarks in the sea.
We employ three different fitting methodologies: multistate constrained Bayesian curve fitting and two versions of the generalized
eigenvalue problem approach.
Within each approach, we verify stability under variation of the fitting range, the numbers of states, and other choices.
The superb consistency between the results of these fitting procedures demonstrates their robustness and accuracy.

Our results suggest a promising outlook for staggered baryon lattice QCD.
As can be seen in Eq.~(\ref{eqn:finalMNbreakdown}), our dominant error arises from the light-quark-mass mistuning on the $0.09$~fm
ensemble, compounded by the continuum extrapolation.
The most direct method to reduce this error would be to generate an ensemble with a better tuned light-quark mass.
Alternatively, an ensemble with slightly heavier light quarks would allow retuning via interpolation.
Further, with three ensembles a $1\sigma$ statistical fluctuation on one of them is not unlikely.
As can be seen in Fig.~\ref{fig:continuum}, the $0.09$~fm ensemble seems to exhibit such a fluctuation.
Another data point at smaller lattice spacing would help alleviate effects from both the mistuning and this potential fluctuation.

After the error from mistuning, the next largest error comes from statistics.
Reducing the statistical error is possible by adding additional configurations, or adopting techniques such as the truncated solver
method~\cite{Jang:2019jkn,Bali:2009hu}, low-mode averaging~\cite{Giusti:2004yp,DeGrand:2004qw}, or all-mode averaging
\cite{Blum:2012uh,Shintani:2014vja}.
Another way to reduce the statistical error is to compute the matrix correlation functions for the $8$ and $8'$~irreps.
In the continuum limit, where taste symmetry is restored, all $N$-like masses should tend to the same point.
Thus, the final result could be improved by combining the information from all three baryon irreps and enforcing a common continuum
limit.
Finally, one could introduce more sophisticated smeared interpolating operators.
We have carried out initial tests with stride-two staggered smearing functions and find them to be promising.

Staggered-baryon methodology can be straightforwardly applied to compute further baryon properties.
The $\Omega$ baryon mass is especially interesting for scale setting in lattice QCD~\cite{Toussaint:2004cj}.
It is long-lived and composed of three strange quarks, so the quark propagators are computationally cheaper than those for light
quarks, and its two-point correlation function has a better signal-to-noise ratio.
Moreover, the $\Omega$ baryon mass is known unambiguously from experiment, unlike the pion decay constant, which relies on
determinations of $|V_{ud}|$ from nuclear beta decay.
Robust and precise scale setting is, of course, crucial as total error budgets for lattice QCD fall below 1\%, which is not only
feasible but, in the case of hadronic-vacuum-polarization contribution to the muon $g-2$, necessary~\cite{Meyer:2018til}.
We have started such work on the ensembles listed in Table~\ref{tab:ensembledetails}), i.e., ensembles that omit QED, which is
especially relevant with a charged particle.

This work also paves the way for all-staggered computations of three-point baryon correlation functions.
Now that we have identified $N$-like states via both the GEVP and multistate Bayesian curve fitting, we can have confidence that the
extracted $N$-like matrix elements do indeed correspond to the physical nucleon.
Especially important for neutrino scattering experiments, for example, is the nucleon axial form factor.
The first step in such a program is to calculate the axial charge, $g_A$, which is just the form factor at zero-momentum transfer.
Because it is precisely known from neutrino beta decay, $g_A$ serves as a calibration point for lattice QCD.
Indeed, we consider the nucleon mass presented here more important as a prerequisite for future all-staggered calculations of
nucleon matrix elements than as a test of (lattice) QCD.

\acknowledgments
We are grateful to the MILC collaboration for the use of the source code adapted to generate the correlators in this study and for
permission to use their 2+1+1-flavor gauge-field ensembles.
Computations for this work were carried out on facilities of the USQCD Collaboration, which are funded by the Office of Science of
the U.~S.\ Department of Energy.
This manuscript has been authored by Fermi Research Alliance, LLC under Contract No.~DE-AC02-07CH11359 with the U.~S.\ Department of
Energy, Office of Science, Office of High Energy Physics.
Brookhaven National Laboratory is supported by the U.~S.\ Department of Energy under Contract No.~DE-SC0012704.

\appendix

\section{Irreducible Representations of GTS}
\label{app:Group}

In this Appendix, we are concerned with the geometric symmetries of staggered fermions, in order to classify physical baryon states.
The properties of physical states under charge conjugation, baryon number, and the exact axial symmetry are straightforward.
Complications arising from the interplay of flavor and taste are deferred to Appendix~\ref{app:irrepident}.

The emergence of four Dirac fermions in the continuum limit stems from the group theory of the shifts, which translate fields by a
single lattice site and multiply fermion fields by a convention- and $x$-dependent staggered sign factor such that
\begin{equation}
    S_\mu S_\nu S_\mu^{-1} S_\nu^{-1} = (-1)^F, \quad \nu\neq\mu,
    \label{eq:SSSS}
\end{equation}
where $F$ is fermion number, and $\mu$ and $\nu$ denote directions of translations.
The shift $S_\mu$ is defined in, for instance, Ref.~\cite{Golterman:1984cy}.
It is a translation by one lattice spacing in the $\mu$ direction, multiplied by a
 sign factor that depends on $x_\nu$, $\nu\neq\mu$, but not on $x_\mu$.
Therefore, $T_\mu\equiv S_\mu^2$ is a normal translation by two lattice sites for all fields.
It is convenient (and permissible, because the translations commute with the shifts) to remove the translation part of the shifts by
introducing $\Xi_\mu\equiv S_\mu T_\mu^{-1/2}$, where $T_\mu^{-1/2}$ is a formal square root of $T_\mu^{-1}$ in any representation
of the symmetry group.
Nowadays one calls the $\Xi_\mu$ the ``taste'' generators (reserving ``shift'' for $S_\mu$ and ``flavor'' for flavor).
They satisfy Eq.~(\ref{eq:SSSS}) and $\Xi_\mu^2=1$; thus, they generate the Clifford group~$\Gamma_4$.

On physical states, $T_4$ is the (two-timeslice~\cite{Osterwalder:1977pc,Sharatchandra:1981si}) time-evolution operator, also known
as the transfer matrix.
The classification of states in Hilbert space hinges on the symmetries that commute with~$T_4$.
These are the spatial translations, all four shifts, and (assuming the same extent in all three spatial directions) the
rotation-reflection symmetries of the cube.
Thus, on a $(2N)^3$ spatial lattice with (anti)periodic boundary conditions, the geometric symmetry group of the staggered-fermion
transfer matrix is~\cite{vandenDoel:1983mf,*Golterman:1984cy,*Golterman:1985dz,Golterman:1984dn,Kilcup:1986dg}
\begin{equation}
    G = (\mathds{Z}_N^3 \times \Gamma_4) \rtimes \text{W}_3,
\end{equation}
where the first two factors are the groups generated by the (two-site) spatial translations $T_i$ and the tastes $\Xi_\mu$.
$\text{W}_3$ is the cubic rotation-inversion symmetry group, generated by $\pi/2$ rotations in the $ij$ plane, $R_{ij}$, and spatial
inversion,~$I_S$.
The last product is semidirect, because $R_{ij}T_j=T_i R_{ij}$, $R_{ij}\Xi_j=\Xi_i R_{ij}$, etc.

Earlier work~\cite{Golterman:1984dn,Kilcup:1986dg} has shown that the problem of finding irreducible representations can be
simplified by grouping the generators judiciously.
In particular, the spatial taste generators can be chosen to be $\Xi_{123}\equiv\Xi_1\Xi_2\Xi_3$ and any two
$\Xi_{ij}\equiv\Xi_i\Xi_j$.
Further, the combination $P\equiv\Xi_4I_S$ commutes with the taste generators as well as with rotations, so in the continuum limit
it is the analog of parity~\cite{Golterman:1984dn,Kilcup:1986dg}.
It is convenient to use $I_S$ as a generator and leave parity until the end; then~\cite{Meyer:2017ddy},
\begin{equation}
    G \cong \left[\left((\mathds{Z}_N^3 \times \mathds{Q}_8) \rtimes \text{SW}_3 \times \text{D}_4\right)/\mathds{Z}_2\right]
        \rtimes \mathds{Z}_2(P),
    \label{eq:GAaron}
\end{equation}
where $\mathds{Z}_2(P)=\{1,P\}$, $\mathds{Q}_8$ is the quaternion group of order~8, $\text{D}_4$ is the dihedral group (also of
order~8), and $\text{SW}_3\subset\text{SO}(3)$ is the cubic rotation group.
The generators of these groups are listed in Table~\ref{tab:genmap}.%
\footnote{Following Ref.~\cite{Golterman:1984dn}, we choose $I_S$ to generate $\text{D}_4$ instead of $\Xi_4$.
Then everything inside the bracket in Eq.~(\ref{eq:GAaron}) is associated with a single timeslice.}
\begin{table}
    \centering
    \caption{The generators of each group appearing in the GTS decomposition of Eq.~(\ref{eq:gtsdecomp}).}
    \label{tab:genmap}
    \begin{tabular}{cc}
        \hline\hline
        Group & Generators \\
        \hline
        $\mathds{Q}_8$ & $\{ \Xi_{12},  \Xi_{23}  \}$ \\
        $\text{SW}_3$  & $\{ R_{12},    R_{23}    \}$ \\
        $\text{D}_4$          & $\{ \Xi_{123}, I_{S}     \}$ \\
        \hline\hline
    \end{tabular}
\end{table}
The $\mathds{Z}_2$ divisor identifies $(-\mathds{1},-\mathds{1}) \in \mathds{Q}_8\times \text{D}_4$ with $(\mathds{1},\mathds{1})$,
leading to the isomorphism $\Gamma_4\cong(\mathds{Q}_8\times \text{D}_4)/\mathds{Z}_2$.

In this paper, we are concerned with the trivial representation of the translation group, namely, zero 3-momentum.
We note in passing, however, that the group theory at nonzero momentum is simpler if the taste generators 
insensitive to rotations are factored as in Eq.~(\ref{eq:GAaron}).

At zero momentum, we are left with the ``geometric rest-frame group''~\cite{Golterman:1984dn}
\begin{equation}
    \text{GRF} = \text{GTS} \times \mathds{Z}_2(P),
    \label{eq:grfdecomp}
\end{equation}
 where the 768-element
 ``geometric timeslice group''~\cite{Golterman:1984dn}
\begin{equation}
    \text{GTS} = \left((\mathds{Q}_8 \rtimes \text{SW}_3) \times \text{D}_4\right)/\mathds{Z}_2.
    \label{eq:gtsdecomp}
\end{equation}
Equation~(\ref{eq:gtsdecomp}) is equivalent to an isomorphism given by Kilcup and Sharpe~\cite{Kilcup:1986dg},
\begin{equation}
    \text{GTS} \cong \left(\text{SW}_4 \times \text{D}_4\right)/\mathds{Z}_2 ,
    \label{eq:gtsdecompKS}
\end{equation}
since $\mathds{Q}_8\rtimes\text{SW}_3$ is isomorphic to $\text{SW}_4$, the rotation group of the four-dimensional hypercube.

Baryon states transform under the ``fermionic'' representations of GTS, namely those that preserve the minus sign in
Eq.~(\ref{eq:SSSS}).
Both $\mathds{Q}_8$ and $\text{D}_4$ have one such irrep, which is two-dimensional and can be expressed as Pauli matrices.
We denote them $\sigma$ and $B$, respectively.
Similarly, fermions obtain a minus sign under $2\pi$ rotations, which is possible with representations of the double cover of
$\text{SW}_3$, $\widetilde{\text{SW}}_3\subset\text{SU}(2)$.
As shown in Table~\ref{tab:octspincontent}, there are three of these~\cite{JOHNSON1982147}, $G_1$, $H=G_1\otimes E$, and
$G_2=G_1\otimes A_2$, where $E$ and $A_2$ are, respectively, the two-dimensional and nontrivial one-dimensional irreps of
$\text{SW}_3$.
The fermionic irreps of GTS are then the tensor products (labeled by their dimension, following Ref.~\cite{Kilcup:1986dg})
\begin{subequations}
    \label{eq:GTSfermionic}
    \begin{align}
        8  &= \sigma \otimes G_1 \otimes B, \\
        8' &= \sigma \otimes G_2 \otimes B, \\
        16 &= \sigma \otimes  H  \otimes B.
    \end{align}
\end{subequations}
From the matrix form of the tensor product, one sees that $\sigma\otimes B$ automatically identifies
$(-\mathds{1},-\mathds{1})\in\mathds{Q}_8\times \text{D}_4$ with $(\mathds{1},\mathds{1})$.

For completeness, we discuss the bosonic representations of GTS, which correspond to even~$F$ in Eq.~(\ref{eq:SSSS}) and no sign for
$2\pi$~rotations.
Because of the $\mathds{Z}_2$ divisor in Eq.~(\ref{eq:GAaron}), these arise from the bosonic representations of all three factors.

$\text{D}_4$ has four 1-dimensional bosonic representations, $A_{\Xi_{123}}^{I_S}$, in which $\pm\Xi_{123}$ and $I_S$ can each be
represented by $\pm1$.
Consequently, for every bosonic irrep of $(\mathds{Q}_8 \rtimes \text{SW}_3)$, four irreps of GTS are induced.
These induced representations are just the tensor products with $A_{\Xi_{123}}^{I_S}$ and, thus, have the same dimension as their
$(\mathds{Q}_8 \rtimes \text{SW}_3)$ counterparts.

To fully classify representations of $(\mathds{Q}_8 \rtimes \text{SW}_3)$, it is easiest to first consider representations of
$\mathds{Q}_8$ and then use the Wigner little-group method to induce the representations of the full group.%
\footnote{For advanced group theory concepts, we refer the reader to Ref.~\cite{zee2016group}.} %
$\mathds{Q}_8$ has four one-dimensional bosonic irreps, which are the trivial representation and three sign representations in which
two of $\Xi_{23}$, $\Xi_{31}$, and $\Xi_{12}$ have character $-1$ (and the third~$+1$).
The trivial representation is in its own orbit, and the latter three for another orbit.
These orbits arise from the way the group elements transform into each other under conjugation with the rotations:
\begin{equation}
    R_{jk}^{-1} \Xi_{ij} R_{jk} \to \pm \Xi_{ik} .
\end{equation}
Physically, the nontrivial bosonic representations act as a 3-vector under rotations.
The vector's direction follows from the signs representing the $\Xi_{ij}$.

The orbits and their little groups, $L\subset\text{SW}_3$, are shown in Table~\ref{tab:littlegroupclass}.
\begin{table}
    \centering
    \caption{
        Structure of GTS irreps~$\gamma$.
        The first column shows the orbits of the $\mathds{Q}_8$ irreps under $\text{SW}_3$.
        The little group $L\subset\text{SW}_3$ and its irreps
         are given in the next two columns.
        (As discussed in the text, the fermionic irrep requires the double cover
          $\widetilde{\rm SW}_3$.)
        The fourth column gives the irrep of the $\text{D}_4$ factor in Eq.~(\ref{eq:GAaron}).
        The next-to-last column gives the dimension $\dim\gamma$ of the induced irreps
        of~GTS.
        The last column gives the number of resulting GTS irreps: in all,
         40 bosonic and 3 fermionic.}
    \label{tab:littlegroupclass}
    \begin{tabular}{cccccc}
        \hline\hline
        $\mathds{Q}_8$ orbit & $L$ & $L$ irreps & $\text{D}_4$ irreps &
            $\dim\gamma$ & $\#(\gamma)$ \\
        \hline
        $\alpha_0$ & $\text{SW}_3$ & $\begin{array}{c}
            A_1, A_2 \\ E \\ T_1, T_2 \end{array}$ & $A_{\Xi_{123}}^{I_S}$ &
            $\begin{array}{c} 1 \\ 2 \\ 3 \end{array}$ &
            $\begin{array}{c} 8 \\ 4 \\ 8 \end{array}$ \\
        \cline{1-6}
        \\[-1.2em] 
        $\vec{\alpha}$ & $\text{D}_4$ &
            $\begin{array}{c}
            A_{R_{23}}^{R^2_{12}}
            \\ B \\ \end{array}$ &
            $A_{\Xi_{123}}^{I_S}$ &
            $\begin{array}{c}  3 \\ 6 \end{array}$ &
            $\begin{array}{c} 16 \\ 4 \end{array}$ \\
        \cline{1-6}
        \\[-1.2em] 
        $\sigma$ & $\widetilde{\rm SW}_3$ & $\begin{array}{c}
            G_1, G_2 \\ H \end{array}$ & $B$ &
            $\begin{array}{c} 8 \\ 16 \end{array}$ &
            $\begin{array}{c} 2 \\  1 \end{array}$ \\
        \hline\hline
    \end{tabular}
\end{table}
Note that the little group~$\text{D}_4$ for the nontrivial 1-dimensional $\mathds{Q}_8$ irreps is generated by,%
\footnote{This~$\text{D}_4$ is not equal to the $\text{D}_4$ of tastes in Eq.~(\ref{eq:GAaron}).} %
e.g., $R_{23}$ and $R_{12}^2$ for the irrep in which the character $\chi(\Xi_{23})=1$ (and $\chi(\Xi_{12})=\chi(\Xi_{31})=-1$).
From this construction, one sees that $(\mathds{Q}_8\times\text{SW}_3)\cong\text{SW}_4$ has 10 bosonic irreps and three fermionic
irreps.
The final step is simple, because GTS is a direct product of $(\mathds{Q}_8\times\text{SW}_3)\cong\text{SW}_4$ with $\text{D}_4$,
but modded out by a $\mathds{Z}_2$, requiring bosonic (fermionic) irreps to be tensored with bosonic (fermionic) irreps.
Thus, GTS has 40 bosonic irreps and three fermionic irreps.

\section{Staggered Lattice Baryon Irrep Identification}
\label{app:irrepident}

The usual strategy to build baryon operators starts with embedding $\text{SU}(2)$ spin and $\text{SU}(3)$ flavor into an
$\text{SU}(6)$ spin-flavor group.
As baryons must obey Fermi statistics, the overall baryon wave function must be antisymmetric.
The antisymmetrization is completely captured by $\text{SU}(3)$ color, so the only needed representations of $\text{SU}(6)$
spin-flavor are those that are overall symmetric.
Decomposition of these symmetric $\text{SU}(6)$ representations back into $\text{SU}(2)_S\times\text{SU}(3)_F$ pairs the symmetric
(mixed-symmetric) representations of $\text{SU}(2)_S$ with the symmetric (mixed-symmetric) representations of $\text{SU}(3)_F$,
giving the usual spin-\onehalf\ octet and spin-\threehalves\ decuplet of physical baryons.

When including the continuum taste symmetry, this can be extended to include the $\text{SU}(4)$ taste symmetry.
Golterman and Smit~\cite{Golterman:1984dn} pursued this strategy without considering flavor, and Bailey~\cite{Bailey:2006zn}
generalized it to include $\text{SU}(3)$ flavor.
Here, we summarize the main steps.

Thus, $\text{SU}(2)_S$, $\text{SU}(3)_F$ and $\text{SU}(4)_T$ are embedded into $\text{SU}(24)$, applying the symmetrization to
combinations of spin, flavor, and taste.
In addition to the usual baryon decuplet and octet, which lie in the symmetric $\text{SU}(4)$ irreps, further states appear in mixed
and asymmetric taste representations combined with mixed and asymmetric spin-flavor representations.
These states have no real-world equivalent, but the $\text{SU}(24)$ embedding demonstrates that they have the same masses and matrix
elements as the physical baryons.

At nonzero lattice spacing, the spin-flavor-taste representations break down into direct sums of GTS irreps.
There are two important consequences.
First, states within a continuum-limit multiplet split into smaller multiplets that differ at order~$a^2$ (or $\alpha_sa^2$ for
tree-level improved actions).
Second, because there are so few GTS irreps, various multiplets can mix, again at order~$a^2$ (or $\alpha_sa^2$).
Excitations of the GTS irreps must, in general, be matched up as the continuum limit is approached to identify higher-spin baryons,
as is familiar elsewhere in spectroscopy~\cite{Edwards:2011jj}.

In the following, the $\text{SU}(N)$ representations are denoted by a number and a subscript, where the number is the dimension
of the representation and the subscript refers to the symmetrization of the representation indices, and can either be symmetric
($S$), mixed-symmetric ($M$), or antisymmetric ($A$).
Subgroups also often carry a subscript for spin ($S$), flavor ($F$), or taste ($T$).
Note that the restriction of a large $\text{SU}(N)$ to smaller $\text{SU}(N)$ subgroups need not be unique.
In all of the following, we use the pattern for $\text{SU}(N)\to \text{SU}(N_1)\times\text{SU}(N_2)$, $N=N_1N_2$, in which an
$\text{SU}(N_1)\times\text{SU}(N_2)$ matrix is the Kronecker product of an $\text{SU}(N_1)$ matrix and an $\text{SU}(N_2)$ matrix.
Thus, this decomposition always starts with and yields only defining representations, i.e., $N\to N_1\otimes N_2$.

Quarks transform under the defining 24-dimensional representation of the $\text{SU}(24)$ embedding group.
The symmetric combination of three fundamental quarks is the representation denoted $2600_S$.
The first step is to separate out the $\text{SU}(2)$ spin group, which yields%
\begin{align}
    \text{SU}(24)_{SFT} &\to \text{SU}(2)_S \times \text{SU}(12)_{FT}, \nonumber\\
        2600_S &\to (4_S, 364_S) \oplus (2_M, 572_M),
    \label{eqn:su24decomp}
\end{align}
where we abbreviate, for example, $4_S\otimes364_S$ by $(4_S,364_S)$.
In Eq.~(\ref{eqn:su24decomp}), $4_S$ ($2_M$) is the usual symmetric (mixed-symmetric) spin $\frac{3}{2}$ ($\frac{1}{2}$)
construction for the baryon decuplet (octet).
Now, however, we have larger multiplets of $\text{SU}(12)_{FT}$.
Because ``flavor'' and ``taste'' are both names for quark species,%
\footnote{Taste and flavor differ crucially at nonzero lattice spacing.} %
the $\text{SU}(12)_{FT}$ symmetry remains, even when these representations are decomposed into
$\text{SU}(3)_F\times\text{SU}(4)_{T}$.
As a consequence, any representation that is formed by decomposing the $(4_S, 364_S)$ representation can be identified with a baryon
from the physical decuplet, and similarly any representations found by decomposing the $(2_M, 572_M)$ irrep can be identified with
baryons from the physical octet.
It is convenient to refer to states in these representations decuplet-like and octet-like, respectively, as a reminder of the
differences with and similarities to the physical decuplet and octet.

Next, the flavor and taste symmetries are separated from each other.
The decomposition of the $572_M$ representation gives
\begin{align}
    \text{SU}(12)_{FT} &\to \text{SU}(3)_F \times \text{SU}(4)_{T}, \nonumber\\
        572_M &\to (8_M, 20_S) \oplus (10_S, 20_M)
    \label{eqn:Ndecomp} \\ & \hspace*{1.5em}
        \oplus (8_M, 20_M)\oplus (8_M, 4_A ) \oplus (1_A, 20_M) .\nonumber
\end{align}
The physical octet is in the taste-symmetric $(8_M, 20_S)$ representation.
The other representations are all a consequence of nontrivial taste symmetry.
As discussed in the main text, they should not be discarded: they are, in fact, useful, in a way similar to the utility of
taste-nonsinglet pions.
Similarly, the decomposition of the $364_S$ representation yields
\begin{align}
    \text{SU}(12)_{FT} &\to \text{SU}(3)_F \times \text{SU}(4)_{T}, \nonumber\\
        364_S &\to (10_S, 20_S) \oplus (8_M, 20_M) \oplus (1_A, 4_A ) .  
    \label{eqn:deltadecomp}
\end{align}
The $(10_S, 20_S)$ is taste symmetric and, thus, identified with the physical decuplet, but the other states are useful once.
Below we relate the $20_S$, $20_M$, and $4_A$ of $\text{SU}(4)_T$ and the $4_S$ and $2_M$ of $\text{SU}(2)_S$ to the irreps of
$\text{GTS}$.

With two equal-mass light quarks and a heavier strange quark, one is interested in the further decomposition from $\text{SU}(3)$
flavor to $\text{SU}(2)$ isospin.
Here, we focus on irreps with zero strangeness.
The $10_S$ and $8_M$ representations of $\text{SU}(3)_F$ in Eqs.~(\ref{eqn:Ndecomp})~and~(\ref{eqn:deltadecomp}) each contain only
one zero strangeness representation:
\begin{align}
    \text{SU}(3)_{F} &\to \text{SU}(2)_F, \nonumber\\
        10_S &\to 4_S \oplus \cdots, \nonumber\\
        8_M  &\to 2_M \oplus \cdots ,
    \label{eq:flavorisospindecomp}
\end{align}
where the ellipses denote representations with nonzero strangeness.
The $4_S$ and $2_M$ are the isospin $\frac{3}{2}$ and $\frac{1}{2}$ representations, respectively.

At this point, we have the group-theoretic ingredients to specify the operators for baryon states labeled by $(S,F,T)$.
To understand the decomposition of these representations into $\text{GTS}$, it is convenient to carry out the decomposition in
several steps.
Each of the subgroups $\mathds{Q}_8$, $\text{SW}_3$, and $\text{D}_4$ that build $\text{GTS}$ has a faithful fermionic
representation generated by Pauli matrices.
As such, identification of each of the subgroups with $\text{SU}(2)$ is useful.%
\footnote{The dihedral group $\text{D}_4\cong\{\pm1,\pm i\sigma_2,\pm\sigma_3,\pm\sigma_1\}\not\subset\text{SU}(2)$, but
$\text{D}_4\subset\Dfour$.
Neither the $\mathds{Z}_4$ factor nor the $\mathds{Z}_2$ identifying $(-\mathds{1},-\mathds{1})$ with $(\mathds{1},\mathds{1})$
affects the structure relating $\text{GTS}$ to $\text{SU}(4)_T\times\text{SU}(2)_S$.} %
One of these factors comes directly from the $\text{SU}(2)_S$ spin in the decomposition of
Eq.~(\ref{eqn:su24decomp}), while the other two are found by decomposing the $\text{SU}(4)$ taste factor into
$\text{SU}(2)_{\mathds{Q}_8}\times\Dfour$.
Under this decomposition, we have the following three representations to consider
\begin{align}
    \text{SU}(4)_{T} &\to \text{SU}(2)_{\mathds{Q}_8} \times \Dfour , \nonumber\\
    20_S &\to (4_S, 4_S) \oplus (2_M, 2_M) , \nonumber\\
    20_M &\to (2_M, 4_S) \oplus (4_S, 2_M) \oplus (2_M, 2_M) , \nonumber\\
    4_A  &\to (2_M, 2_M) .
    \label{eq:tastedecomp1}
\end{align}
For brevity, we do not introduce a label for the $\mathds{Z}_4$ quantum number.
The nontrivial element is $\pm i$ ($\pm1$) in fermionic (bosonic) representations, with the sign modded out by~$\mathds{Z}_2$.

To mimic $\text{GTS}$, the quaternion factor $\text{SU}(2)_{\mathds{Q}_8}$ should be in a semidirect product with something
corresponding to the lattice rotation group, which we denote $\text{SU}(2)_{\text{SW}_3}$.
In the continuum limit, however, spin and taste commute, namely $\text{SU}(2)_S\times\text{SU}(4)_T$; cf.\
Eqs.~(\ref{eqn:su24decomp}) and~(\ref{eqn:Ndecomp}).
It is possible to arrive and the desired structure by noting
\begin{align}
    \text{SO}(4) &\cong \text{SU}(2)_{\mathds{Q}_8} \times \text{SU}(2)_S , \nonumber \\
                 &\cong \text{SU}(2)_{\mathds{Q}_8}\rtimes \text{SU}(2)_{\text{SW}_3} ,
    \label{eq:SO4SU2SU2}
\end{align}
where $\text{SO}(4)\supset\text{SW}_4$ of Eq.~(\ref{eq:gtsdecompKS}).
If the generators of $\text{SU}(2)_{\mathds{Q}_8}$ and $\text{SU}(2)_S$ are $\bm{\tau}$ and $\bm{\Sigma}$, respectively, then the
generators of $\text{SU}(2)_{\text{SW}_3}$ are $\bm{\sigma}\equiv\bm{\Sigma}+\bm{\tau}$.
Although $\bm{\tau}$ and $\bm{\Sigma}$ commute, one finds the desired behavior of the tastes under lattice rotations:
$[\sigma_i,\tau_j]=2i\varepsilon_{ijk}\tau_k$.

In summary, to mimic $\text{GTS}$ with $\text{SU}(2)$ groups,
\begin{equation}
    \text{GTS} \subset \left(\text{SU}(2)_{\mathds{Q}_8} \times \text{SU}(2)_S \times \Dfour \right)/\mathds{Z}_2;
    \label{eq:GTSASK}
\end{equation}
the last $\mathds{Z}_2$ is the same as the $\mathds{Z}_2$ factor in Eq.~(\ref{eq:gtsdecomp}).
The remainder of this section focuses on decomposing the various $\text{SU}(2)$ group factors down into their discrete lattice
subgroups, being mindful that $\text{SU}(2)_{\mathds{Q}_8} \times \text{SU}(2)_S$ in Eq.~(\ref{eq:GTSASK}) is isomorphic to
$\text{SU}(2)_{\mathds{Q}_8}\rtimes \text{SU}(2)_{\text{SW}_3}$, as shown in Eq.~(\ref{eq:SO4SU2SU2}).
In this way we derive the full map from $\text{SU}(24)$,
 where it is easiest to construct operators
obeying Fermi statistics, to the $\text{GTS}$ symmetry of staggered lattice fermions.

The decompositions of all irreps to this point have resulted in just two $\text{SU}(2)$ representations: $2_M$ and $4_S$.
It is important to keep track of which subgroup factor each representation belongs to.
In the interest of clarity, the subduction of these subgroup factors are listed for each subgroup factor individually,
with a guide for assembling the individual subduction patterns into 
$\text{GTS}$~irreps.
The end result of this assembly yields the subduction of Eq.~(\ref{eq:GTSASK}), summarized in Eq.~(\ref{eq:decomptogts}).

The $\Dfour$ factor is separated from the other products by a direct product, and so may be considered independently.
Subduction to $\text{D}_4$ can yield only one fermionic representation, and so all fermionic representations must subduce to
multiples of this irrep.
This gives
\begin{align}
    \Dfour &\to \text{D}_4, \nonumber \\
        2_M &\to B, \nonumber \\
        4_S &\to B \oplus B.
    \label{eq:su2d4decomp}
\end{align}
Thus, when $4_S$ of \Dfour\ appears in Eq.~(\ref{eq:su2d4decomp}), the irreps subduced from the other groups appear twice in 
the subduction to~$\text{GTS}$.

Eq.~(\ref{eq:GTSASK}) demonstrates that the taste and spin representations may be considered separately in the continuum.
However, the presence of the semidirect product in Eq.~(\ref{eq:gtsdecomp}), $\mathds{Q}_8\rtimes \text{SW}_3$, means that the
lattice generators of rotations mix up the discrete taste transformations.
It is instructive to trace the subduction from the continuum $\text{SU}(2)_{\mathds{Q}_8} \rtimes \text{SU}(2)_{\text{SW}_3}$ down
to the discrete subgroup $\mathds{Q}_8\rtimes \text{SW}_3$ to see how the spin and taste degrees of freedom become mixed up by the
discretization.

The first step in the subduction is to identify how the direct product is converted to a semidirect product in
Eq.~(\ref{eq:SO4SU2SU2}).
The semidirect product is enforced by replacing $\bm{\Sigma}$ with $\bm{\sigma}$, to arrive at the rotation group
$\text{SU}(2)_{\text{SW}_3}$, which acts on both spin and taste.
Since the two groups in Eq.~(\ref{eq:SO4SU2SU2}) are isomorphic, their irreps must be in a one-to-one correspondence.
In addition, to preserve the representations' dimensions, the semidirect product cannot mix different irreps, and the little group
is nothing but the entire $\text{SU}(2)_{\text{SW}_3}$ group in all cases.
This means that the mapping between irreps of the direct product and the semidirect product is trivial,
\begin{align}
    \text{SU}(2)_{\mathds{Q}_8} \times \text{SU}(2)_{S} &\to \text{SU}(2)_{\mathds{Q}_8} \rtimes \text{SU}(2)_{\text{SW}_3},
    \nonumber\\
        (2_M, 2_M) &\to (2_M, 2) ,\nonumber\\
        (2_M, 4_S) &\to (2_M, 4) ,\nonumber\\
        (4_S, 2_M) &\to (4_S, 2) ,\nonumber\\
        (4_S, 4_S) &\to (4_S, 4) ,
\end{align}
where for clarity below, we omit the second subscript when referring to the semidirect product.

To understand the decomposition of the semidirect product group, it is not sufficient to break the two $\text{SU}(2)$ subgroups into
$\mathds{Q}_8$ and $\text{SW}_3$, respectively, but the separate breakings provide a useful ingredient.
The $\mathds{Q}_8$ group factor has only one fermionic irrep, $\sigma$, and so all fermionic irreps of $\text{SU}(2)_{\mathds{Q}_8}$
must break into copies of that irrep,
\begin{align}
    \text{SU}(2)_{\mathds{Q}_8} &\to \mathds{Q}_8 ,\nonumber\\
        2_M &\to \sigma ,\nonumber\\
        4_S &\to \sigma \oplus \sigma .
\end{align}
The $\text{SW}_3$ factor has three fermionic irreps, and the decomposition of the relevant $\text{SU}(2)_{\text{SW}_3}$ irreps is
simple,
\begin{align}
    \text{SU}(2)_{\text{SW}_3} &\to \text{SW}_3, \nonumber\\
        2 &\to G_1, \nonumber\\
        4 &\to H .
 \label{eq:su2sw3decomp}
\end{align}

The key point is that the semidirect product also induces a rotation of the copies of the $\sigma$ irreps into
each other.
This additional transformation acts as an irrep of the $\text{SW}_3$ rotation group, and is combined as a tensor product with the
irrep resulting from the direct decomposition of the $\text{SU}(2)_{\text{SW}_3}$ factor.
We can write this as an additional irrep factor belonging to the $\text{SW}_3$ group instead of as an uninformative multiplicative
factor on the number of irreps,
\begin{align}
    \text{SU}(2)_{\mathds{Q}_8} &\to \mathds{Q}_8 (\rtimes \text{SW}_3), \nonumber\\
        2_M &\to \sigma (\otimes A_1) , \nonumber\\
        4_S &\to \sigma (\otimes E  ) .
\end{align}
Combining this together with the irreps in Eq.~(\ref{eq:su2sw3decomp}), we get the full decomposition
\begin{align}
    \text{SU}(2)_{\mathds{Q}_8} \rtimes \text{SU}(2)_{\text{SW}_3} &\to \mathds{Q}_8 \rtimes \text{SW}_3, \nonumber\\
        (2_M, 2) &\to (\sigma, A_1 \otimes G_1) = (\sigma,G_1), \nonumber\\
        (2_M, 4) &\to (\sigma, A_1 \otimes H  ) = (\sigma,H  ), \nonumber\\
        (4_S, 2) &\to (\sigma, E   \otimes G_1) = (\sigma,H  ), \nonumber\\
        (4_S, 4) &\to (\sigma, E   \otimes H  ) \nonumber\\
        &\hspace*{1em}= (\sigma,G_1) \oplus (\sigma,G_2) \oplus (\sigma,H) .
 \label{eq:su2q8sw3decomp}
\end{align}

We have now completed the decomposition of Eq.~(\ref{eq:GTSASK}).
Combining the identifications of the irreps (Eq.~(\ref{eq:GTSfermionic})) with the decompositions in
Eqs.~(\ref{eq:su2d4decomp})~and~(\ref{eq:su2q8sw3decomp}), we get
\begin{align}
    (\text{SU}(2)_{\mathds{Q}_8} \times \text{SU}(2)_{S} \times &\Dfour ) /\mathds{Z}_2 \to \text{GTS}, \nonumber\\
        (2_M, 2_M, 2_M) &\to  8_N, \nonumber\\
        (4_S, 2_M, 2_M) &\to 16_N, \nonumber\\
        (2_M, 4_S, 2_M) &\to 16_\Delta, \nonumber\\
        (4_S, 4_S, 2_M) &\to 8_\Delta  \oplus 8'_\Delta \oplus 16_\Delta , \nonumber\\
        (2_M, 2_M, 4_S) &\to 2\times ( 8_N ), \nonumber\\
        (4_S, 2_M, 4_S) &\to 2\times (16_N ) ,\nonumber\\
        (2_M, 4_S, 4_S) &\to 2\times (16_\Delta), \nonumber\\
        (4_S, 4_S, 4_S) &\to 2\times (8_\Delta \oplus 8'_\Delta \oplus 16_\Delta ) .
 \label{eq:decomptogts}
\end{align}
Since the particle content, either $N$ or $\Delta$, is determined by the $\text{SU}(2)_{S}$ irrep, an additional subscript has been
added to the $\text{GTS}$ irreps.
Additionally, the two choices of $\Dfour$ irrep result only in a different multiplicity in the number of irreps, not in the irreps
that appear.

\section{Staggered Baryon Operators in the 16~Irrep}
\label{appendix:staggeredop}

In this appendix, we give specific details about the operators for the 16~irrep.
The 16 components are related by $\text{GTS}$ symmetry, and here we construct an interpolating operator for each component.
As described in Sec.~\ref{subsec:OpConstruct}, these 16 components split into two sets of eight components.
We conventionally construct the 16-irrep elements such that the corresponding baryon operators at the origin are eigenstates under a
$\pi/2$ $z$-axis rotation with eigenvalue~$s=\pm1$.
The remaining 16-irrep elements can be obtained by applying shifts to take these two rotation-eigenstate operators to each of the
remaining unit cube sites.
As also discussed in Sec.~\ref{subsec:OpConstruct}, there are four different classes of operators that one can construct in the
$I=\frac{3}{2}$ 16~irrep.
No class of the 16~irrep can be transformed into another by a $\text{GTS}$ symmetry.
Each class thus serves as a different operator construction, and each effectively gives a different overlap with the nucleon
wave function.

In the following, we give the operator coefficients
$\mathcal{O}^{S,16,C}_{s\vec{D},\vec{A}\vec{B}\vec{C}}$ appearing in Eq.~(\ref{eq:baryop}).
Here, $S$ denotes the totally-symmetric representation of both spin-taste and isospin, to distinguish these operators from
mixed-symmetry representations which may be studied in the future; $C$ denotes the class introduced by Golterman and
Smit~\cite{Golterman:1984dn} (in the 16~irrep, $C\in\{2,3,4,6\}$); and $\vec{D}$ is an
unspecified index that can have one of eight different values.
As the eight different $\vec{D}$ components are related by a shift symmetry, we fix $\vec{D}$ and only give a single operator within
this set.
The other seven operators within this set can be generated with the nontrivial shift symmetry transformations.

\begin{figure*}[t]
    \centering
    \includegraphics[trim=65 340 45 340,clip,width=0.95\textwidth]{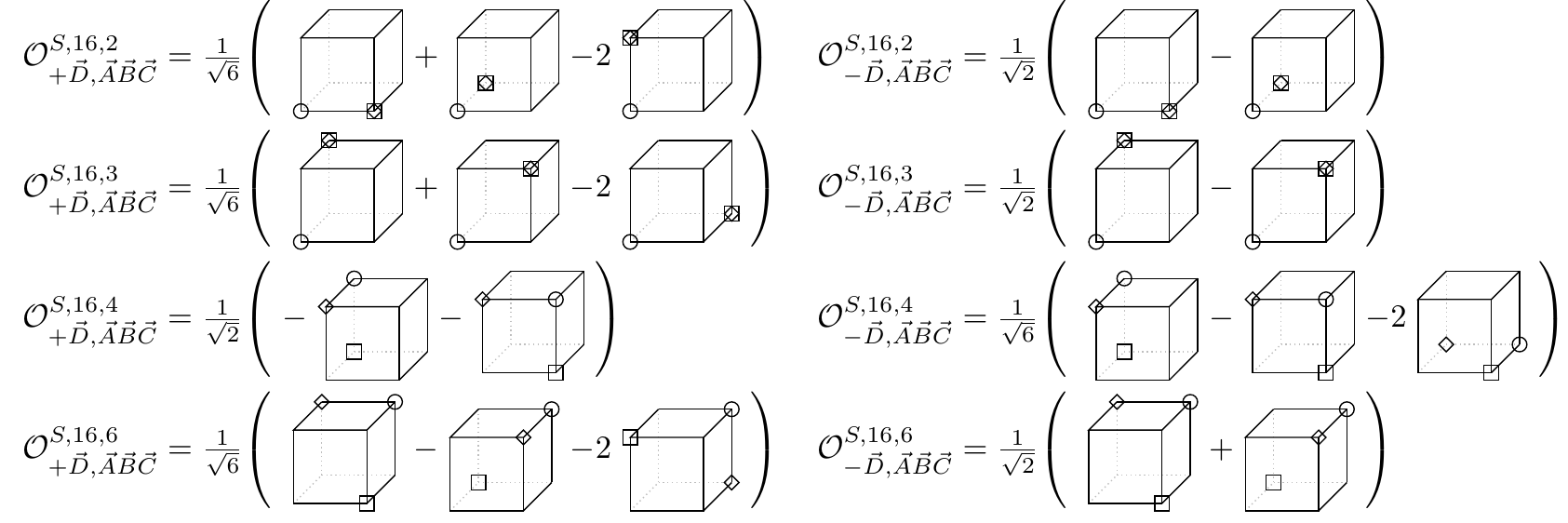}
    \caption{ Diagrams of the operators listed in Eqs.~(\ref{eq:GBsxtpStw})--(\ref{eq:GBsxtdSsx}).
    As discussed in the main text, this representation of GTS permits four classes
    of operators that can be constructed: classes~2, 3, 4, and 6.
    The locations of the three quarks within the unit cube are depicted with a
    circle, square, and diamond.}
    \label{fig:boxes}
\end{figure*}

Equations~(\ref{eq:GBsxtpStw})$-$(\ref{eq:GBsxtdSsx}) give all the operators we use that are unrelated by a shift symmetry, e.g.,
one choice for each value of $s$ and $C$.
A hat over multiple letters is shorthand for the sum of unit vectors in each of those directions, e.g., and
$\widehat{xy}=\hat{x}+\hat{y}$.
Quarks on the site $\widehat{xz}$ conventionally appear with an extra minus sign so they respect the shift and rotation symmetry
operations.
The operators are written with only the positive directions, but symmetrization over all combinations of positive and negative
directions is implied.
These operators are also shown diagrammatically in Fig.~\ref{fig:boxes}.
We have found empirically that correlators with the class-3 operator $\mathcal{O}^{S,16,3}$ are noisier than the others and so we
have excluded them from the analysis.
This operator can only couple to the $N$-like state at order $\alpha_sa^2$
 (or order $a^2$ in actions without the Naik term).
Unlike the other operator classes, the class-3 operator has all three quarks on even sites, which can only be subduced from the
completely symmetric $4_S$ representation of $\Dfour$.
For the isospin-$\frac{3}{2}$ operators in the 16 irrep, the $N$-like state comes from a $2_M$ irrep of $\Dfour$, and so its
contribution to the class-3 operator must vanish in the continuum limit upon restoration of taste symmetry.

\widetext
\begin{align}
 \GBsxtpStw \\
 \GBsxtdStw \\
 \GBsxtpSth \\
 \GBsxtdSth \\
 \GBsxtpSfr \\
 \GBsxtdSfr \\
 \GBsxtpSsx \\
 \GBsxtdSsx
\end{align}

\section{Additional Data}
\label{appendix:moredata}

\begin{figure*}
	\centering
	\includegraphics[width=\columnwidth]{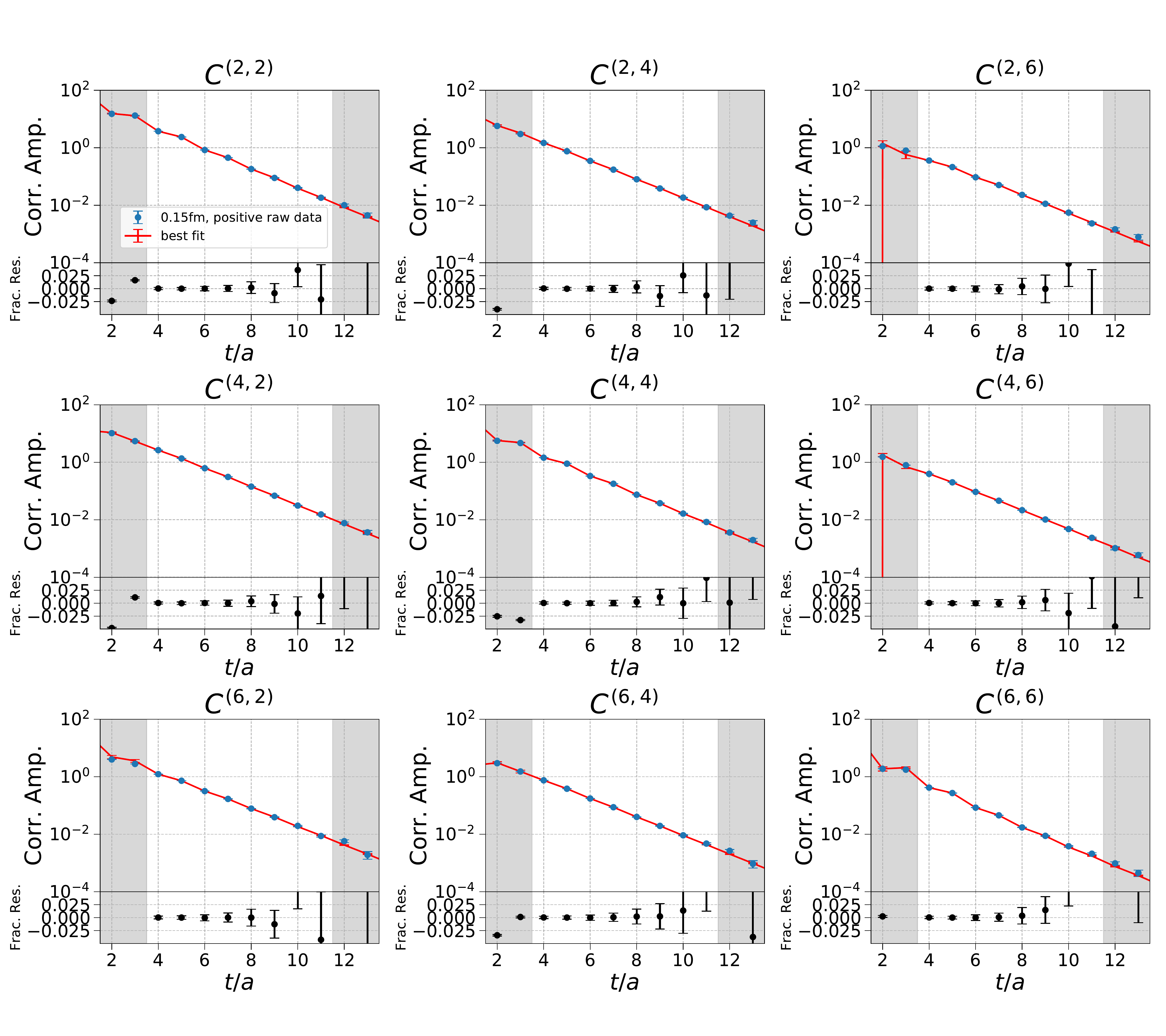}
	\caption{Similar to Fig.~\ref{fig:corr012} but for the $0.15$~fm ensemble.}
	\label{fig:corr015}
\end{figure*}

\begin{figure*}[p]
	\centering
	\includegraphics[width=\columnwidth]{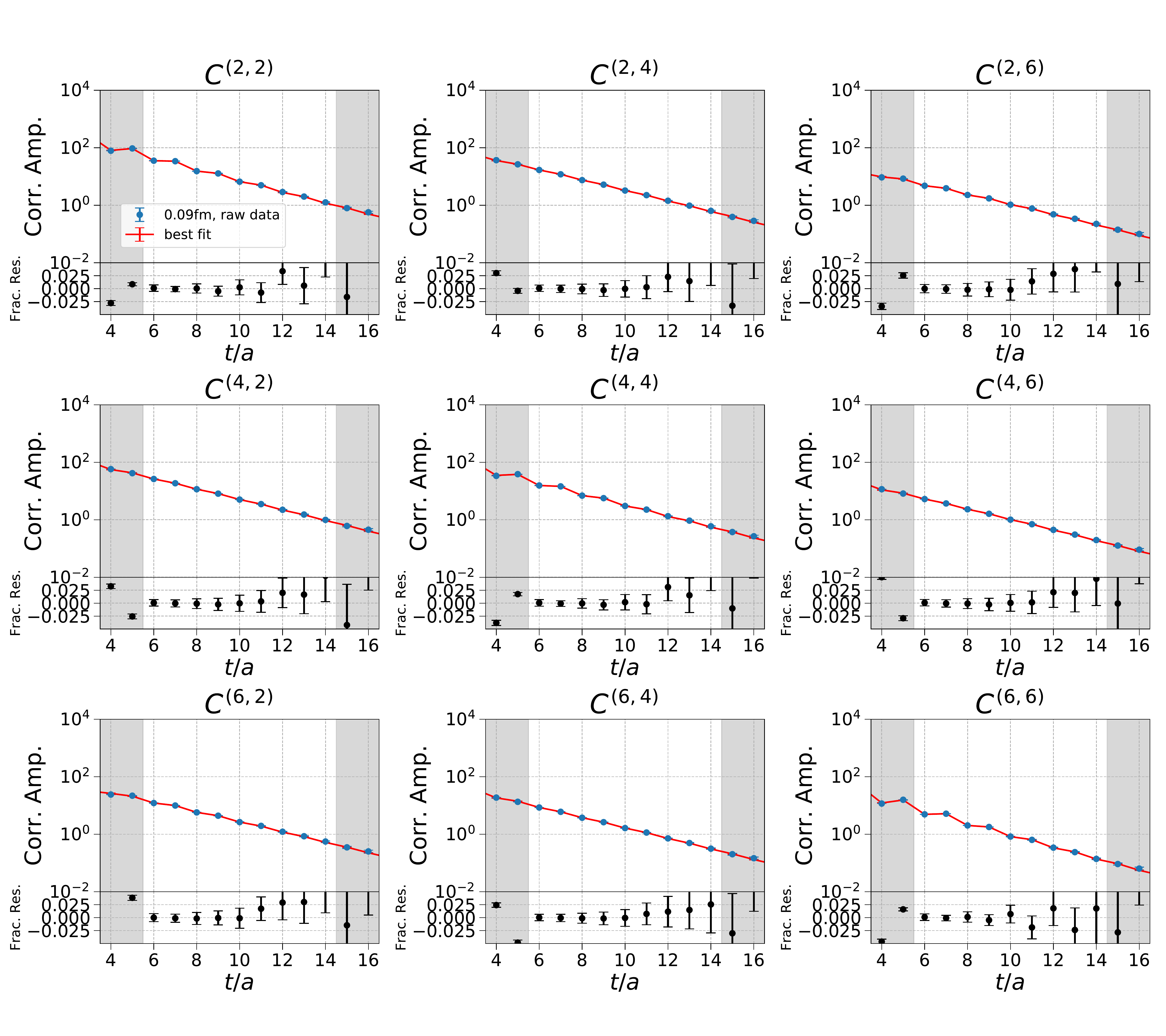}
	\caption{Similar to Fig.~\ref{fig:corr012} but for the $0.09$~fm ensemble.}
	\label{fig:corr009}
\end{figure*}

\begin{figure*}[p]
	\centering
	\includegraphics[width=\columnwidth]{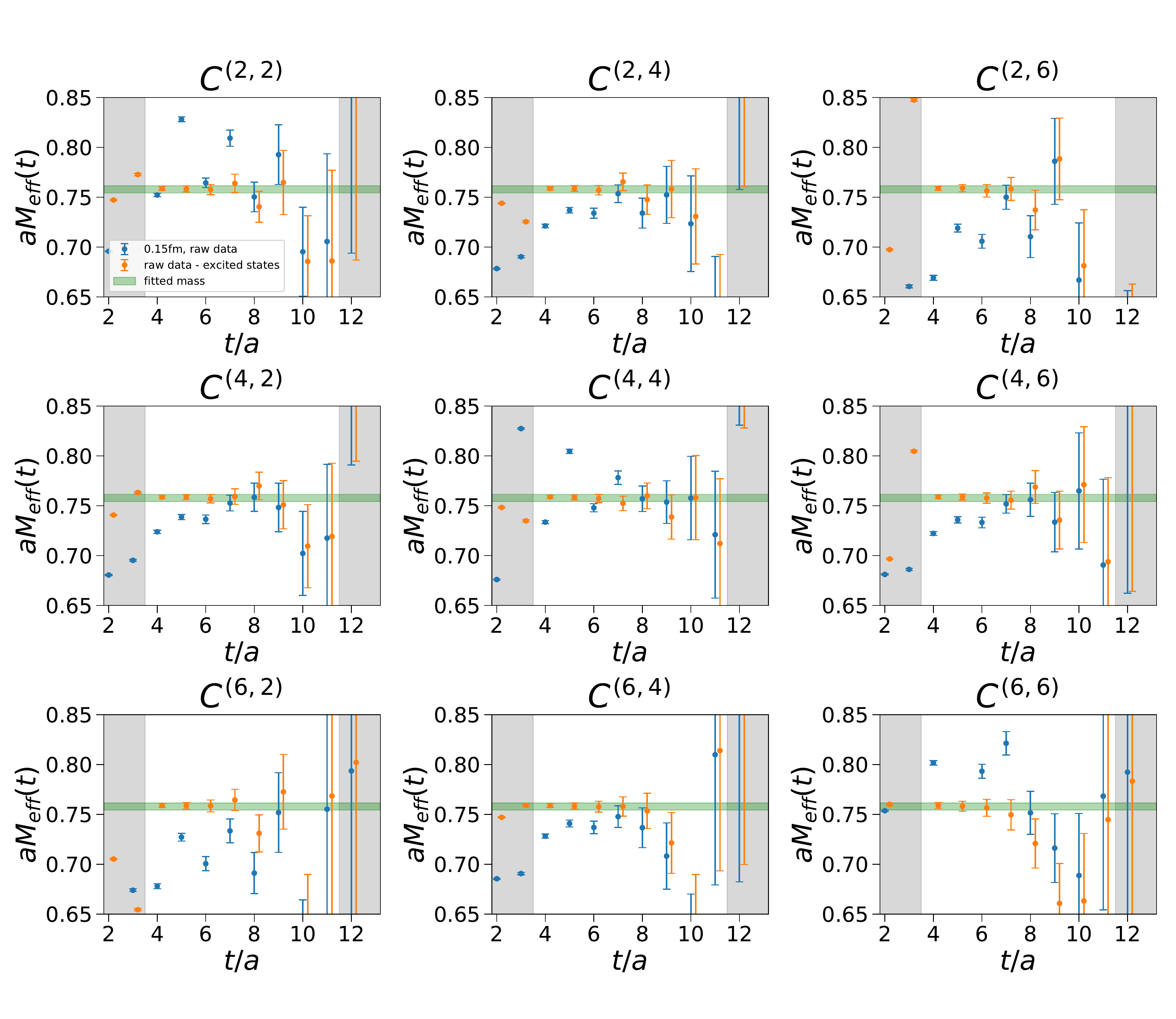}
	\caption{Similar to Fig.~\ref{fig:meff012} but for the $0.15$~fm ensemble.}
	\label{fig:meff015}
\end{figure*}

\begin{figure*}[p]
	\centering
	\includegraphics[width=\columnwidth]{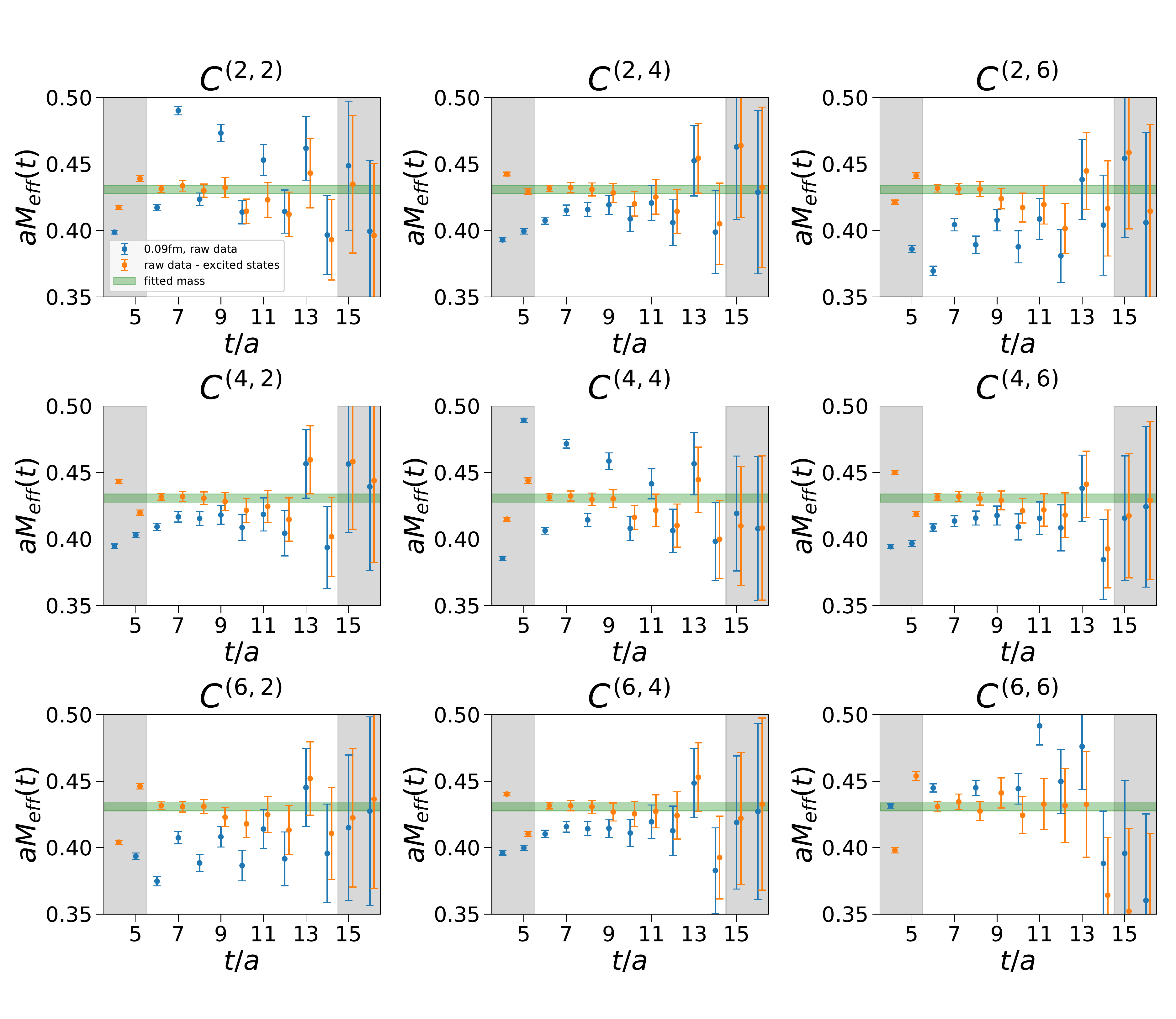}
	\caption{Similar to Fig.~\ref{fig:meff012} but for the $0.09$~fm ensemble.}
	\label{fig:meff009}
\end{figure*}

\begin{figure*}
	\centering
	\includegraphics[width=0.47\columnwidth]{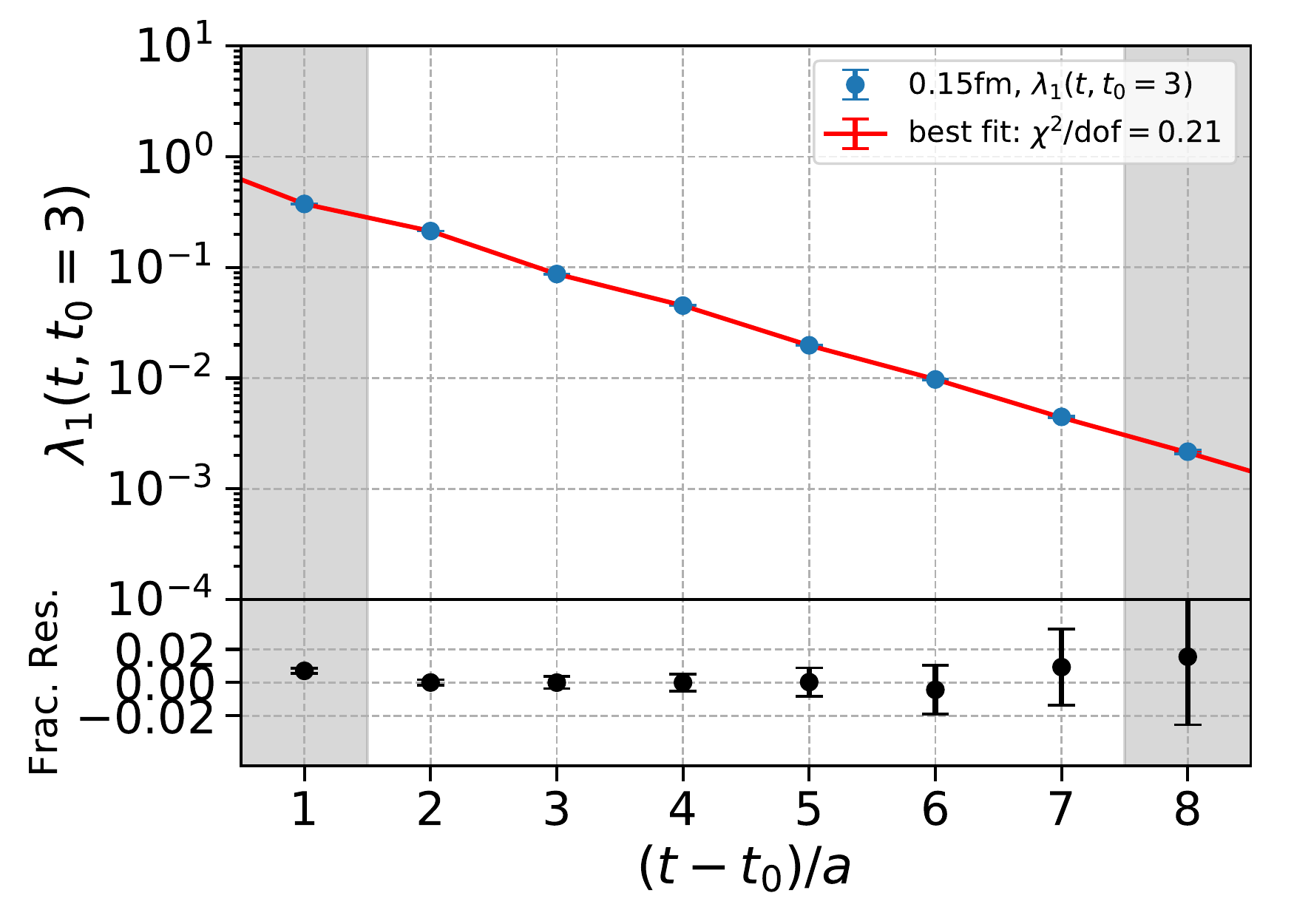} \hfill
    \includegraphics[width=0.47\columnwidth]{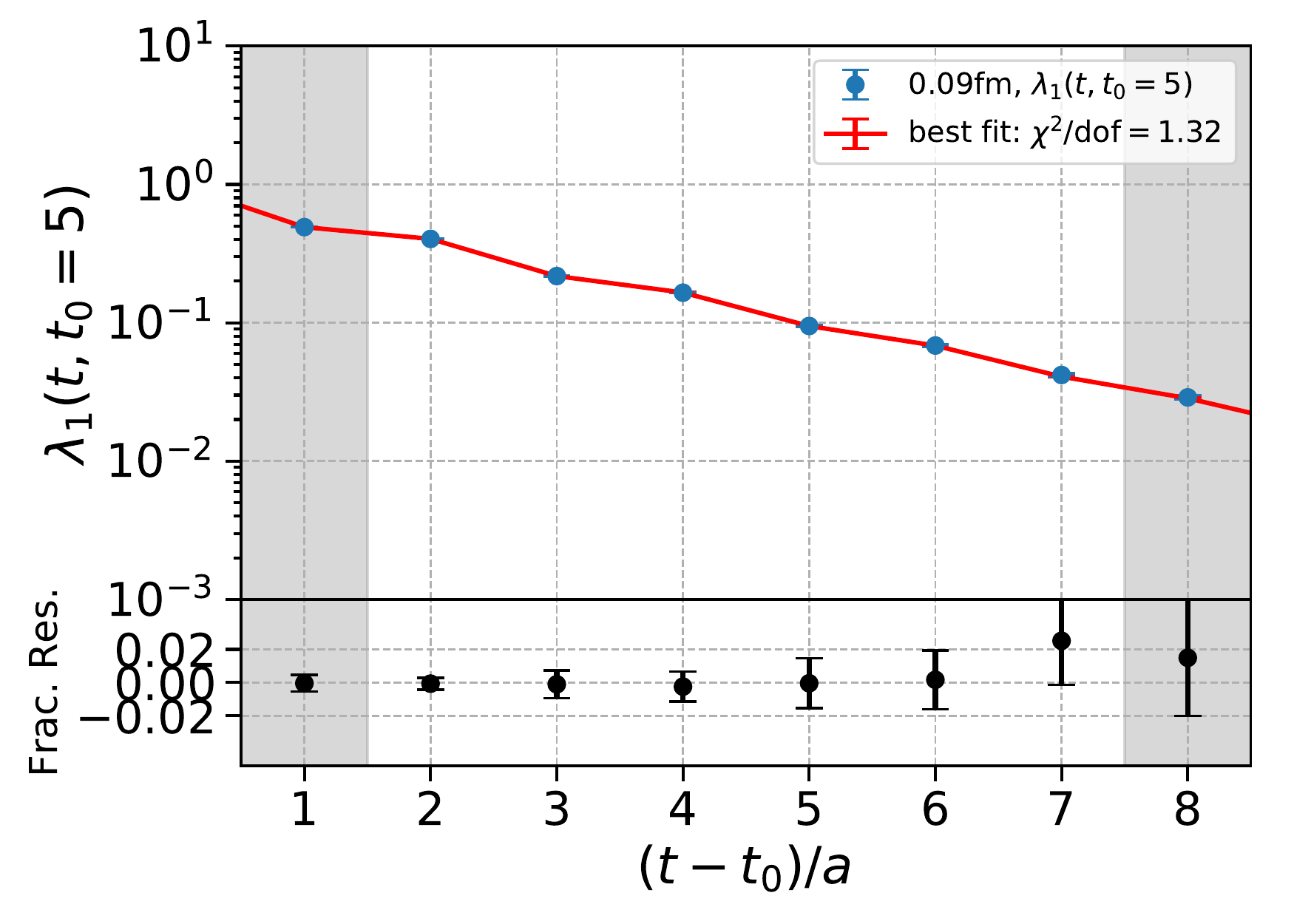}
	\caption{Similar to Fig.~\ref{fig:gevp012} but for the $0.15$~fm ensemble (left) and the $0.09$~fm ensemble (right).}
	\label{fig:gevp-other}
\end{figure*}

In this Appendix, we provide additional data for the other ensembles that are not given in the main text.
Table~\ref{tab:detailpos} gives detailed prior and posterior information from the Bayesian fits of Sec.~\ref{sec:DataAnalysis}.
Figures~\ref{fig:corr015}~and~\ref{fig:corr009} give the fit curves and residuals from the joint fit with Bayesian priors, analogous
to Fig.~\ref{fig:corr012}.
Figures~\ref{fig:meff015}~and~\ref{fig:meff009} give the effective masses of correlators both before and after subtraction of
excited states, analogous to Fig.~\ref{fig:meff012}.
Figure~\ref{fig:gevp-other} gives the lowest eigenvalue after applying the GEVP in Eq.~(\ref{eq:gevpright}), analogous to
Fig.~\ref{fig:gevp012}.
\begin{table*}[hp]
    \centering
    \setlength{\tabcolsep}{5pt}

    \caption{Bayesian fit priors and posteriors from fits to Eqs.~(\ref{eq:corr_raw}), (\ref{eq:deltaprime}), 
        and~(\ref{eq:corr_final}) on the $0.15$~fm, $0.12$~fm, and $0.09$~fm ensembles.
        All posterior uncertainties are estimated with $1000$ bootstrap samples.
        The symbol ``$\delta$'' denotes that the mass difference $M_i - M_{i-1}$ is fit instead of the absolute mass~$M_i$.
        For the even-parity sector, $M_1=M_N$, and the masses in order of increasing $i$ are $M_{\Delta'_1}$, $M_{\Delta'_2}$, and
        $M_{r,1}$.
        In the odd-parity sector, the masses $M_i$ correspond directly to $M_{-,i}$.}
    \label{tab:detailpos}
\begin{tabular}{lllllllll}
\hline\hline
  & $aM_N$ & $a\delta M_{\Delta'_1}$ & $a\delta M_{\Delta'_2}$ & $a\delta M_{r,1}$ & $aM_{-,1}$ & $a\delta M_{-,2}$ & $a\delta M_{-,3}$ &
    $a\delta M_{-,4}$ \\
\hline
$0.15$~fm prior     & $0.715 (40)$ & $0.220 (76)$ & $0.114 (38)$ & $0.30 (15)$ & $1.06 (15)$ & $0.150(75)$ & $0.150 (75)$ & $0.30 (15)$ \\
$0.15$~fm posterior & $0.7582(30)$ & $0.157 (18)$ & $0.107 (12)$ & $0.311(75)$ & $0.938(39)$ & $0.121(10)$ & $0.132(29)$  & $0.303(52)$ \\
$0.12$~fm prior     & $0.572 (30)$ & $0.176 (60)$ & $0.06  (3)$  & $0.24 (12)$ & $0.85 (12)$ & $0.12 (6)$  & $0.12 (6)$   & $0.24 (12)$ \\
$0.12$~fm posterior & $0.5954(27)$ & $0.210 (31)$ & $0.057 (13)$ & $0.236(33)$ & $0.783(44)$ & $0.128(36)$ & $0.117(27)$  & $0.264(47)$ \\
$0.09$~fm prior     & $0.430 (22)$ & $0.132 (45)$ & $0.028 (28)$ & $0.18 (9)$  & $0.64 (9)$  & $0.090(45)$ & $0.090 (45)$ & $0.18 (9)$  \\
$0.09$~fm posterior & $0.4308(30)$ & $0.143 (22)$ & $0.033 (16)$ & $0.191(42)$ & $0.627(21)$ & $0.089(21)$ & $0.088(22)$  & $0.182(23)$ \\
\hline\hline
\end{tabular}
\end{table*}

\clearpage

\endwidetext

\bibliographystyle{apsrev4-1}
\bibliography{main,usqcd-wp}

\end{document}